\documentclass[a4,11pt]{article}

\usepackage[utf8x]{inputenc}
\usepackage{ucs}
\usepackage{xcolor} 
\usepackage[numbers]{natbib}
\usepackage{amsmath}
\usepackage{amsfonts}
\usepackage{amssymb}
\usepackage{latexsym}
\usepackage{graphicx}
\usepackage{lineno}

\usepackage[position=bottom]{subfig}
\usepackage[margin=2.0cm]{geometry}
\usepackage{enumitem}
\usepackage[title]{appendix}
\usepackage{hyperref}
\hypersetup{
    colorlinks = true,
    urlcolor   = black,
    citecolor  = black,
}

\DeclareTextFontCommand\textsfbi{\usefont{OT1}{phv}{b}{it}}
\DeclareMathAlphabet\mathsfbi            {OT1}{phv}{b}{it}

\definecolor{myblue}{rgb}{0 0 1}
\definecolor{Oxblood}{rgb}{0.6 0 0}
\definecolor{myred}{rgb}{0.7412 0.2588 0.2588}
\definecolor{mypurple}{rgb}{0.4 0 1}
\usepackage{tikz,siunitx}
\usetikzlibrary{shapes.geometric,shapes.symbols}
\usepackage{gensymb}
\newrobustcmd*{\hexagram}[1]{\tikz{\draw[thick, draw=#1, fill=white] (0cm, 0cm) -- (0.02887cm, -0.05cm) -- (0.08661cm, -0.05cm) -- (0.05723cm, -0.1cm) -- (0.08661cm, -0.15cm) -- (0.02887cm, -0.15cm) -- (0cm, -0.2cm) -- (-0.02887cm, -0.15cm) -- (-0.08661, -0.15) -- (-0.05723, -0.1) -- (-0.08661, -0.05) -- (-0.02887, -0.05) -- cycle;}}
\newrobustcmd*{\emptytriangle}[1]{\tikz{\draw[thick, draw=#1] (0,0) --
(0,-0.18cm) -- (0.1559cm, -0.09cm) -- cycle;}}
\newrobustcmd*{\emptydiamond}[1]{\tikz{\draw[thick, draw=#1] (0,0) --
(0.0684cm,-0.103cm) -- (0cm, -0.205cm) -- (-0.0684cm,-0.103cm) -- cycle;}}

    \makeatletter
\def\@fnsymbol#1{\ensuremath{\ifcase#1\or \dagger\or \ddagger\or
   \mathsection\or \mathparagraph\or \|\or **\or \dagger\dagger
   \or \ddagger\ddagger \else\@ctrerr\fi}}
    \makeatother

\begin{document}


\normalsize	
\author
{Paria Makaremi-Esfarjani%
  \thanks{Email address for correspondence:
    paria.makaremiesfarjani@mail.mcgill.ca}$\,$, Andrew J. Higgins\thanks{Email address for correspondence:
    andrew.higgins@mcgill.ca} $\,$, Alireza Najafi-Yazdi\\
  Department of Mechanical Engineering, McGill University,\\ Montr\'{e}al, Qu\'{e}bec, Canada}

\date{}

\title{A level set-based solver for two-phase incompressible flows: Extension to magnetic fluids}

{\let\newpage\relax\maketitle}

\begin{abstract}
\small
Development of a two-phase incompressible solver for magnetic flows in the magnetostatic case is presented. The proposed numerical toolkit couples the Navier--Stokes equations of hydrodynamics with Maxwell's equations of electromagnetism to model the behaviour of magnetic flows in the presence of a magnetic field. To this end, a rigorous implementation of a second-order two-phase solver for incompressible nonmagnetic flows is introduced first. This solver is implemented in the finite-difference framework, where a fifth-order conservative level set method is employed to capture the evolution of the interface, along with an incompressible solver based on the projection scheme to model the fluids. The solver demonstrates excellent performance even with high density ratios across the interface (Atwood number $\approx 1$), while effectively preserving the mass conservation property. Subsequently, the numerical discretization of Maxwell's equations under the magnetostatic assumption is described in detail, utilizing the vector potential formulation. The primary second-order solver for two-phase flows is extended to the case of magnetic flows, by incorporating the Lorentz force into the momentum equation, accounting for high magnetic permeability ratios across the interface. The implemented solver is then utilized for examining the deformation of ferrofluid droplets in both quiescent and shear flow regimes across various susceptibility values of the droplets. The results suggest that increasing the susceptibility value of the ferrofluid droplet can affect its deformation and rotation in low capillary regimes. In higher capillary flows, increasing the magnetic permeability jump across the interface can further lead to droplet breakup as well. The effect of this property is also investigated for the Rayleigh--Taylor instability growth in magnetic fluids.
\end{abstract}

Incompressible flow\\
Magnetostatic\\
Magnetic flows\\
Level set method\\
Ferrofluid droplets\\
Rayleigh--Taylor instability\\

\section{\label{sec:introduction} Introduction}
Modelling multi-phase (interfacial) flows involves simulating systems with the presence of two or more immiscible fluids with different physical properties and distinguishable interfaces. Multi-phase flows, particularly two-phase flows, have garnered substantial interest in various applications, including spray atomization~\citep{Desjardins2008_1}, bubbly flows~\citep{Clift2005}, and nuclear reactors~\citep{radman2021}. The complexity of analytically and experimentally studying the physics of two-phase flows~\citep{prosperetti2009} has necessitated the development of accurate, cost-effective, and consistent numerical methods. Numerous studies have been conducted to establish a comprehensive numerical toolkit in the field. 
Despite these efforts, developing a numerical solver to study two-phase flows remains a challenging task. The difficulty originates from modelling fluid property discontinuities across thin interfaces, particularly when large density ratios are present. The inability to accurately capture discontinuities in fluid properties can result in numerical instabilities. Errors arising from the inadequate discretization of fluid property discontinuities can become particularly pronounced for large density ratios, thereby limiting simulations to low density ratios. However, most realistic problems of interest involve large density ratios, such as the formation and dynamics of bubbles, molten metal flows in atmospheric air, gas entrainment in liquid phases, and the aerodynamic effects of gas on the liquid phase~\citep{bussmann2002}. Enforcing mass and momentum conservation is also crucial to obtain physical results and to avoid numerical instabilities in simulations. Therefore, the numerical solver must be able to provide consistent mass and momentum exchange across interfaces throughout the simulation. It is essential for the two-phase solver to be able to address topology changes of the interface and accommodate a wide range of time and length scales as well. In numerical solvers for two-phase flows, it is crucial to properly couple the governing equations of fluids with an appropriate interface-tracking method.

Existing techniques for simulating two-phase flows can be grouped into four categories: the lattice Boltzmann method (LBM), smoothed particle hydrodynamics (SPH), two-fluid, and one-fluid models. The first two methods examine the behaviour of the fluid by representing it as a collection of particles~\citep{li2022fractional, lai2023lattice, fonty2019mixture, cui2021multiphase}, while the last two approaches assume the fluid as a continuum medium that can be described by solving the Navier--Stokes equations. The two-fluid method treats the two phases as separate fluids that interact with each other. This model has demonstrated success in simple problems but has been shown to be inadequate for more complex scenarios~\citep{prosperetti2009}. In this study, we adopt the one-fluid formulation and implement it on a fixed Eulerian grid to develop a two-phase incompressible numerical solver. The Navier--Stokes equations are solved for the entire computational domain, accounting for the density and viscosity jump at the interface while implicitly imposing appropriate boundary conditions across the interface separating the two fluid regions. Among different numerical schemes for treating high density ratios across the interface and modelling surface tension forces, the ghost fluid method (GFM)~\citep{fedkiw1999} and the continuum surface force (CSF)~\citep{brackbill199} method stand out as robust solutions. The GFM is based on a generalized Taylor series expansion and explicitly accounts for the density jump at the interface; as a result, it is not sensitive to the amplitude of the density jump. The surface tension force is also directly incorporated in the pressure jump condition, leading to a sharp numerical treatment of this singular term~\citep{Desjardins2008_1}. However, in the CSF approach, instead of including the surface tension force directly in the pressure jump condition, this force is represented as a volumetric force spread over a few grid points surrounding the interface. While this approach may result in a slightly less accurate interface representation, particularly in cases with small front structures, it is generally considered to be less numerically challenging. Furthermore, discretizing the viscous terms using the GFM can be difficult and complex to implement numerically, making it less desirable for some applications~\citep{Desjardins2008_1}. Thus, many researchers employ the CSF approach to discretize the viscous term. In this study, we used the CSF method to model both the surface tension and the viscosity terms. Albeit slightly less accurate as compared to GFM for surface tension modelling, the CSF approach is more straightforward to implement and can provide robust and accurate results.

The available methods to numerically transport an interface can be divided into two categories: interface-tracking and interface-capturing~\citep{mirjalili2017}. One well-known approach in the interface-tracking category is the front-tracking method introduced by Unverdi and Tryggvason \cite{unverdi1992}. This method involves breaking down the fluid interface into discrete material points, referred to as front-tracking points, which are then transported using a moving mesh that follows a Lagrangian approach. While this approach benefits from purely Lagrangian transport, it faces difficulties in preserving liquid volume due to the requirement for frequent mesh rearrangements~\citep{Desjardins2008_1}. Additionally, parallelization of the front-tracking method presents a significant challenge. Furthermore, any break-up or merging of the interface should be addressed manually due to the inability of this technique to inherently handle topology changes. As a result, front-tracking methods are not well-suited for simulations with frequent topological changes, such as primary atomization~\citep{Desjardins2008_1}. 

Interface-capturing methods such as the volume-of-fluid (VOF) method~\citep{scardovelli1999} and level set method~\citep{sethian1999} implicitly capture the interface and can robustly address complex topological changes in the simulation. The VOF method employs a liquid volume fraction transport equation to depict the interface, ensuring mass conservation. However, since the VOF scalar is discontinuous across the interface, specific numerical treatments are required for the discretization of the transport equation. The discontinuous nature of the VOF scalar presents difficulties in computing interface properties such as normal and curvature values as well.

The level set method, introduced by Sethian \cite{sethian1999} in the field of image processing and computer graphics, represents an interface implicitly using the iso-level of a smooth function, i.e., the signed distance function. The smoothness of the level set function is maintained with the re-initialization process, and the Eulerian scalar transport equation can be solved using high-order numerical schemes. In addition, parallelization of the solver can be accomplished efficiently, and interface characteristics such as normal and curvature values are easily calculable due to the smoothness of the level set function. Despite all the mentioned advantages of the level set method, this method does not inherently conserve mass during the simulations, leading to potentially significant errors. Various hybrid methods have been introduced to overcome the stated drawback of the level set method, such as the coupled-level-set and volume-of-fluid method (CLSVOF) by Sussmann and Puckett \cite{sussman2000}. The CLSVOF method incorporates the mass conservation property of the VOF method with the smoothness of the level set function~\citep{meng2022}. Another hybrid method is the hybrid particle level set method (HPLS) proposed by Enright \textit{\textit{et al.}}~\cite{enright2002}. This method updates the interface location computed using the Eulerian transport equation through the use of Lagrangian markers, resulting in improved mass conservation. Although all of these hybrid methods improved the mass conservation property of the original level set method, they lack the main benefits of the original level set method, i.e., the cost-effective, straightforward implementation of the Euler transport equations using different existing high-order schemes~\citep{mirjalili2017}. 

Furthermore, several studies have explored the use of mesh refinement techniques to mitigate errors in mass conservation. For instance, Herrmann \cite{herrmann2008} proposed the refined level set grid (RLSG) method, in which the level set equation is solved on an auxiliary high-resolution grid. Another approach is the standard arbitrary mesh refinement (AMR) method, in which the mesh is made finer near the interface~\citep{gibou2018, chen2023}. Although mesh refinement techniques offer improved mass conservation, they can be computationally expensive, difficult to implement in parallel systems, and constrained by small time steps due to the finer mesh resolution. 

Studies by Olsson and Kreiss~\cite{olsson2005} and later Olsson \textit{\textit{et al.}}~\cite{olsson2007} addressed the conservation issue of the classical level set method by proposing a modification while maintaining its simplicity. They replaced the traditional signed distance level set function with the diffuse interface profile defined by the hyperbolic tangent function and solved the transport and re-initialization equations in a conservative form. This approach showed an improvement in mass conservation by an order of magnitude compared to results using the signed distance function~\citep{olsson2005}. In this study, we will use the conservative level set (CLS) approach to capture the interface between two flows. 

The implemented interface-capturing scheme should then be coupled with a proper incompressible flow solver to simulate the physics of two-phase incompressible flows. Here, we will utilize the projection method introduced by Chorin \cite{chorin1997} to model the behaviour of incompressible flows. In this approach, the momentum equation is split into two parts. The first one solves the momentum equation while ignoring the pressure term to calculate the intermediate velocity field, which does not necessarily satisfy the divergence-free constraint. Subsequently, the second equation uses pressure to project the intermediate velocity field into a divergence-free velocity field.

Magnetic fields exert a considerable influence on the behaviour of conducting fluids. The motion of these fluids in the presence of magnetic fields is described through the coupling of the Navier--Stokes equations with Maxwell's equations of electromagnetics. This coupling gives rise to a set of equations known as the magnetohydrodynamics (MHD) equations.
The interaction between electromagnetic fields and incompressible conducting fluids finds applications in fusion reactors, the metallurgical industry, MHD generators, and aluminum reduction cells~\citep{davidson2002}. To investigate the physics of these problems, a two-phase MHD solver is required. While numerous numerical studies have been conducted on simulating one-phase incompressible or compressible MHD problems~\citep{jiang1999,wu2007,makaremi2022}, the existing two-phase MHD solvers are highly limited due to their complexity and multi-physics nature. For instance, Huang \textit{et al.}~\cite{huang2002} developed a three-dimensional free-surface MHD solver to simulate the evolution of the liquid lithium film free-surface due to the existing magnetic forces in a fusion reactor known as NSTX (National Spherical Torus Experiment). Their implemented MHD model is based on the magnetic field induction equation, and the free boundary is tracked using the concept of a fractional volume of fluid method. On the other hand, Gao \textit{et al.}~\cite{gao2004} simulated the motion of a liquid lithium droplet under a strong non-uniform magnetic field in a vacuum environment without the influence of gravity. In their approach, the VOF model is incorporated to capture the interface, and the CFS method is used to account for the surface tension. Later, Tagawa~\cite{tagawa2006} developed a numerical solver to investigate the movement of a falling droplet of liquid metal into a pool of liquid metal under a uniform magnetic field in the cylindrical geometry. The primary level set approach of Sussman \textit{et al.}~\cite{Sussman1994} has been used in the study by Tagawa~\citep{tagawa2006} to capture the interface without the re-initialization step. In that study, the mass conservation, convergence, and consistency of the solver have not been verified for other benchmarks. Additionally, various studies have been also conducted to simulate the two-phase MHD flows in the finite-element framework. For example,  Yang \textit{et al.}~\cite{yang2019} proposed a diffuse interface model to numerically simulate the two-phase MHD flows and studied the performance of their solver for two-phase Hartmann flows, which is the MHD version of the classical Poiseuille flows.

Despite various numerical efforts in MHD flows in compressible liquid, there remains a need for a general and systematic implementation of a numerical framework with a higher order of accuracy. This is crucial for effectively capturing the formation of instabilities at the interface, which is of high importance in different applications, particularly when dealing with abrupt changes in magnetic properties across the interface. In this study, the magnetostatic case of Maxwell's equations is studied and integrated into the implemented two-phase incompressible solver. Two-phase magnetostatic solvers are widely employed to simulate the deformation of ferrofluid droplets in various flow fields, with applications in different fields, including biomedicine and rheology~\citep{afkhami2008field, afkhami2010deformation,majidi2022}. In the presence of a magnetic field in two-phase magnetic flows, magnetic permeability experiences a discontinuity across the interface, leading to the induction of the Lorentz force. This force significantly influences the evolution of the interface. Therefore, it is essential to incorporate the role of the Lorentz force into the governing equations. The proposed two-phase magnetostatic solver adequately addresses the magnetic permeability jumps across the interface, imposes proper boundary conditions for the magnetic field at the interface, and satisfies the divergence-free constraint for the magnetic field.

The objective of this study is two-fold. First, it primarily aims to present a detailed second-order numerical toolkit for simulating the physics of two-phase incompressible flows and its extension to magnetic flows. The mathematical formulation and numerical grid implementation will be given in Sec.~\ref{sec:griddiscretization}. While investigating the surface instabilities necessitates the use of higher-order numerical solvers, most numerical studies in this area are limited to first-order accuracy. Thus, a fifth-order mass-conservation level set approach is presented in Sec.~\ref{sec:levelset}, which includes a conservative re-initialization step to minimize the mass loss. The accuracy and robustness of the implemented level set solver are also analyzed using several benchmarks existing in the literature. A high-order conservative incompressible solver based on the projection method is discussed in Sec.~\ref{sec:incompressible} to solve the momentum equation under the divergence-free constraint of the velocity field. The level set and incompressible solvers are then coupled in Sec.~\ref{sec:twophase}, and the detailed implementation of a second-order two-phase incompressible solver is demonstrated. This numerical approach solves the governing equations in a conservative form and effectively handles high density ratios and viscosity jumps across the interface without introducing numerical instabilities. Furthermore, a consistent method for calculating the interface curvature is utilized, which is cost-effective and straightforward to implement. The robustness and accuracy of this two-phase solver are examined through three benchmarks. This effort is followed by introducing a two-phase magnetohydrodynamics solver under the magnetostatic assumption, achieved by extending the implemented second-order two-phase solver to the magnetic case. To the knowledge of the authors, studies on developing a high-order two-phase solver for magnetic flows are limited, and there is no specific study focusing on the development of a two-phase magnetostatic solver in the finite-difference framework using a high-order level set method to capture the interface between two fluids. Thus, in Sec.~\ref{sec:conductingTwophase}, a procedure for adding the magnetic terms to the solver is established. The introduced solver successfully accounts for significant magnetic permeability variations across the interface while satisfying the divergence-free condition of the magnetic field. The performance of the solver is evaluated by proposing three test cases: The deformation of a ferrofluid droplet in quiescent and shear flow regimes and the magneto-Rayleigh--Taylor instability in magnetic fluids.

The second and principal contribution of this paper is the investigation of sheared ferrofluid droplet deformation for various values of the droplet's susceptibility in both low and high capillary flow regimes. While previous studies have examined the effects of various factors on droplet deformation, such as the viscosity ratio between the droplet and the surrounding medium and the strength and direction of the imposed magnetic field, exploring deformations for different susceptibility values remains a crucial avenue to explore. Therefore, in Sec.~\ref{sec:conductingTwophase}, this paper also investigates the impact of droplet susceptibility values on its deformation, rotation, and potential breakup.

\section{\label{sec:griddiscretization} Grid arrangement and mathematical formulation}
\begin{figure}
\centering
\includegraphics[scale=2.5]{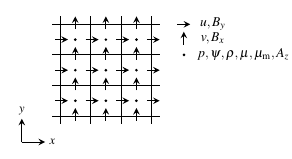}
\caption{\label{fig:grid} Staggered grid system in Cartesian coordinates. In the staggered grid arrangement, the values of scalar fields such as pressure ($p$), level set function ($\psi$), density ($\rho$), dynamic viscosity ($\mu$), magnetic permeability ($\mu_\mathrm{m}$), and $z-$component of the vector potential ($A_z$), are defined at cell centers. Velocity components, $u$ and $v$, as well as magnetic field components, $B_x$ and $B_y$, are defined at cell faces.}
\end{figure}

The use of a staggered grid in an incompressible solver ensures the accurate coupling between the velocity and pressure fields. The staggered arrangement, as illustrated in Fig.~\ref{fig:grid}, eliminates oscillations in the pressure field and avoids the checker-board problem, a common issue for incompressible numerical simulations~\citep{morinishi1998}. In this computational grid system, scalar values such as pressure are defined at the cell centers and velocity components are defined at the cell faces. As a result, the continuity equation is solved at cell center points, while the momentum equation corresponding to each velocity component is defined at cell faces.

 In this study, we have employed similar notation of the conservative centred high-order finite-difference scheme of Morinishi \textit{et al.}~\cite{morinishi1998} and Desjardins \textit{et al.}~\cite{Desjardins2008_2}, briefly introduced in this section for the sake of completeness. According to their notation, the second-order finite-difference operator with the stencil size $n$ for a variable $\phi$ in the $x_1-$direction in the computational domain $\mathbf{x}=\left(x_1, x_2, x_3 \right)$ is defined as 
\begin{equation}
    \frac{\delta_{n} \, \phi}{\delta_{n} \, x_1}=\frac{\phi\left(x_1+n \, \Delta x/2, \, x_2, \, x_3\right) - \phi\left(x_1-n \, \Delta x/2, \, x_2, \, x_3\right)}{n \, \Delta x}.
\end{equation}
The second-order differential operator with respect to the $x_2-$ and $x_3-$directions, $\frac{\delta_{n} \, \phi}{\delta_{n} \, x_2}$ and $\frac{\delta_{n} \, \phi}{\delta_{n} \, x_3}$, can be defined in the same manner.
The second-order interpolation of a quantity $\phi$ defined on the computational domain $\mathbf{x}=\left(x_1, x_2, x_3 \right)$ with the stencil size $n$ in the $x_1-$direction is given as
\begin{equation}
    {\overline{\phi}}^{n \, x_1} = \frac{\phi\left(x_1+n \, \Delta x/2, \, x_2, \, x_3\right)+ \phi\left(x_1-n \, \Delta x/2, \, x_2, \, x_3\right)}{2},
\end{equation}
and is defined similarly in the $x_2-$ and $x_3-$directions.

The $n^{\mathrm{th}}-$order central finite-difference operator in $x_i-$direction is defined as 
\begin{equation}
    \frac{\delta_{n{\mathrm{th}}} \, \phi}{\delta_{n{\mathrm{th}}} \, x_i}=\sum_{l=1}^{n/2} \alpha_l \frac{\delta_{\left(2l-1 \right)} \, \phi}{\delta_{\left(2l-1 \right)} \, x_i},
\end{equation}
where weight values, $\alpha_l$, are calculated as
\begin{equation}
    \sum_{l=1}^{n/2} \left(2l-1\right)^{2 \left(k-1\right)} \alpha_l = \delta_{kl},  \;  \;  k \in [1, \, n/2].
\end{equation}
Additionally, the $n^{\mathrm{th}}-$order interpolation in the $x_i-$direction is given as
\begin{equation}
    {\overline{\phi}}^{n{\mathrm{th}} \, \, x_i} = \sum_{l=1}^{n/2} \alpha_l \, {\overline{\phi}}^{\left(2l-1\right) \, x_i}.
\end{equation}
The introduced $n^{\mathrm{th}}-$order finite-difference and interpolation schemes will be later used to discretize the governing equations.

\section{\label{sec:levelset} Implementation of level set}
The traditional level set function introduced by Sethian \cite{sethian1999} is a smooth signed distance function given as 
\begin{equation}
    |\phi\left(\mathbf{x},t \right)| = |\mathbf{x}-\mathbf{x}_\Gamma|,
    \label{eq:ls1}
\end{equation}
where variable $\mathbf{x}_\Gamma$ denotes the closest point on the interface from point $\mathbf{x}$. The positive and negative values of $\phi(\mathbf{x},t)$ indicate the location of a point relative to the interface, with convention determining which side is positive or negative. Therefore, the zero iso-contour of the defined signed distance function, $\phi(\mathbf{x},t)$=0, corresponds to the interface itself. The level set motion under the velocity field $\mathbf{u}$ can be described using the following transport equation
\begin{equation}
     \frac{\partial \phi}{\partial t} + \mathbf{u} \cdot \nabla \phi = 0.
     \label{eq:ls2}
\end{equation}
Advecting the interface employing Eq.~(\ref{eq:ls2}) for a few time steps can cause the $\phi$ function to lose its signed distance property, becoming distorted and losing its smoothness, thereby causing issues in the simulation~\citep{Desjardins2008_1}. To prevent this issue, different re-initialization methods have been introduced to reconstruct the $\phi$ function to be a smooth signed distance function during the simulation. One of the well-known re-initialization techniques is solving a Hamilton--Jacobi equation, introduced by Sussman \textit{et al.}~\cite{Sussman1994}, which can be solved using high-order numerical discretization schemes and recover the distance function accurately. This method is proven to have various limitations, thoroughly discussed in the literature~\citep{Desjardins2008_1,sethian1999}, which are outside the scope of this paper. However, the main disadvantage of using this approach for simulating two-phase flows is that both the transport equation and re-initialization process fail to conserve the volume of the region enclosed by the zero iso-contour. This can result in mass gain or loss in numerical simulations, leading to unphysical results. In this study, the conservative level set method (CLS) is employed to address this issue.

In the CLS approach~\citep{olsson2005}, the interface between two immiscible flows is defined using a diffuse profile in the form of a hyperbolic tangent function, $\psi$, as below
\begin{equation}
    \psi(\mathbf{x}, t_0) = \frac{1}{2} \left(\tanh \left( \frac{\phi(\mathbf{x},t_0)}{2 \epsilon} \right)  +1 \right),
    \label{eq:ls3}
\end{equation}
where $\epsilon$ is a parameter to indicate the interface thickness and is commonly defined as a function of the mesh resolution. In the hyperbolic tangent definition, the iso-contour 0.5, $\psi \left(\mathbf{x}, t\right)$=0.5, specifies the location of the interface. For an incompressible flow, $\nabla \cdot \mathbf{u}$ = 0, the transport equation can then be re-written as
\begin{equation}
    \frac{\partial \psi}{\partial t} + \nabla \cdot (\mathbf{u} \psi) = 0.
    \label{eq:ls4}
\end{equation}
In order to recover the hyperbolic tangent form of the level set profile, maintain the interface thickness, and prevent diffusion and smearing of the interface during the simulation, a re-initialization step should be introduced. The derived conservative re-initialization step by Olsson~\cite{olsson2005} is given as
\begin{equation}
    \frac{\partial \psi}{\partial \tau} + \nabla \cdot \left(\psi \left(1 - \psi\right) \mathbf{n} \right) = \nabla \cdot \left(\epsilon \nabla \psi \right),
    \label{eq:ls5}
\end{equation}
where the variable $\tau$ is the pseudo-time, $\mathbf{n}$ is the normal vector at $\tau=0$, calculated as $\mathbf{n} = \nabla \psi/|\nabla \psi|$, and the equation is solved until convergence is reached. In the proposed re-initialization step, the compression flux, $\psi \left(1 - \psi \right) \mathbf{n}$, is included to maintain the resolution of the interface and sharpen the profile which may smear due to the numerical diffusion occurring during the simulation of the transport equation. Also, in order to make sure that the level set profile remains of thickness $\epsilon$ and avoids the formation of discontinuities at the interface, the diffusion flux, $\epsilon \nabla \psi$, with a small amount of viscosity is added to the re-initialization equation. In Eq.~(\ref{eq:ls5}), the diffusion in the normal direction to the interface would be balanced by the compression term. However, diffusion might also occur in the direction tangential to the interface, causing the interface to move~\citep{olsson2007}. For that reason, in the later study by Olsson \textit{et al.}~\cite{olsson2007}, the diffusion term has been modified as $\nabla \cdot \left(\epsilon \left(\nabla \psi \cdot \mathbf{n} \right) \mathbf{n} \right)$ to avoid any tangential movement of the interface due to diffusion, improving the re-initialization process.

\subsection{\label{sec:lsTransport} High-order level set transport}

Employing the notation introduced in Section ~\ref{sec:griddiscretization}, the $n^\mathrm{th}-$order level set transport can be discretized as
\begin{equation}
    \nabla \cdot \left(\mathbf{u} \psi \right) = \sum_{i=1}^{3} \left(\frac{\delta_{2{\mathrm{nd}}}}{\delta_{2\mathrm{nd}}  \, x_i} \left[u_i \overline{\psi}^{n\mathrm{th} \, \, x_i} \right] \right)
\end{equation}
where $\overline{\psi}^{n\mathrm{th} \, \, x_i}$ is a $n^\mathrm{th}-$order interpolation of the variable $\psi$ to the cell face in $i-$direction and $u_i$ is the $i^{\mathrm{th}}-$component of the velocity vector. Instead of using central interpolation schemes, it is better to use an upwind-biased scheme to prevent numerical oscillations appearing around $\psi=0.5$. Thus, different upwind-biased approaches can be employed in order to calculate interpolated $\psi$ values at the cell faces, such as WENO-type schemes~\cite{liu1994, jiang1996, su2018} or High Order Upstream Central (HOUC) schemes~\citep{Nourgaliev2007}. The use of higher order schemes improves the conservation of the transported level set, reducing the need for re-initialization and yielding more accurate results. In this study, we utilize the fifth-order WENO interpolation method, described in detail in Appendix~\ref{appendixA}. This approach is non-oscillatory and aims to mitigate unwanted oscillations, especially near sharp gradients, providing a smooth solution at each time step. The third-order total variation diminishing (TVD) Runge--Kutta scheme is used for the temporal integration, presented in Appendix~\ref{appendixB}. Employing TVD spatial and temporal numerical schemes suppresses the formation of unwanted oscillations in the numerical simulation of the $\psi$ profile, which is one of the main considerations in solving Eq.~(\ref{eq:ls4}).

\subsection{\label{sec:lsReinit} Level set conservative re-initialization}
As mentioned earlier, it is essential to maintain a consistent thickness of the hyperbolic tangent profile during the transport of the level set function. However, numerical schemes may diffuse the interface, resulting in a violation of mass conservation in the simulation. To address this issue, the re-initialization step is incorporated into the level set transport equation. The implemented re-initialization equation in this study, given as  
\begin{equation}
    \frac{\partial \psi}{\partial \tau} + \nabla \cdot \left(\psi \left(1 - \psi\right) \mathbf{n} \right) = \nabla \cdot \left(\epsilon \left(\nabla \psi \cdot \mathbf{n}\right) \mathbf{n} \right),
    \label{eq:reinitt1}
\end{equation}
is taken from the work by Olsson \textit{et al.}~\cite{olsson2007}, although their discretization method is based on the finite-element approach. Therefore, we will establish a robust, consistent finite-difference approach inspired by the study of Desjardins \textit{et al.}~\cite{Desjardins2008_1} to numerically discretize Eq.~(\ref{eq:reinitt1}). To proceed, we denote the diffusive and compressive fluxes as $\mathbf{F}_\mathrm{D} = \epsilon \left(\nabla \psi \cdot \mathbf{n}\right) \mathbf{n}$ and $\mathbf{F}_\mathrm{C}= \psi \left(1-\psi\right) \mathbf{n}$, respectively. According to the analysis shown by Desjardins \textit{et al.}~\cite{Desjardins2008_1}, employing a more compact computational stencil to discretize Eq.~(\ref{eq:reinitt1}) will lead to a more accurate and robust reconstruction of the interface while eliminating the appearance of spurious oscillations; hence, the second-order discretization is employed. For the sake of clarity, we rewrite Eq.~(\ref{eq:reinitt1}) for a two-dimensional case as below
\begin{eqnarray}
\frac{\partial \psi}{\partial \tau} + \overbrace{\frac{\partial}{\partial x} \left(\psi \left(1-\psi \right) n_x \right) +
    \frac{\partial}{\partial y} \left(\psi \left(1-\psi \right) n_y \right)}^{\mathrm{compression}} =  
    \underbrace{\epsilon \frac{\partial}{\partial x} \left(\frac{\partial \psi}{\partial x} {n_x}^2 + \frac{\partial \psi}{\partial y} n_x n_y \right) + \epsilon \frac{\partial}{\partial y} \left(\frac{\partial \psi}{\partial x} n_x n_y + \frac{\partial \psi}{\partial y} {n_y}^2 \right)}_{\mathrm{diffusion}}.
\end{eqnarray}
In order to update the cell center $\psi$ values, compression and diffusion fluxes should be calculated at cell faces, and components of the normal vector should be found at cell faces. To this end, normal values at cell faces, i.e., face normals~\citep{Desjardins2008_1}, are calculated, determining $x-$ and $y-$components of the normal vector at cell faces in both $x-$ and $y-$directions.
The $x_i-$component of $\psi$ gradient across the $x_i-$face is given as
\begin{equation}
    \left(\nabla \psi \right)_i^{x_i-\mathrm{face}} = \frac{\partial \psi}{\partial x_i} \Bigg|_{x_i- \mathrm{face}}= \frac{\delta_{\mathrm{2nd}} \psi}{\delta_{\mathrm{2nd}} \, x_i},
\end{equation}
while for the $x_j-$component, a second-order interpolation of $\psi$ in the $x_i-$direction is needed, and the face gradient value is calculated as
\begin{equation}
    \left(\nabla \psi \right)_j^{x_i-\mathrm{face}} = \frac{\partial \psi}{\partial x_j} \Bigg|_{x_i- \mathrm{face}}= \frac{\delta_{\mathrm{2nd}} {\overline{\psi}}^{2\mathrm{nd} \, x_i}}{\delta_{\mathrm{2nd}} \, \, x_j}.
\end{equation}
Normalized face gradient values will correspond to normal vector values at cell faces, defining the normal vector at $x_i-$face as $\mathbf{n}^{x_i}=\left(\nabla \psi \right)^{x_i-\mathrm{face}}/|\left(\nabla \psi \right)^{x_i-\mathrm{face}}|$. Therefore, the discrete version of diffusion and compression terms can be written as
\begin{equation}
    \nabla \cdot \left(\epsilon \left(\nabla \psi \cdot \mathbf{n}\right) \mathbf{n} \right) = \epsilon \sum_{i=1}^{3} \left(\frac{\delta_{\mathrm{2nd}}} {\delta_{\mathrm{2nd}} \, x_i} \left[\mathbf{n}_i ^ {x_i} \left(\mathbf{n}^{x_i} \cdot \left(\nabla \psi \right)^{x_i-\mathrm{face}} \right) \right]\right),
\end{equation}
and
\begin{equation}
    \nabla \cdot \left(\psi \left(1 - \psi\right) \mathbf{n} \right) = \sum_{i=1}^{3} \left(\frac{\delta_{\mathrm{2nd}}} {\delta_{\mathrm{2nd}} \, x_i} \left[\mathbf{n}_i ^ {x_i} {\overline{\psi \left(1-\psi \right)}}^{2\mathrm{nd} \, x_i}\right]\right),
\end{equation}
respectively. According to \cite{olsson2007}, the re-initialization equation converges quickly. In their study, it has been shown that for the case of $\Delta x \sim \Delta t \sim \epsilon$, the solution of the conservative re-initialization will converge within one or two steps. 

Desjardins \textit{et al.}~\cite{Desjardins2008_1} demonstrated that by employing the length scale analysis, the proper $\Delta \tau$ and the number of steps needed to obtain the steady state solution of the re-initialization equation can be determined. They showed that by taking the CFL number of the convection equation, Eq.~(\ref{eq:ls4}), to be $n$ times greater than the CFL number corresponding to the compression term in the re-initialization equation, Eq.~(\ref{eq:reinitt1}), $n \, \mathrm{CFL}_\mathrm{comp} = \mathrm{CFL}_{\mathrm{conv}}$, the solution of the re-initialization equation converges after $n$ steps. Thus, there is no requirement to evaluate the convergence criteria during the simulation. Our simulations utilize this approach and produce decent, robust results for the re-initialization step, further explained in the following section.

\subsection{\label{sec:lsTests} Level set test cases}

In Appendix~\ref{appendixC}, the numerical order of accuracy of the implemented level set solver is demonstrated through the rotating circle test case. In this section, the accuracy, consistency, and robustness of the level set solver along with the re-initialization procedure are verified and discussed using two additional test cases, i.e., the circle in a deformation field and Zalesak's disk problems.

\subsubsection{Circle in a deformation field}
\begin{figure*}
\centering
\subfloat[]{\includegraphics[width = 0.9\textwidth]{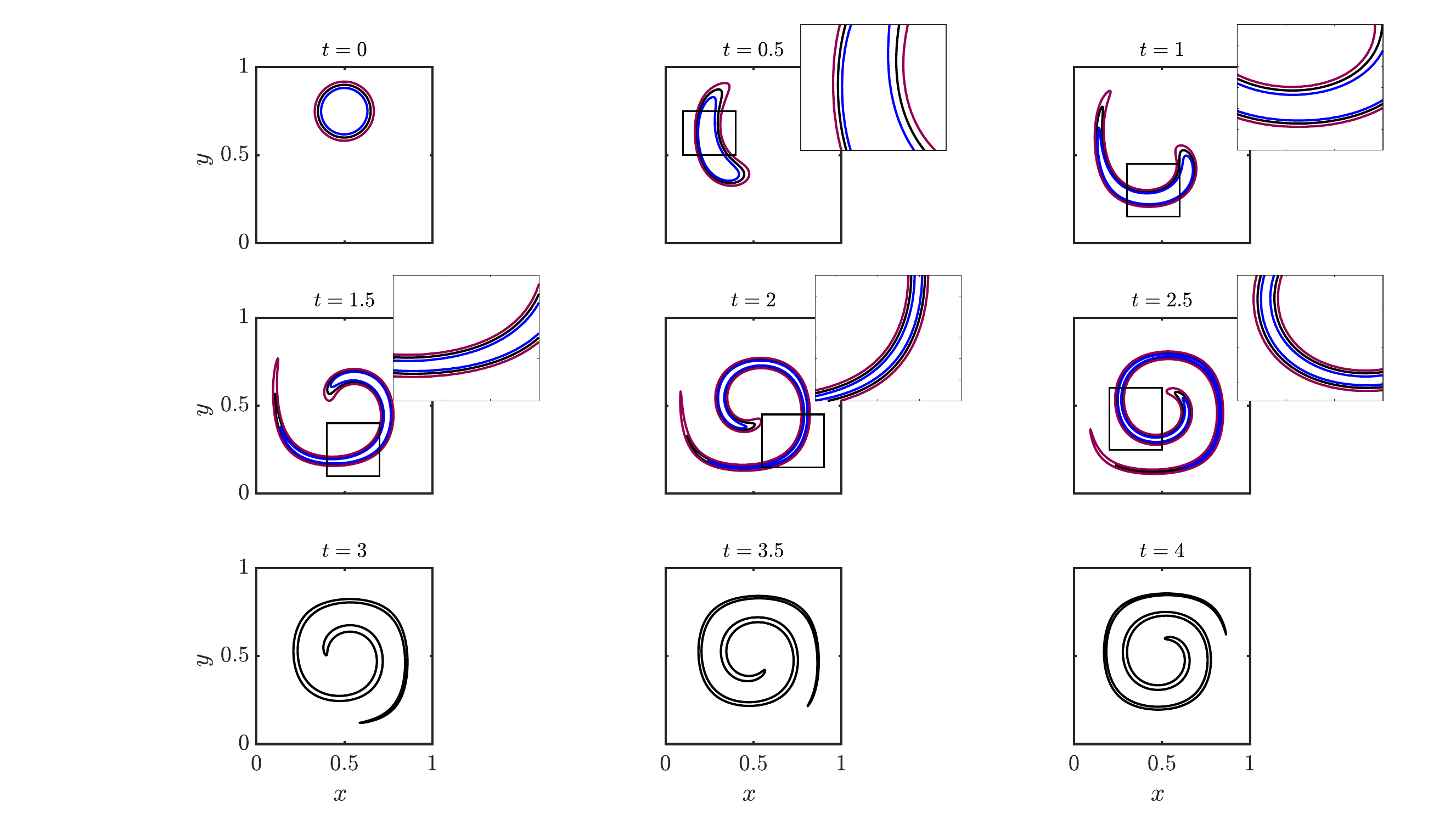}}
\\
\subfloat[]{\includegraphics[trim = 0 0 0 0, width = 0.8\textwidth]{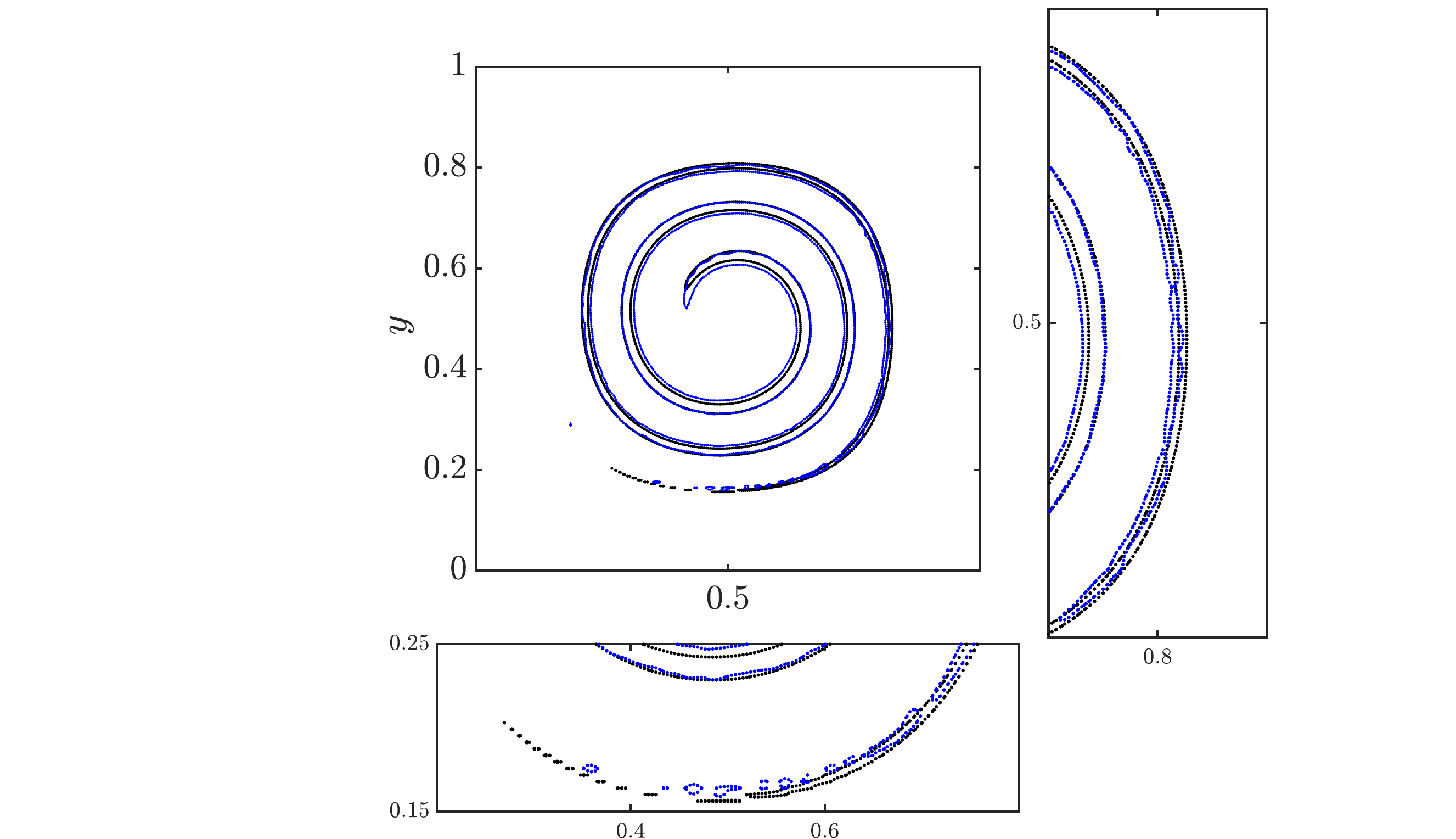}}
\caption{\label{fig:LSTest2} (\textit{a}) The interface location, $\psi=0.5$ (in black), of the single vortex in the velocity field defined by Eqs.~(\ref{eq:LSTest2_u}-\ref{eq:LSTest2_v}), from $t=0$ to $t=4.0$, with the grid resolution of $256 \times 256$ and CFL$=0.5$. For $t=0$ to $2.5$, $\psi=0.05$ and $\psi=0.95$ contours are also plotted in blue and purple, respectively. (\textit{b})  Interface location of the vortex in the deformation field, black: with re-initialization process applied every five time steps, and blue: without re-initialization step.}
\end{figure*}

The main purpose of this test case is to evaluate the ability of the solver to properly resolve thin filament structures, mainly appearing in stretching and tearing flows~\citep{enright2002}. Here, the initial center of the circle interface is located at $(0.5, 0.75)$, with the radius of $r=0.15$. The interface thickness equals to $\epsilon=(\Delta x)^{0.7}/2$, and the simulation is performed on the computational domain $[x, y] \in [0, 1] \times [0, 1]$, with the grid resolution of $256 \times 256$. The velocity field is defined as
\begin{equation}
     u = \sin \left(\pi x \right) \sin \left(\pi x \right) \sin \left(2 \pi y \right),
     \label{eq:LSTest2_u}
\end{equation}
and
\begin{equation}
     v = - \sin \left(\pi y \right) \sin \left(\pi y \right) \sin \left(2 \pi x \right),
     \label{eq:LSTest2_v}
\end{equation}
causing the circle interface to stretch out into a long, thin fluid element that continuously wraps around itself. The CFL values are set as $0.5$ and $0.25$ for transport and re-initialization equations, respectively, and the re-initialization process is applied every five steps. Figure~\ref{fig:LSTest2}(a) displays the evolution of the interface, $\psi=0.5$, at eight different time steps: $t = 0.5, 1, 1.5, 2, 2.5, 3, 3.5$, and 4 in black. Figure~\ref{fig:LSTest2}(a) visually confirms that the implemented solver is capable of sustaining thin, elongated filament structures of the interface for a long simulation time. The calculated maximum area loss during the simulation, from $t = 0$ to $4$, is equal to $2.8\%$, demonstrating the area-conserving character of the level set solver, even after stretching the vortex for a long time and causing the trailing ligament thickness to become of the order of the mesh resolution.
Additionally, in order to evaluate how the thickness of the transition layer changes, $\psi=0.05$ and $\psi=0.95$ contours are also depicted in blue and purple, respectively, for the first six time steps, as for the final stages of the vortex stretching, the thickness becomes too thin, and visualization of the transition layer is a bit difficult. According to the presented contours, the thickness of the transition layer remains almost constant even after drastic changes in the interface. However, at the tail of the stretched vortex, it can be seen that the transition layer is not constant. This is mainly due to the fact that the thickness of the stretched circle becomes similar to the thickness of the interface. This behavior was also observed in other studies like the one by Olsson and Kreiss~\cite{olsson2005} and is called a pinch-off effect. The pinch-off is a numerical effect and can be prevented if the interface's thickness is smaller than the distance between two interfaces.

Figure~\ref{fig:LSTest2}(b) presents the interface location of the vortex at $t=4.5$ in black, while the blue iso-contour depicts the solution at the same time without applying the re-initialization step during the simulation, from $t=0$ to $4.5$. Figure~\ref{fig:LSTest2}(b) makes evident that incorporating the re-initialization process in the simulation leads to a smoother solution with better area conservation. The zoomed-in insets in Fig.~\ref{fig:LSTest2}(b) better illustrate that employing the re-initialization step during the simulation results in a better reconstruction of the thin filaments and prevent the tearing of the interface in the solution, leading to a more conservative solution.  

\begin{figure*}
\subfloat[]{\includegraphics[width = 0.95\textwidth]{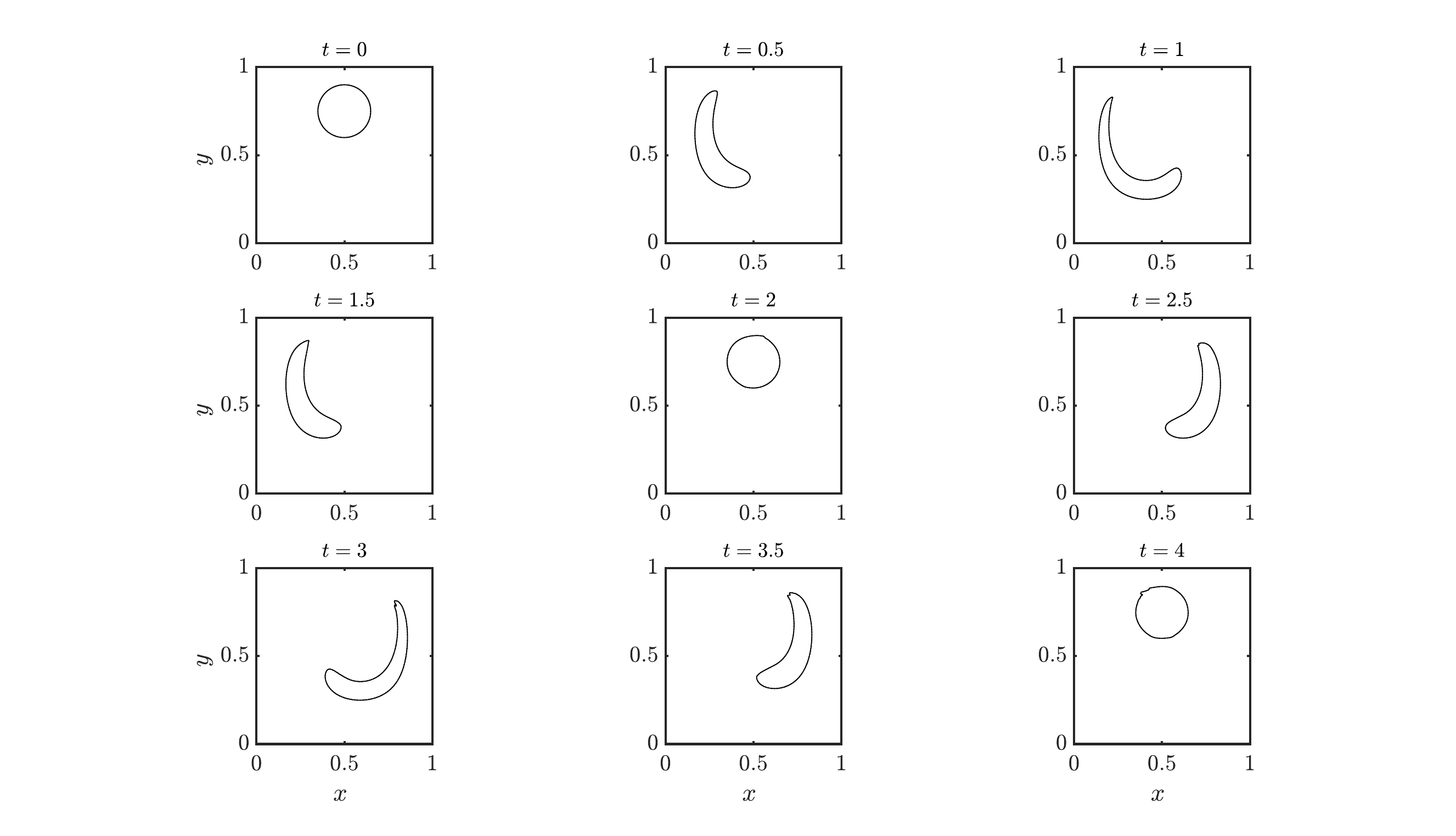}}\\ \hspace*{-0.5cm}
\subfloat[]{\includegraphics[width = 0.62\textwidth]{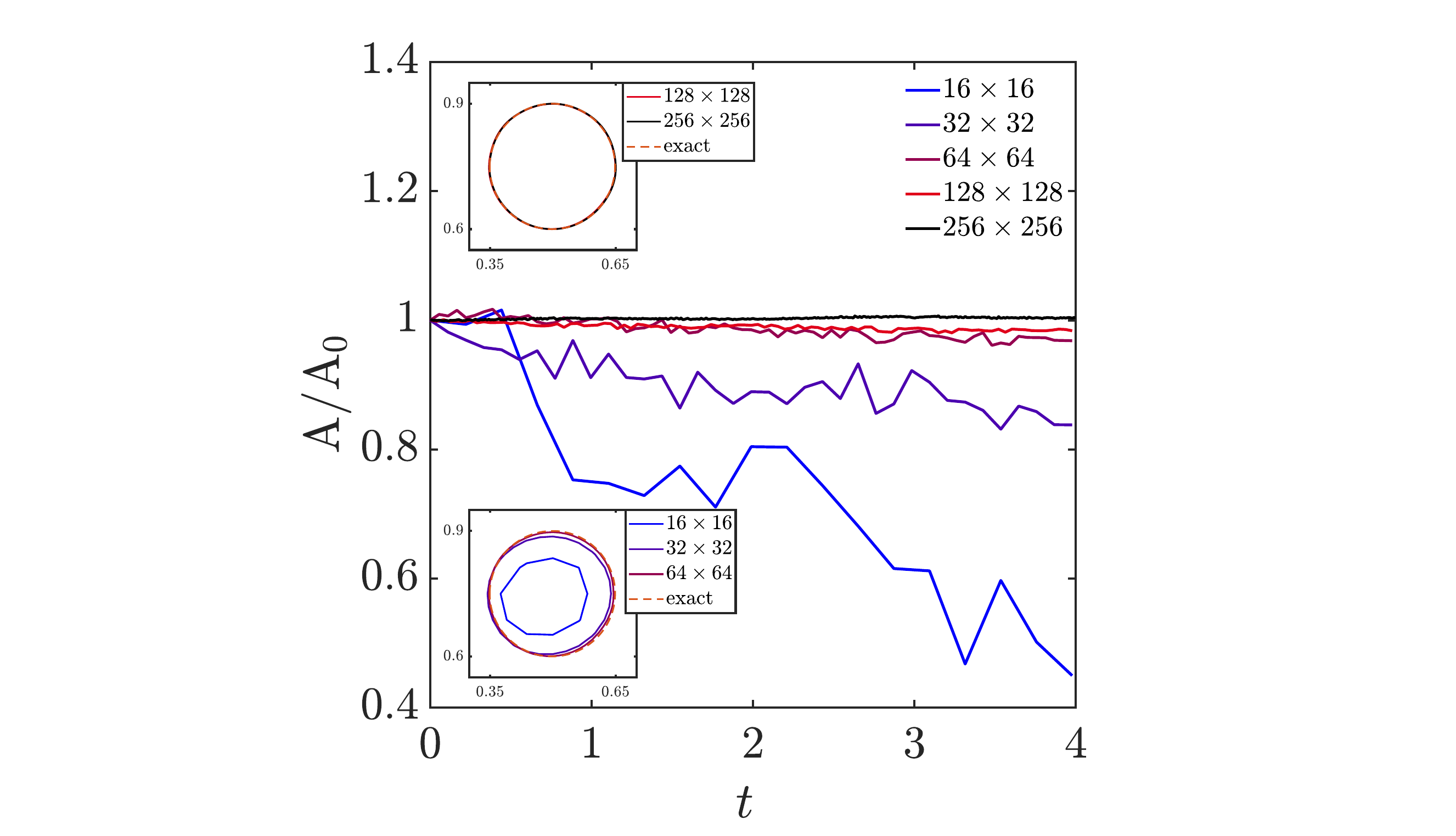}}\hspace*{-2.5cm}
\subfloat[]{\includegraphics[width = 0.62\textwidth]{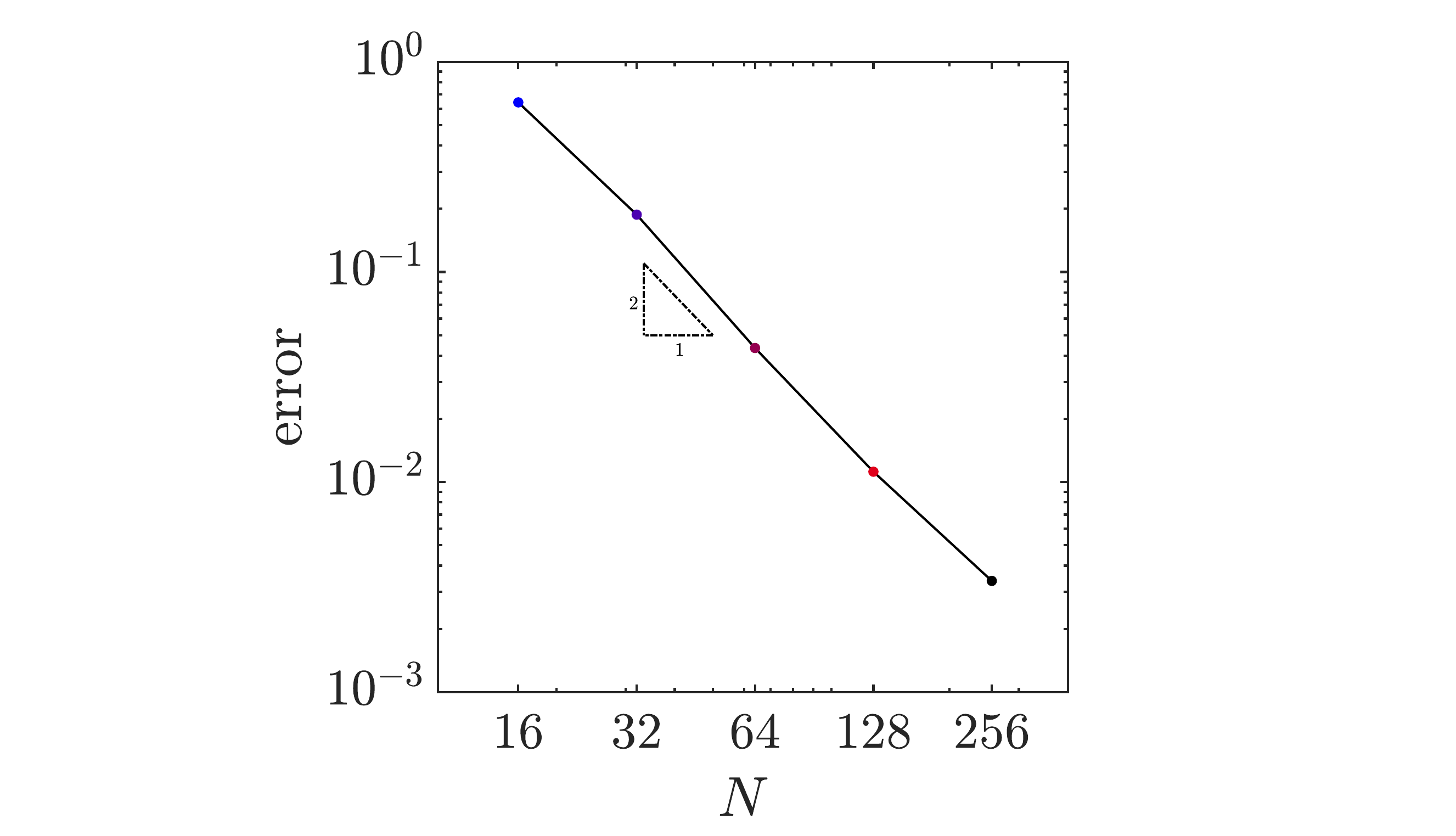}}
\caption{\label{fig:LSTest2_2} (\textit{a})  The interface location of the circle in the vortex field with the periodicity of $T=2$ at nine different time steps, $t=0, 0.5, 1, 1.5, 2, 2.5, 3, 3.5$, and $4$. The simulation is performed in the two-dimensional computational domain, $[x,y] \in [0, 1] \times [0, 1]$, with the grid resolution of $256 \times 256$, and CFL$=0.5$. The re-initialization step is applied every five time steps. (\textit{b}) The temporal evolution of the normalized area for the circle in the vortex field is compared for five different mesh resolutions of $16 \times 16$, $32 \times 32$, $64 \times 64$, $128 \times 128$, and $256 \times 256$. Additionally, the interface location of the circle, $\psi=0.5$, is shown for all five different grid resolutions. (\textit{c}) Order of convergence analysis of the area loss property. The $L_\infty$ error values of the area loss are computed for five different grid resolutions and plotted against the grid size.}
\end{figure*}

To achieve a better quantitative analysis and error calculation, the vortex field can be reversed in time by multiplying the velocity components by $\cos \left(\pi t /T \right)$, where $T$ denotes the time that the circle returns to its initial state, and, therefore, the interface location is known. In our simulation, the periodicity is set to $T=2$, and the simulation is run from $t=0$ to $t=4$, hence, the interface goes through two complete rotations. Figure~\ref{fig:LSTest2_2}(a) illustrates the interface from $t=0$ to $4$ with increments of $0.5$, showing that the circle interface goes through two complete rotations and returns to its original place at $t=4$. The calculated root mean square (rms) of the error for the level set field at $t=2$ and $4$, are $4.8\times10^{-3}$ and $7\times10^{-3}$, respectively.

The area loss during the simulation and the convergence of the numerical result for increasing mesh resolutions are studied by repeating this test case for four other resolutions, $16 \times 16$, $32 \times 32$, $64 \times 64$, and $128 \times 128$. In order to investigate the mass conservation property of the implemented level set, ${A}/{A}_0$ is calculated, where ${A}_0$ represents the initial area and ${A}$ is the computed area at each time step, for all five grid resolutions. In Fig.~\ref{fig:LSTest2_2}(b), the conservative behaviour of the implemented level set is demonstrated by a decrease in area loss as the mesh is refined and the value of ${A}/{A}_0$ becomes closer to one as the grid resolution is increased. Figure~\ref{fig:LSTest2_2}(b) also compares the numerical and analytical results for different mesh resolutions. Furthermore, increasing the grid resolution leads to the convergence of the numerical result to its analytical counterpart. The conservative property noticeably improved when the mesh resolution is increased from $16 \times 16$ to $32 \times 32$. For mesh resolutions $64 \times 64$, $128 \times 128$, and $256 \times 256$, the calculated $\psi=0.5$ contour is close to the analytical interface, and for $128 \times 128$ and $256 \times 256$, the computed interface is indistinguishable from the exact one. In order to evaluate the accuracy of the solver, the $L_\infty$ error of the area loss during the simulation from $t=0$ to $4$ is calculated for all five mesh resolutions. In Fig.~\ref{fig:LSTest2_2}(c), the calculated error values are plotted against the computational grid size using a logarithmic scale. According to Fig.~\ref{fig:LSTest2_2}(c),  the error decreases as the mesh is refined, and a second-order convergence is achieved. This convergence analysis indicates that the solver is able to maintain second-order accuracy for the mass conservation property even for more pronounced interface deformations over prolonged periods of simulation. It is worth noting that the complexity of the interface evolution may necessitate more frequent re-initialization steps to ensure optimal mass conservation during the simulation. The re-initialization step is second-order accurate and can affect the global order of accuracy of the solver. That is why the convergence rate here is expected to be less than that of the test case presented in Appendix~\ref{appendixC}.

\subsubsection{Zalesak's disk}

\begin{figure*}
\includegraphics[ width = 1\textwidth]{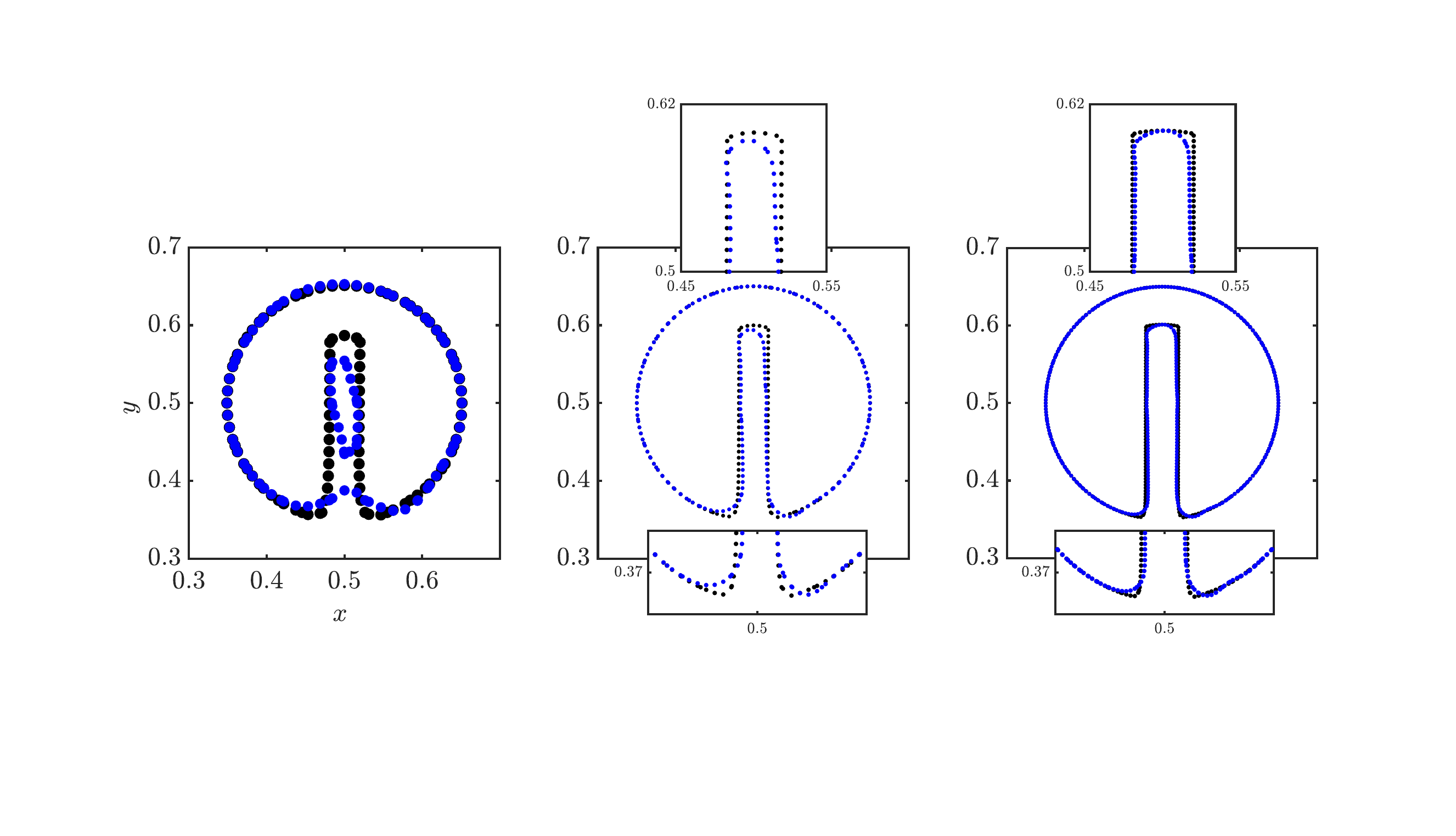}
\vspace{-2.5cm}
\caption{\label{fig:LSTest3_1} The interface of the Zalesak's disk, $\psi=0.5$, for three different grid resolutions, $64 \times 64, 128 \times 128$, and $256 \times 256$, from left to right. The black contour shows the initial interface while the blue one represents the calculated interface at $t=1$. The numerical solution converges to the exact solution by increasing the mesh resolution, accurately capturing the corners and structure of the interface.}
\end{figure*}

The Zalesak's disk problem, widely used in the literature to indicate the robustness of a solver towards diffusion errors~\citep{mizuno2022finite}, is studied in this section. In this test case, the ability of the solver to transport sharp corners and thin structures without introducing noticeable diffusion errors can be examined. This test case includes a notched circle in solid body rotation under the velocity field of $u=2\pi \left(y-0.5 \right)$ and $v=-2\pi \left(x - 0.5 \right)$. Initially, a notched circle of radius $r=0.15$ is centred at $(0.5, 0.5)$ with a notch of width $0.05$ and height $0.25$ in the computational domain of $(x, y) \in [0, 1] \times [0, 1]$. At $t=1$, the disk goes through one complete rotation and returns to its primary location. The level set thickness and CFL values are similar to the previous case, and a Dirichlet boundary condition is imposed at all boundaries. Figure~\ref{fig:LSTest3_1} displays the solution at $t=1$ for three different grid resolutions $64 \times 64$, $128 \times 128$, and $256 \times 256$. As can be seen in Fig.~\ref{fig:LSTest3_1}, increasing the grid resolution reduces the diffusion of the interface, and the final solution is close to the initial description of the interface, recovering the edges and corners of the disk more accurately (see zoomed-in insets). The calculated area loss for the three grid resolutions from coarse to fine are $0.43\%$, $0.25\%$, and $0.07\%$, respectively, showing the convergence of the solution by increasing the grid resolution.

\section{\label{sec:incompressible}Implementation of incompressible Navier--Stokes solver}

In this section, the implementation of a high-order incompressible solver is presented. The density of each fluid particle does not change as it moves in the incompressible flow regime. Therefore, the mass conservation equation simplifies to a divergence-free condition for the velocity field, which must be satisfied while solving the momentum equation. To this end, the projection method is adopted in our algorithm, and two test cases are examined to verify the accuracy and robustness of the implemented solver.

\subsection{\label{sec:incompressibleEqs} Governing equations}
The incompressible form of the Navier--Stokes equations are given as
\begin{subequations}
\begin{equation}
    \frac{\partial \mathbf{u}}{\partial t} + \nabla \cdot \left(\mathbf{u} \mathbf{u} \right) = -\frac{1}{\rho} \nabla p + \frac{1}{\rho} \nabla \cdot \mu \left(\nabla \mathbf{u} + \nabla \mathbf{u}^\mathrm{T} \right) + \mathbf{g},
    \label{eq:incomp1}
\end{equation}
\begin{equation}
   \nabla \cdot \mathbf{u} = 0,
   \label{eq:incomp2}
\end{equation}
\end{subequations}
where $\mathbf{u}$ is the velocity vector, $\rho$ is the density field, and $p$ is the pressure field. Variables $\mu$ and $\mathbf{g}$ denote the dynamic viscosity and gravitational acceleration, respectively. It is noteworthy to mention that the momentum equation is solved in a conservative form, ensuring momentum conservation and avoiding unphysical numerical solutions. The presented incompressible numerical solver employs a spatial staggered-variable formulation, described in Section~\ref{sec:griddiscretization}. Compared to a collocated formulation, staggering offers the benefit of localized derivative stencils in space that enhance the precision of the stencil. In order to solve the momentum equation, the pressure gradient is required at cell faces, where the velocity values are defined. In the staggered arrangement, the second-order pressure gradient can be calculated using pressure values that are one cell apart. However, in the collocated formulation, the pressure values that span three cells are needed to compute the second-order pressure derivative. The localized stencils accessible in staggered approaches are considerably more precise than the broader stencils that would be used in a collocated approach on the same mesh~\citep{pierce2001}. Furthermore, the staggered grid system offers a strong coupling between the pressure and velocity field when a Poisson equation is solved for the pressure compared to the collocated grids~\citep{morinishi1998}. As a result, the unphysical solution obtained in the collocated grid system due to weak coupling is prevented.

\subsection{\label{sec:incompressibleProjection} Projection method}

The main difficulty of numerically solving the incompressible Navier--Stokes equations is the lack of an explicit time-derivative term in the continuity equation. Consequently, the velocity divergence-free constraint must be satisfied by implicitly coupling the mass equation with the pressure term in the momentum equation~\citep{kim1985}. To this end, this study incorporates the projection method initially introduced by Chorin~\cite{chorin1997}, which is based on the Helmholtz--Hodge decomposition that states any vector field can be decomposed into two components: solenoidal (divergence-free), and irrotational (curl-free). In this operator splitting approach, the momentum equation is divided into two separate equations, given as
\begin{subequations}
\begin{equation}
     \frac{{\Tilde{\mathbf{u}}}^{n+1} - \mathbf{u}^n}{\Delta t} = - \nabla \cdot \left(\mathbf{u}{^n} \mathbf{u}{^n} \right) + \frac{1}{\rho} \nabla \cdot \mu \left(\nabla \mathbf{u}{^n} + (\nabla \mathbf{u}{^n})^\mathrm{T} \right) + \mathbf{g},
     \label{eq:incomp3}
\end{equation}
and
\begin{equation}
     \frac{\mathbf{u}^{n+1} - \Tilde{\mathbf{u}}^{n+1}}{\Delta t} + \frac{\nabla p^{n+1}}{\rho} = 0,
     \label{eq:incomp4}
\end{equation}
\end{subequations}
where $\Tilde{\mathbf{u}}$ denotes the intermediate velocity which does not necessarily satisfy the divergence-free constraint. Adding Eq.~(\ref{eq:incomp3}) and Eq.~(\ref{eq:incomp4}) recovers the original momentum equation, i.e., Eq.~(\ref{eq:incomp1}). Equation~(\ref{eq:incomp3}), also known as the intermediate step, is straightforward to solve since only one unknown is present in the equation, that is, the intermediate velocity, $\Tilde{\mathbf{u}}$. However, Eq.~(\ref{eq:incomp4}) has two unknown variables and cannot be solved in its present form. To address this issue, by taking the divergence of Eq.~(\ref{eq:incomp4}) and knowing that the velocity field solution at time step $n+1$ should be divergence-free, $\nabla \cdot \mathbf{u}^{n+1}=0$, the following equation will be obtained
\begin{equation}
    \nabla \left(\frac{1}{\rho} \nabla p^{n+1} \right) = \frac{1}{\Delta t} \nabla \cdot \Tilde{\mathbf{u}}^{n+1},
    \label{eq:incomp5}
\end{equation}
which is a Poisson equation for pressure. Since the density is constant, Eq.~(\ref{eq:incomp5}) can be simplified as
\begin{equation}
    \Delta p^{n+1} = \frac{\rho}{\Delta t} \nabla \cdot \Tilde{\mathbf{u}}^{n+1}.
    \label{eq:incomp6}
\end{equation}
In this way, the intermediate velocity field, which was initially computed without forcing incompressibility, is projected into the divergence-free field, satisfying Eq.~(\ref{eq:incomp2}). Solving the Poisson equation for pressure results in the solution for the pressure field. Lastly, the velocity field at time step $n+1$ can be updated by rearranging Eq.~(\ref{eq:incomp4}) to be read as
\begin{equation}
      \mathbf{u}^{n+1} = \Tilde{\mathbf{u}}^{n+1} - \left(\frac{\Delta t}{\rho} \right) \nabla p^{n+1}.
      \label{eq:incomp7}
\end{equation}
Here, the projection method is explained using the first-order Euler scheme to discretize time derivative terms. The projection method can be easily extended to other numerical temporal-integration schemes, such as the third-order Runge--Kutta method employed in this study.

Using the notation introduced in Section~\ref{sec:griddiscretization}, the $n^{\mathrm{th}}-$order spatial discretization of the diffusion term for the $x_1-$component can be written as

\begin{eqnarray}
\label{eq:diffusion}
     \left[\nabla \cdot \mu \left(\nabla \mathbf{u} + \nabla \mathbf{u}^T \right) \right]_{x_{1}}^{n\mathrm{th-order}} =  \frac{\delta _{n\mathrm{th}}}{\delta_{n\mathrm{th}} \, x_1} \left[ 2 \mu \left(\frac{\delta_{n\mathrm{th}} \, u_1}{\delta_{n\mathrm{th}} \, x_1} \right) \right] 
     + \frac{\delta_{n\mathrm{th}}}{\delta_{n\mathrm{th}} \, x_2} \left[ \overline{\overline{\mu}^{2\mathrm{nd} \, x_1}}^{2\mathrm{nd} \, x_2} \left(\frac{\delta_{n\mathrm{th}} \, u_1}{\delta_{n\mathrm{th}} \, x_2} + \frac{\delta_{n\mathrm{th}} \, u_2}{\delta_{n\mathrm{th}} \, x_1} \right)  \right] \nonumber \\
      + \frac{\delta_{n\mathrm{th}}}{\delta_{n\mathrm{th}} \, x_3} \left[ \overline{\overline{\mu}^{2\mathrm{nd} \, x_1}}^{2\mathrm{nd} \, x_2} \left(\frac{\delta_{n\mathrm{th}} \, u_1}{\delta_{n\mathrm{th}} \, x_3} + \frac{\delta_{n\mathrm{th}} \, u_3}{\delta_{n\mathrm{th}} \, x_1} \right)  \right],
\end{eqnarray}
where $\mathbf{u} = \left(u_1, u_2, u_3 \right)$. In this study, the second-order finite-difference scheme is implemented to discretize the diffusion term. Hence, the $x_1-$ and $x_2-$components of the diffusion term, $\nabla \cdot \mu \left(\nabla \mathbf{u} + \nabla \mathbf{u}^\mathrm{T} \right)$, can be computed as
\begin{subequations}
 \begin{eqnarray}
     \left[\nabla \cdot \mu \left(\nabla \mathbf{u} + \nabla \mathbf{u}^\mathrm{T} \right) \right]_{x_1}^{2\mathrm{nd-order}} = \frac{\delta_2 \, T_{x_1x_1}}{\delta_2 \, x_1} 
     +  \frac{\delta_2 \, T_{x_1x_2}}{\delta_2 \, x_2}, 
 \end{eqnarray}   
 and
 \begin{eqnarray}
     \left[\nabla \cdot \mu \left(\nabla \mathbf{u} + \nabla \mathbf{u}^\mathrm{T} \right) \right]_{x_2}^{2\mathrm{nd-order}} =\frac{\delta_2 \, T_{x_1x_2}}{\delta_2 \, x_1} 
     + \frac{\delta_2 \, T_{x_2x_2}}{\delta_2 \, x_2}, 
 \end{eqnarray}  
 where
\begin{equation}
     T_{x_1x_1} = 2 \mu \left(\frac{\delta_2 \, u_1}{\delta_2 \, x_1} \right),
\end{equation}
\begin{equation}
      T_{x_1x_2} = \mu \left( \frac{\delta_2 \, u_1}{\delta_2 \, x_2}  + \frac{\delta_2 \, u_2}{\delta_2 \, x_1} \right),
\end{equation}
and
\begin{equation}
    T_{x_2x_2} = 2 \mu \left(\frac{\delta_2 \, u_2}{\delta_2 \, x_2} \right).
\end{equation}
\end{subequations}
The discretization can be simply extended to the three-dimensional case, $\mathbf{x}=(x_1, x_2, x_3)$.

Most previous numerical studies have employed a second-order discretization for the convection term, $\nabla \cdot (\mathbf{uu})$, in the context of two-phase flows. 
This choice is made because discontinuities are expected to arise in the velocity fields across the interface, and transitioning to a higher-order finite difference scheme introduces numerical challenges. In this study, instead of utilizing second-order interpolation, we adopted fifth-order WENO interpolation for discretizing the convection term. This approach has been demonstrated to yield approximately third-order accuracy for the single-phase incompressible solver (please refer to Appendix~\ref{appendixD}) and provides second-order accuracy for the two-phase solver. This improvement is notable when compared to other second-order based numerical discretizations. The employed discretization of $\nabla \cdot (\mathbf{uu})$ term for $\mathbf{x}=(x_1, x_2)$ in this study is given as
\begin{subequations}
    \begin{eqnarray}
        \left[ \nabla \cdot \left(\mathbf{u}\mathbf{u} \right) \right]_{x_1} = \frac{\delta_2}{\delta_2 \, x_1} \left(\overline{u_1}^{5^{\mathrm{th}} \mathrm{WENO} \, \, x_1} \, \, \overline{u_1}^{5^{\mathrm{th}} \mathrm{WENO} \, \, x_1} \right) + \frac{\delta_2}{\delta_2 \, x_2} \left(\overline{u_1}^{5^{\mathrm{th}} \mathrm{WENO} \, \, x_2} \, \, \overline{u_2}^{5^{\mathrm{th}} \mathrm{WENO} \, \, x_1} \right), 
    \end{eqnarray}
    and
    \begin{eqnarray}
        \left[ \nabla \cdot \left(\mathbf{u}\mathbf{u} \right) \right]_{x_2} = \frac{\delta_2}{\delta_2 \, x_1} \left(\overline{u_1}^{5^{\mathrm{th}} \mathrm{WENO} \, \, x_2} \, \, \overline{u_2}^{5^{\mathrm{th}} \mathrm{WENO} \, \, x_1} \right) + \frac{\delta_2}{\delta_2 \, x_2} \left(\overline{u_2}^{5^{\mathrm{th}} \mathrm{WENO} \, \, x_2} \, \, \overline{u_2}^{5^{\mathrm{th}} \mathrm{WENO} \, \, x_2} \right).
    \end{eqnarray}
\end{subequations}

\color{black}
The Poisson equation for the pressure is discretized using the second-order central finite-difference scheme. The solution of the Poisson equation is computed at each iteration of the simulation by employing Krylov methods and multi-grid preconditioning implemented in the PETSc package library~\citep{petsc}. 

\begin{figure}
\centering
\includegraphics[width = 0.50\textwidth]{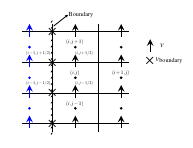}
\caption{\label{fig:boundary} The proper boundary condition for the tangential velocity is implemented by employing ghost points located outside of the computational domain, shown in blue. The tangential velocity of these ghost points is specified to enable the linear interpolation to produce the intended tangent velocity at the wall.}
\end{figure}

In the staggered grid arrangement shown in Fig.~\ref{fig:grid}, boundary conditions for the velocity components normal to the boundary can be readily implemented. However, imposing the proper boundary condition for the tangential components of the velocity can be more challenging. Thus, ghost points are used to apply the desired boundary condition for velocity components tangential to the boundary~\citep{tryggvason2011}. For example, consider a case where the no-slip boundary condition is needed to be implemented for the velocity component in the $y-$direction, $v$, at the left boundary (see Fig.~\ref{fig:boundary}). In order to update the value of $v_{i, j+1/2}$, which has a half-cell offset from the left boundary, and to impose a proper boundary, the value of $v_{i-1, j+1/2}$ is needed that is located outside of the domain. The tangential velocity at this ghost point can be obtained by using linear interpolation as $ v_{\mathrm{boundary}}=\left(v_{i, j+1/2} + v_{i-1, j+1/2} \right)/2$, where $v_{\mathrm{boundary}}$ is the velocity value at the boundary, equal to the wall tangential velocity for the no-slip case. Therefore, the value at the ghost point can be calculated as $v_{i-1, j+1/2} = 2 v_{\mathrm{boundary}} - v_{i, j+1/2}$, and the no-slip boundary condition is imposed. The same approach can be taken for other boundary conditions as well. For instance, for the slip boundary condition where the derivative of the tangential velocity is zero, the value of the ghost point is given as $v_{i, j+1/2} = v_{i-1, j+1/2}$. In Appendix~\ref{appendixD}, the order of accuracy and performance of the introduced incompressible solver are examined. Interested readers can refer to this section for more details.

\section{\label{sec:twophase} Implementation of two-phase solver for nonmagnetic flows}

This section presents the methodology implemented for simulating incompressible two-phase flows. The introduced solver is based on coupling the conservative level set (CLS) approach, detailed in Section~\ref{sec:levelset} with the incompressible solver presented in Section~\ref{sec:incompressible}. In two-phase flows, material properties such as density and viscosity experience a jump across the interface, requiring special numerical treatments to avoid the appearance of numerical instabilities. In addition, proper boundary conditions must be imposed across the interface to obtain accurate physical results. This section, thus, discusses these issues and introduces a complete solution procedure for modelling two-phase flows.

\subsection{\label{sec:BCs} One-fluid formulation}
The governing equation for the one-fluid formulation approach to describe the two-phase incompressible flows is given as
\begin{equation}
    \frac{\partial \mathbf{u}}{\partial t} + \nabla \cdot \left(\mathbf{u} \mathbf{u} \right) = -\frac{1}{\rho} \nabla p + \frac{1}{\rho} \nabla \cdot \mu \left(\nabla \mathbf{u} + \nabla \mathbf{u}^\mathrm{T} \right) + \mathbf{g} + \frac{1}{\rho} \mathbf{F}_\mathrm{v},
    \label{eq:twophasenonconduct}
\end{equation}
where $\mathbf{F}_\mathrm{v}$ represents any volume force that may be present. The solution of Eq.~(\ref{eq:twophasenonconduct}) should also satisfy the incompressibility constraint. In each phase, the material properties are constant, that is, $\rho=\rho_{\mathrm{l}}$ and $\mu=\mu_{\mathrm{l}}$ for the liquid phase, while $\rho=\rho_{\mathrm{g}}$ and $\mu=\mu_{\mathrm{g}}$ for the gas phase. However, across the thin interface, denoted by $\Gamma$, fluid properties experience a jump that can be written as $[\rho]_\Gamma=\rho_\mathrm{l} - \rho_\mathrm{g}$ and $[\mu]_\Gamma=\mu_\mathrm{l} - \mu_\mathrm{g}$ for the density and dynamic viscosity, respectively. Since there is no mass exchange between the two phases, the normal component of the velocity should be continuous across the interface, i.e.,
\begin{equation}
    [\mathbf{u} \cdot \mathbf{n}]_\Gamma = \mathbf{u}_{\mathrm{l}} \cdot \mathbf{n} - \mathbf{u}_{\mathrm{g}} \cdot \mathbf{n} = 0,
\end{equation}
where $\mathbf{n}$ is the normal vector to the interface. For viscous flows, the tangential component of the velocity should also be equal for the two phases at the interface. Thus, the velocity field should be continuous across the interface, and the boundary condition for the velocity can be written as
\begin{equation}
    [\mathbf{u}]_\Gamma = 0.
\end{equation}
Additionally, applying the momentum conservation principle to a control volume located at the interface leads to the following boundary condition for the pressure jump
\begin{equation}
    [\left(-p\mathbf{I} + \mu \left(\nabla \mathbf{u} + \nabla \mathbf{u}^\mathrm{T} \right) \right) \cdot \mathbf{n}]_\Gamma = 0,
\end{equation}
where $\mathbf{I}$ is the identity tensor. If the surface tension force is considered, the boundary condition for the pressure jump across the interface is modified as
\begin{equation}
    [\left(-p\mathbf{I} + \mu \left(\nabla \mathbf{u} + \nabla \mathbf{u}^\mathrm{T} \right) \right) \cdot \mathbf{n}]_\Gamma = \sigma \kappa \mathbf{n},
\end{equation}
where variable $\sigma$ denotes the surface tension coefficient. The curvature of the interface, $\kappa$, is calculated as $\kappa = - \nabla \cdot \mathbf{n} = - \nabla \cdot \left(\nabla \psi/|\nabla \psi| \right)$. In the present study, the corresponding jump condition in the pressure gradient is modelled by employing the continuum surface force (CSF) method of Brackbill \textit{et al.}~\cite{brackbill199}.
The CSF approach defines surface tension as a volume force spreading over the finite interface width, expressed as $F_{\sigma} = \sigma \kappa \nabla \psi$. Therefore, proper calculation of the interface curvature is essential for accurate and robust surface tension modelling, which will be discussed in more detail in the next section. The viscous term is also discretized using the CSF method.

\subsection{\label{sec:twophaseDiscretization} Projection method and discretization}
As previously introduced, the projection method is employed to solve the incompressible Navier--Stokes equations. This method involves two steps, known as prediction and correction.  During the prediction step, Eq.~(\ref{eq:twophasenonconduct}) is solved, ignoring the pressure term, to advance the velocity field $\mathbf{u}^n$ to an intermediate velocity $\Tilde{\mathbf{u}}^{n+1}$. In the correction step, the intermediate velocity is projected to its divergence-free solution, $\mathbf{u}^{n+1}$, using the solution of the pressure Poisson equation. The projection method has been discussed in Sec.~\ref{sec:incompressibleProjection} for the single incompressible flow case. However, for two-phase flows, special considerations and numerical treatments must be taken into account. The discretization of the convective term, $\nabla \cdot \left(\mathbf{u} \mathbf{u} \right)$, can be performed similarly to the one demonstrated earlier in Sec.~\ref{sec:incompressibleProjection}.
The main challenges associated with treating the interface discontinuities in two-phase incompressible flows include properly modelling the viscosity discontinuity across the interface and accurately calculating the curvature. The viscosity term in the momentum equation must be appropriately discretized, particularly in the presence of a dynamic viscosity jump, to ensure the accurate calculation of kinetic energy dissipation near the interface. Additionally, 
a robust and accurate method to assess the evolution of the interface curvature should be employed to model the surface tension force. Another issue in two-phase flow simulations pertains to the numerical discretization of the pressure gradient term. The pressure gradient component in Eq.~(\ref{eq:twophasenonconduct}) includes the density term as well, and since the density has a jump across the interface, specific consideration is required. We will conclude this section by discussing these issues in more detail and presenting the discretization used to model viscosity, surface tension, and pressure jump terms.

\subsubsection*{Viscosity and surface tension modelling}
The viscosity term, $\nabla \cdot \mu \left(\nabla \mathbf{u} + \nabla \mathbf{u}^\mathrm{T} \right)$, can be discretized using Eq.~(\ref{eq:diffusion}), which displays the general form of the $n^\mathrm{th}-$order spatial discretization of this term, considering a non-constant dynamic viscosity field. The implemented two-phase solver employs the second-order finite-difference scheme and second-order linear interpolation to discretize this term. Furthermore, the density and viscosity are assumed to smoothly vary over the interface~\citep{olsson2005}. Therefore, the density and viscosity fields can be represented using the level set function as
\begin{equation}
\label{eq:density_update}
    \rho \left(\mathbf{x}, t \right) = \rho_\mathrm{g} + (\rho_\mathrm{l} - \rho_{\mathrm{g}}) \psi \left(\mathbf{x}, t \right)
\end{equation}
and
\begin{equation}
\label{eq:viscosity_update}
    \mu \left(\mathbf{x}, t \right) = \mu_\mathrm{g} + (\mu_\mathrm{l} - \mu_{\mathrm{g}}) \psi \left(\mathbf{x}, t \right),
\end{equation}
respectively. As a result, the discontinuity in the density and viscosity fields is smoothed out across the interface within a few layers of cells, which is the function of the interface thickness. Thus, numerical instabilities that may appear due to the sharp jump in fluid properties are avoided in the solution.   

According to the CSF model introduced by Brackbill \textit{et al.}~\cite{brackbill199}, the surface tension force per unit interfacial area between two fluids with a constant surface tension coefficient, $\sigma$, can be written as
\begin{equation}
    \mathbf{F}_{\mathrm{sa}}\left(\mathbf{x}_\Gamma \right) = \sigma \kappa\left(\mathbf{x}_\Gamma \right) \mathbf{n}\left(\mathbf{x}_\Gamma\right),
\end{equation}
where $\mathbf{x}_\Gamma$ is an arbitrary point located on the interface. The introduced interfacial surface tension force can be recast as 
\begin{equation}
    \mathbf{F}_{\mathrm{sv}} (\mathbf{x}) = \sigma \overbrace{\left(- \nabla \cdot \frac{\nabla \psi}{|\nabla \psi|} \right)}^{\kappa} \nabla \psi, 
\end{equation}
which represents the surface tension volume force at any point $\mathbf{x}$~\citep{olsson2005}. The introduced force $\mathbf{F}_{\mathrm{sv}} (\mathbf{x})$ results in the same total force as that of $\mathbf{F}_\mathrm{sa} (\mathbf{x}_\Gamma)$, but it is distributed over the width of the interface. This approximation is only valid for small interface thicknesses. Therefore, it is essential to avoid excessively wide interfaces and to maintain a constant thickness for the interface during the simulation, which has been addressed in the implementation of the level set (see Sec.~\ref{sec:lsReinit}). In order to calculate the surface tension force, the solver should be able to robustly and accurately compute the curvature value. Employing high-order numerical schemes to calculate curvature results in oscillations appearing in the curvature field, leading to an unphysical solution for the velocity field, known as spurious currents. To tackle this issue, first- or second-order schemes are usually used to calculate curvature values. Various methods have been developed, such as utilizing height functions or implementing the least-squares minimization approach~\citep{boniou2022}, to formulate a robust and consistent framework for curvature calculation. However, most of these approaches are computationally expensive and difficult to implement. In this study, curvature computation is based on using the calculated face normals introduced in Sec.~\ref{sec:lsReinit}. Therefore, the curvature field for the two-dimensional case is given as
\begin{equation}
\label{eq:curvature}
    \kappa = \frac{\delta_{2\mathrm{nd}} \, (\mathbf{n}_x)^{x-\mathrm{face}}}{\delta_{2\mathrm{nd}} \, x} + \frac{\delta_{2\mathrm{nd}} \, (\mathbf{n}_y)^{y-\mathrm{face}}}{\delta_{2\mathrm{nd}} \, y}.
\end{equation}
Equation (\ref{eq:curvature}) is second-order accurate and can be easily implemented, requiring no additional computational effort. Various test cases have been conducted to evaluate the results obtained from the Eq.~(\ref{eq:curvature}) curvature calculation, demonstrating its robustness. Notably, the results exhibit no discernible oscillations, which will be expounded upon in the subsequent section. Finally, for the two-dimensional case of $\mathbf{x}=(x,y)$, the $x-$ and $y-$components of the surface tension force spatial discretization can be written as

\begin{subequations}
\begin{equation}
    [\sigma \kappa \nabla \psi]_{x-\mathrm{comp}} =  \sigma \overline{\left(\frac{\delta_{2\mathrm{nd}} \, (\mathbf{n}_x)^{x-\mathrm{face}}}{\delta_{2\mathrm{nd}} \, x} + \frac{\delta_{2\mathrm{nd}} \, (\mathbf{n}_y)^{y-\mathrm{face}}}{\delta_{2\mathrm{nd}} \, y} \right)}^{2\mathrm{nd} \, x}  \frac{\delta_{2\mathrm{nd}} \, \psi}{\delta_{2\mathrm{nd}} \, x}
\end{equation}
\text{and}
\begin{equation}
        [\sigma \kappa \, \nabla \psi]_{y-\mathrm{comp}}  = \sigma \overline{\left(\frac{\delta_{2\mathrm{nd}} \, (\mathbf{n}_x)^{x-\mathrm{face}}}{\delta_{2\mathrm{nd}} \, x} + \frac{\delta_{2\mathrm{nd}} \, (\mathbf{n}_y)^{y-\mathrm{face}}}{\delta_{2\mathrm{nd}} \, y} \right)}^{2\mathrm{nd} \, y}  \frac{\delta_{2\mathrm{nd}} \, \psi}{\delta_{2\mathrm{nd}} \, y},
\end{equation}
\end{subequations}
respectively.

\subsubsection*{Poisson equation}
In the two-phase incompressible solver, since the density field is not constant in the computational domain, the Poisson equation cannot be simplified as Eq.~(\ref{eq:incomp6}), and the variable coefficients should be considered while discretizing the Poisson equation. Therefore, the second-order discretization of the Poisson equation can be written as

\begin{eqnarray}
\label{eq:Poisson}
    \nabla \left(\frac{1}{\rho} \nabla p \right) \Bigg|_ {i,j} = \frac{\partial}{\partial x} \left(\frac{1}{\rho} \frac{\partial p}{\partial x} \right) \Bigg|_ {i,j}+ \frac{\partial}{\partial y} \left(\frac{1}{\rho} \frac{\partial p}{\partial y} \right) \Bigg|_ {i,j} 
    = \frac{\left(\frac{1}{\rho} \frac{\partial p}{\partial x} \right)_ {i+1/2,j} - \left(\frac{1}{\rho} \frac{\partial p}{\partial x} \right)_ {i-1/2,j}}{\Delta x}
    + \frac{\left(\frac{1}{\rho} \frac{\partial p}{\partial y} \right)_{i,j+1/2} - \left(\frac{1}{\rho} \frac{\partial p}{\partial y} \right)_{i,j-1/2}}{\Delta y} \nonumber \\
    = \frac{\left(\frac{2}{\rho_{i+1, j} + \, \rho_{i,j}} \frac{p_{i+1,j} - p_{i,j}}{\Delta x} \right) - \left(\frac{2}{\rho_{i, j} + \, \rho_{i-1,j}} \frac{p_{i,j}-p_{i-1,j}}{\Delta x} \right)}{\Delta x} 
    + \frac{\left(\frac{2}{\rho_{i, j+1} + \, \rho_{i,j}} \frac{p_{i,j+1} - p_{i,j}}{\Delta y} \right) - \left(\frac{2}{\rho_{i, j} + \, \rho_{i,j-1}} \frac{p_{i,j}-p_{i,j-1}}{\Delta y} \right)}{\Delta y}. 
\end{eqnarray}
It is noteworthy to mention that second-order interpolation is used to calculate density values at cell faces, e.g., $(1/\rho)_{i+1/2,j}=2/\left(\rho_{i+1,j} + \rho_{i,j} \right)$. The same interpolation should also be applied while calculating the $1/\rho$ coefficients for pressure gradient, viscosity, and surface tension terms in Eq.~(\ref{eq:twophasenonconduct}). Furthermore, it is imperative to use the same finite-difference scheme when computing the gradient of pressure and level set gradient for determining the surface tension force. This ensures that the gradient operator provides the proper force balance between pressure gradient and surface tension across the interface, thereby reducing the formation of unphysical spurious velocities in the solution~\citep{boniou2022}.

\subsection{\label{sec:procedure} Solution procedure}

The complete solution procedure for the two-phase incompressible solver can be summarized as follows:
\begin{enumerate}[]
    \item The conservative level set (CLS) approach is used to implicitly advance the location of the interface from $t^n$ to $t^{n+1}$, employing the velocity field at $t^n$.
    \item The updated location of the interface, $\psi^{n+1}$, is then utilized to obtain the density and viscosity fields at $t^{n+1}$. Thus, $\rho^{n+1}$ and $\mu^{n+1}$ fields are computed using Eqs.~(\ref{eq:density_update}) and (\ref{eq:viscosity_update}).
    \item The intermediate velocity field at time step $t^{n+1}$ is calculated by solving Eq.~(\ref{eq:twophasenonconduct}), while ignoring the pressure term.
    \item The obtained intermediate velocity is then projected into a divergence-free field by solving the Poisson equation for the pressure based on the discretization introduced in Eq.~(\ref{eq:Poisson}).
    \item The correct velocity field is calculated at time $n+1$ by solving Eq.~(\ref{eq:incomp7}). The obtained velocity at $t^{n+1}$ is then employed to advance the level set profile for the next time step.
\end{enumerate}

\noindent Based on the CFL condition, the stability constraint for the time step due to convection, viscosity, and surface tension terms is given as~\citep{boniou2022, finster2023}

\begin{equation}
\label{eq:cfl}
    \Delta t \leq \mathrm{min} \left(\frac{\Delta x}{\mathrm{max}(||\mathbf{u}||)}, \sqrt{\frac{{\Delta x}^3 \left(\rho_\mathrm{l} + \rho_\mathrm{g} \right)}{4 \pi \sigma}}, \frac{1}{4} \frac{{\Delta x}^2}{ \mathrm{max}(\nu_\mathrm{l}, \nu_\mathrm{g})} \right).
\end{equation}

\noindent Usually, the surface tension limits the time step; however, for large density ratios, the viscosity term may be more restrictive.

\subsection{\label{sec:twophaseTests} Two-phase solver test cases}

To investigate the accuracy, robustness, and performance of the implemented two-phase solver, three test cases are studied: the static droplet, damping surface wave, and Rayleigh–Taylor instability. The static droplet test case, which aims to evaluate the capability of the two-phase solver to accurately calculate curvature and model surface tension forces, is presented in Appendix~\ref{appendixE}, and interested readers are encouraged to refer to this section for more detail. The latter two test cases are examined in this section. The damping surface wave benchmark is employed to investigate the order of accuracy of the implemented solver and the interaction between viscosity and surface tension terms in a simulation. Lastly, the Rayleigh--Taylor instability is investigated to assess the performance of the solver in handling the evolution of a complex interface and also in treating cases where a high-density ratio exists across the interface.

\subsubsection{Damping surface wave}
The viscous damping of a surface wave is a well-known test case in the literature, widely used to assess the capability of the implemented solver to accurately model the interaction between viscosity and surface tension forces. In this test case, two superimposed fluids with density $\rho_1$ and $\rho_2$ are separated by a sinusoidal interface with wavelength $\lambda$ and amplitude ${A}_0$, and thus the perturbed interface profile is given as 
\begin{equation}
    y = y_0 - {A}_0 \cos(2 \pi x/ \lambda).
\end{equation}
For the case where both fluids have the same kinematic viscosity, $\nu$, and constant surface tension, $\sigma$, the analytical solution for the evolution of the wave amplitude with time is derived by Prosperetti \cite{prosperetti1981} by employing the initial value theorem. The initial value solution is obtained as~\citep{prosperetti1981}
\begin{eqnarray}
{A}_\mathrm{exact}(t) = \frac{4(1-4\beta) k^4 \nu^2}{8(1-4\beta) k^4 \nu^2 + {\omega_0}^2} \, {A}_0 \, \mathrm{erfc} \left(\sqrt{\nu k^2 t} \right) 
+ \sum\limits_{i=1}^{4} \frac{z_i}{Z_i} \left(\frac{{\omega_0}^2 \, A_0}{{z_i}^2 - \nu k^2} \right) \exp [\left({z_i}^2 - \nu k^2 \right) t] \mathrm{erfc}\left(z_i \sqrt{t} \right),
\end{eqnarray}
where $z_i$ are the roots of
\begin{eqnarray}
    z^4 - 4 \beta \sqrt{k^2 \nu} \, z^3 + 2 \left(1-6\beta \right) k^2 \nu \, z^2 
    + 4(1-3\beta) (k^2 \nu)^{\frac{3}{2}} \, z + (1-4\beta) k^4 \nu^2 + {\omega_0}^2 = 0.
\end{eqnarray}
The dimensionless parameter $\beta$ is defined as $\beta=\rho_1 \rho_2/\left(\rho_1 + \rho_2 \right)^2$, while the inviscid frequency of the wave oscillation is given by $\omega_0 = \sqrt{\frac{\sigma k^3}{\rho_1 + \rho_2}}$, and $Z_i = \Pi_{\substack{j=1 \\ j\neq i}} ^4 \left(z_j - z_i \right)$. \\

\begin{figure*}
\includegraphics[width = 1\textwidth]{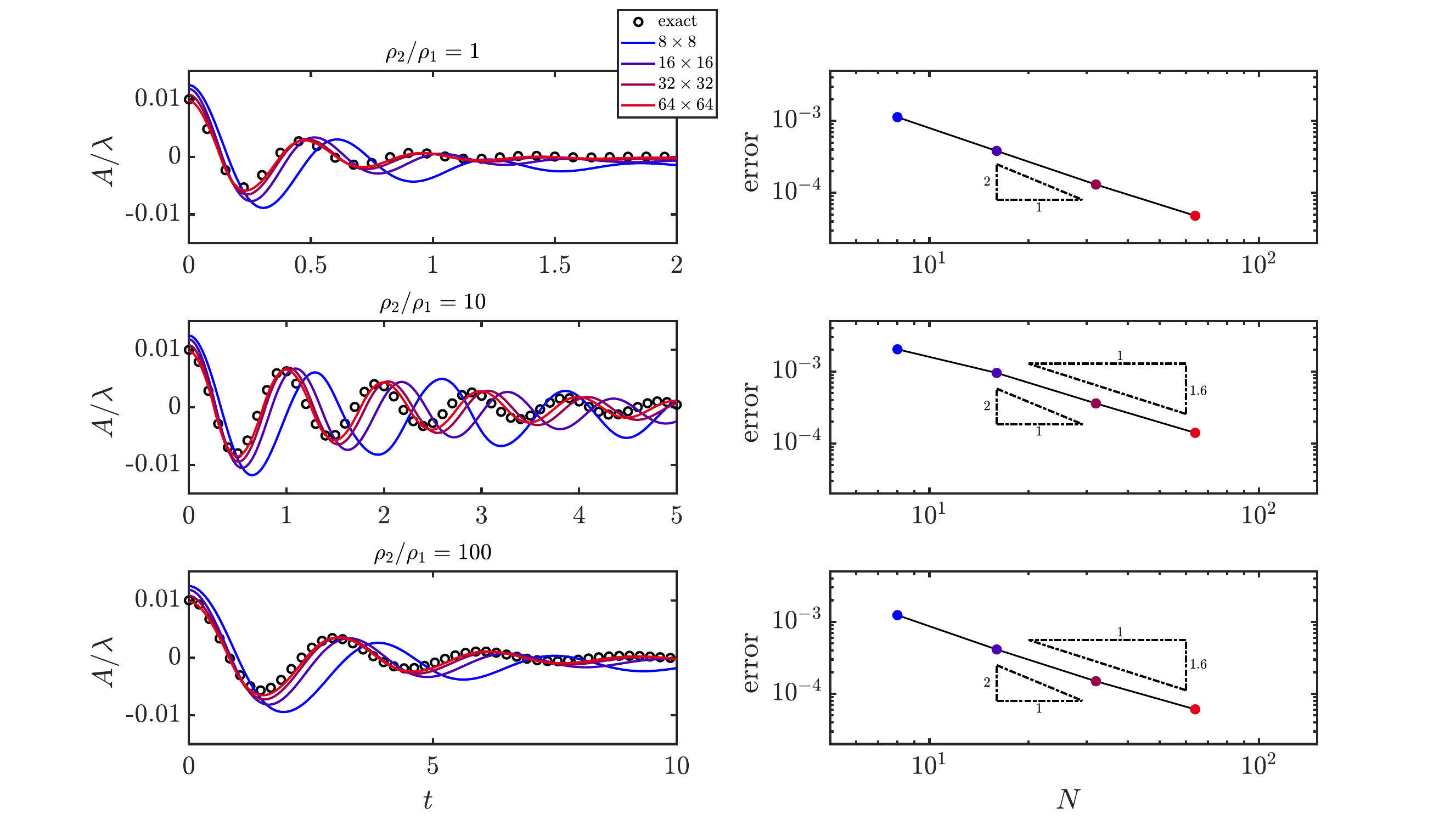} 
\caption{\label{fig:CapillaryWave} The solutions of the damped surface wave problem for three density ratios of [top]~unity, [middle]~10, and [bottom]~100, are represented for four different mesh resolutions of $8\times8$, $16\times16$, $32\times32$, and $64\times64$. The accuracy analysis of the implemented two-phase solver is also illustrated for all three density ratios.}
\end{figure*}

Here, the simulation is performed in a computational domain of $[0, 1] \times [0, 1]$, with periodic boundary conditions along the $x-$direction and slip wall boundary along the $y-$direction, and $y_0$ is set to $y_0=0.5$. The wavelength of the perturbation is set to $\lambda= 1$, and the initial amplitude of the wave is $A_0=0.01 \lambda$. Three scenarios are investigated, namely, density ratios of 1, $10$, and $1000$, for four different grid resolutions of $8 \times 8$, $16 \times 16$, $32 \times 32$, and $64 \times 64$, with the interface thickness being set to $\epsilon=\Delta x/2$. 
For the first case of unity density ratio, the surface tension coefficient is $\sigma = 2$, with the constant kinematic viscosity for both fluids being set to $\nu=0.0647$, and $\rho_1 = \rho_2 = 1$. The simulation was run for all different mesh resolutions with the constant time step $\Delta t= 0.0005$. Figure~\ref{fig:CapillaryWave}[top-left] displays the time evolution of the normalized wave amplitude, i.e., ${A}/\lambda$, for all four meshes, along with the analytical solution. This figure visually confirms that, as mesh resolution is increased, the numerical solution converges to the analytical one. For a better quantitative comparison, the rms value of the error in the wave amplitude is plotted against the mesh resolution in Fig.~\ref{fig:CapillaryWave}[top-right].

As can be concluded from the computed error values, although for the coarsest mesh resolution, the solver is able to capture the proper behaviour of the surface wave, a small error in the numerical oscillation period of the wave has caused a noticeable error value for this mesh. As the mesh resolution is increased, more accurate results are obtained, and according to Fig.~\ref{fig:CapillaryWave}[top-right], close to a second-order convergence rate is observed. It is noteworthy to mention that for the case of $\rho_2/\rho_1=1$, there is no dynamic viscosity or momentum jump across the interface. Therefore, the numerical errors solely originate from the solution of the level set transport equation and the curvature computation. However, by increasing the density ratio, due to the existence of the density and dynamic viscosity jump across the interface, both convection and viscosity terms of the momentum equation affect the solution. Therefore, the numerical solver is also evaluated for density ratios of 10 and 100 to investigate whether the numerical model is capable of addressing these jump conditions and momentum transfer across the interface in the presence of a surface tension force. For both density ratios of $10$ and $100$, the kinematic viscosity of the fluids is set to $\nu=0.00647$, and time steps are set to $\Delta t=0.001$ and $0.003$, respectively. As expected, as the mesh resolution increases, the numerical results become more accurate for both cases, as shown in Fig.~\ref{fig:CapillaryWave}[middle-left, bottom-left]. According to the convergence study, the approximate convergence rate of 1.6 is achieved for both cases, see Fig.~\ref{fig:CapillaryWave}[middle-right, bottom-right], demonstrating that in the presence of a density and viscosity jump, the solver tends to preserve its order of accuracy.

\subsubsection{Rayleigh--Taylor instability}

The Rayleigh--Taylor instability occurs when a layer of liquid is superimposed to another less dense liquid layer in such a way that by interchanging the fluids, the energy of the system can be reduced. The Rayleigh--Taylor instability has been widely studied in classical hydrodynamics and is studied as a benchmark to evaluate the performance of the implemented two-phase solver for simulating problems containing the highly nonlinear and multi-scale nature of fluid dynamics. To simulate the Rayleigh--Taylor instability, we consider a two-dimensional rectangular domain $[x, y] \in [0, 1] \times [0,  4]$, with a fluid interface parallel to the horizontal axis at $y_0 = 2.0$. The fluid interface is initialized with a small sinusoidal perturbation, whose wavelength and amplitude are $2\pi$ and 0.1, respectively. Following the study by Huang \textit{et al.}~\cite{huang2020}, the material properties are set to $\rho_1 = 1$, $\rho_2 = 3$, $\mu_1 = \mu_2 = 0.01$, $\sigma=10^{-12}$, and the gravity is acting downwards with a magnitude of unity. The simulation is performed for four mesh resolutions of $16\times64$, $32\times128$, $64\times256$, and $128\times512$. For all cases, the time step is set to $\Delta t=5 \times 10^{-4}/\sqrt{\mathrm{At}}$, where the Atwood number is defined as $\mathrm{At}=\left(\rho_2 - \rho_1 \right)/ \left(\rho_1 + \rho_2 \right)$. 

\begin{figure*}
\subfloat[]{\includegraphics[width = 0.95\textwidth]{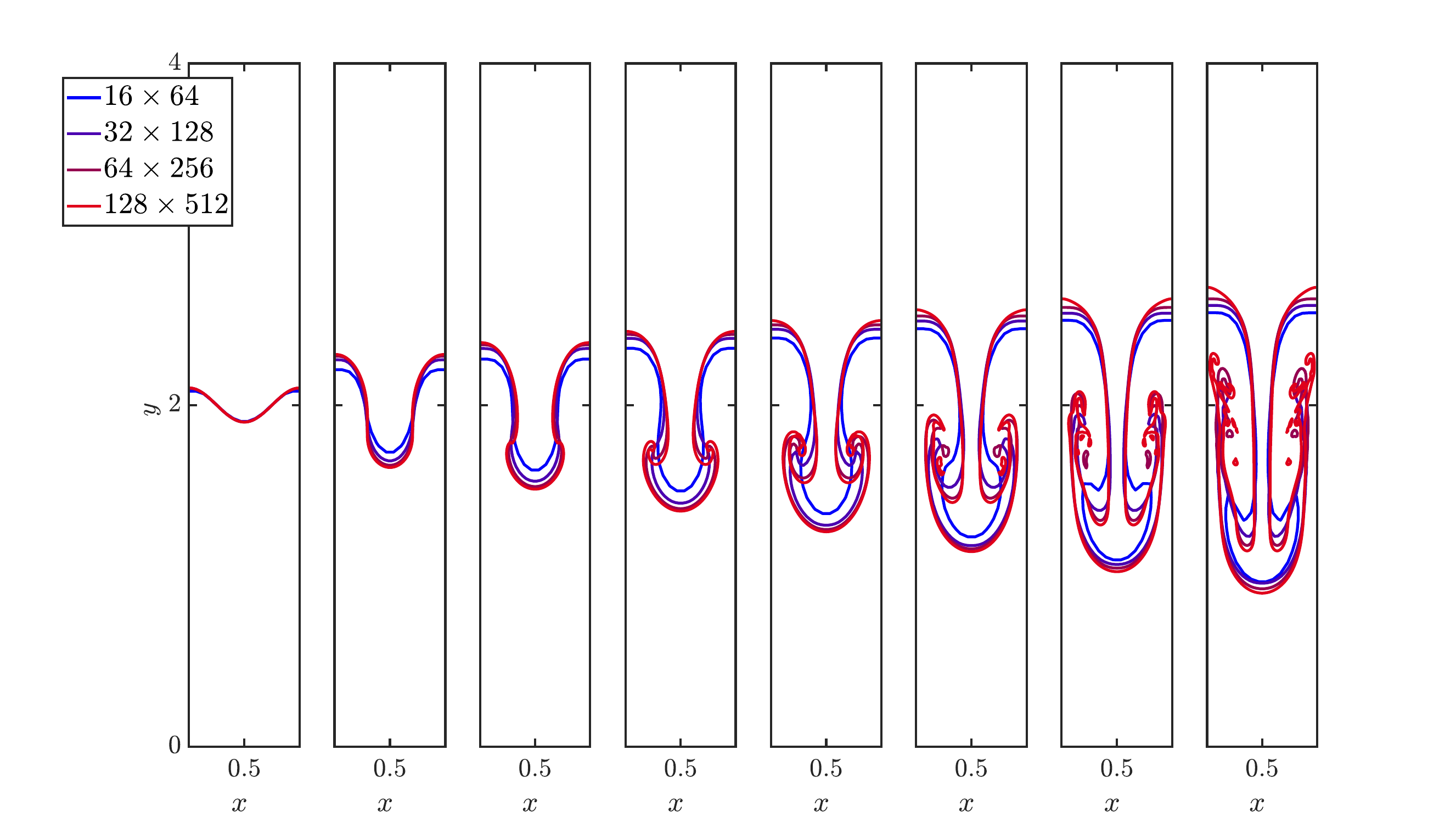}}\\ \hspace*{-2.4cm}
\subfloat[]{\includegraphics[width = 0.62\textwidth]{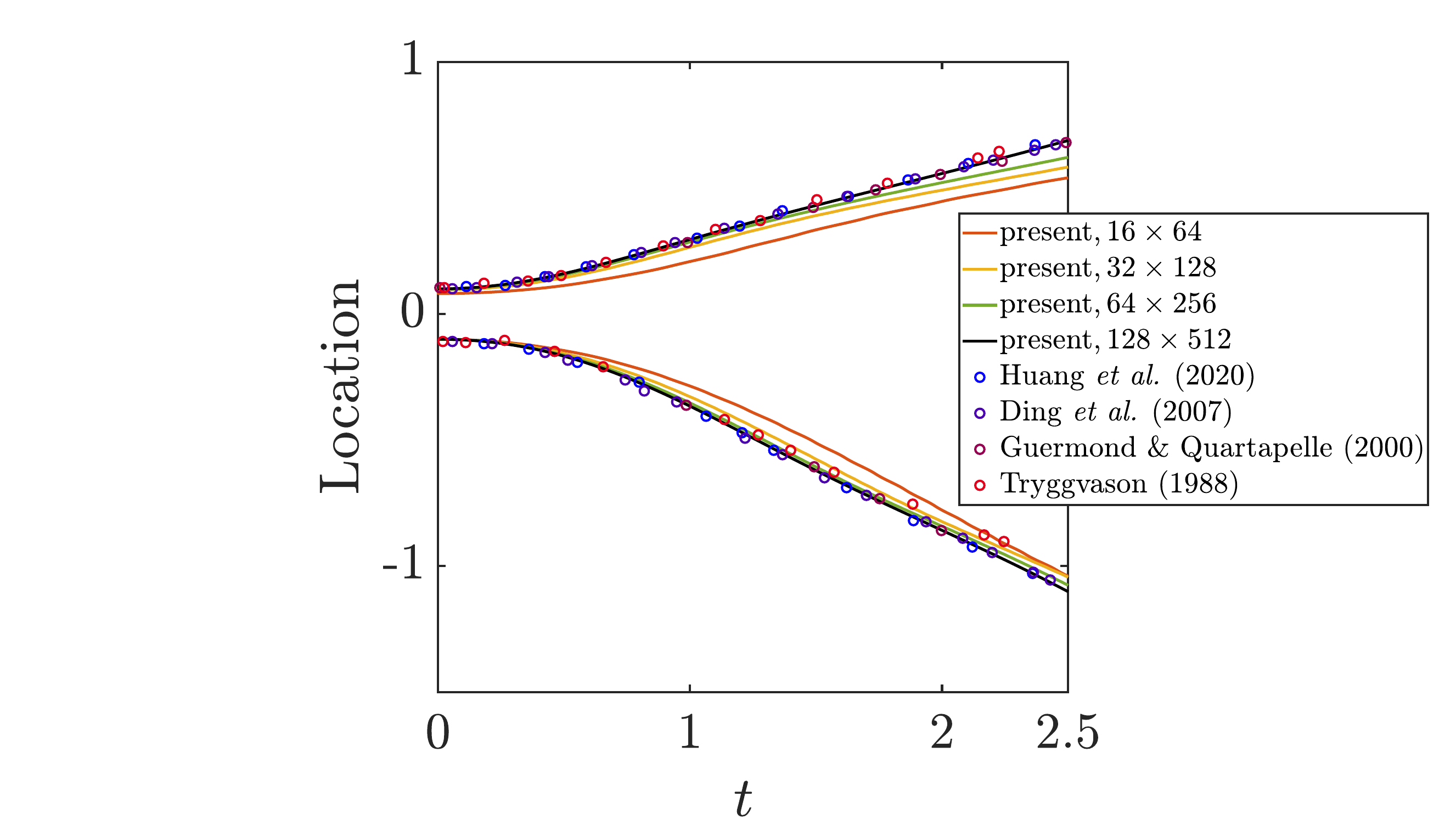}}\hspace*{-0.02cm}
\subfloat[]{\includegraphics[width = 0.62\textwidth]{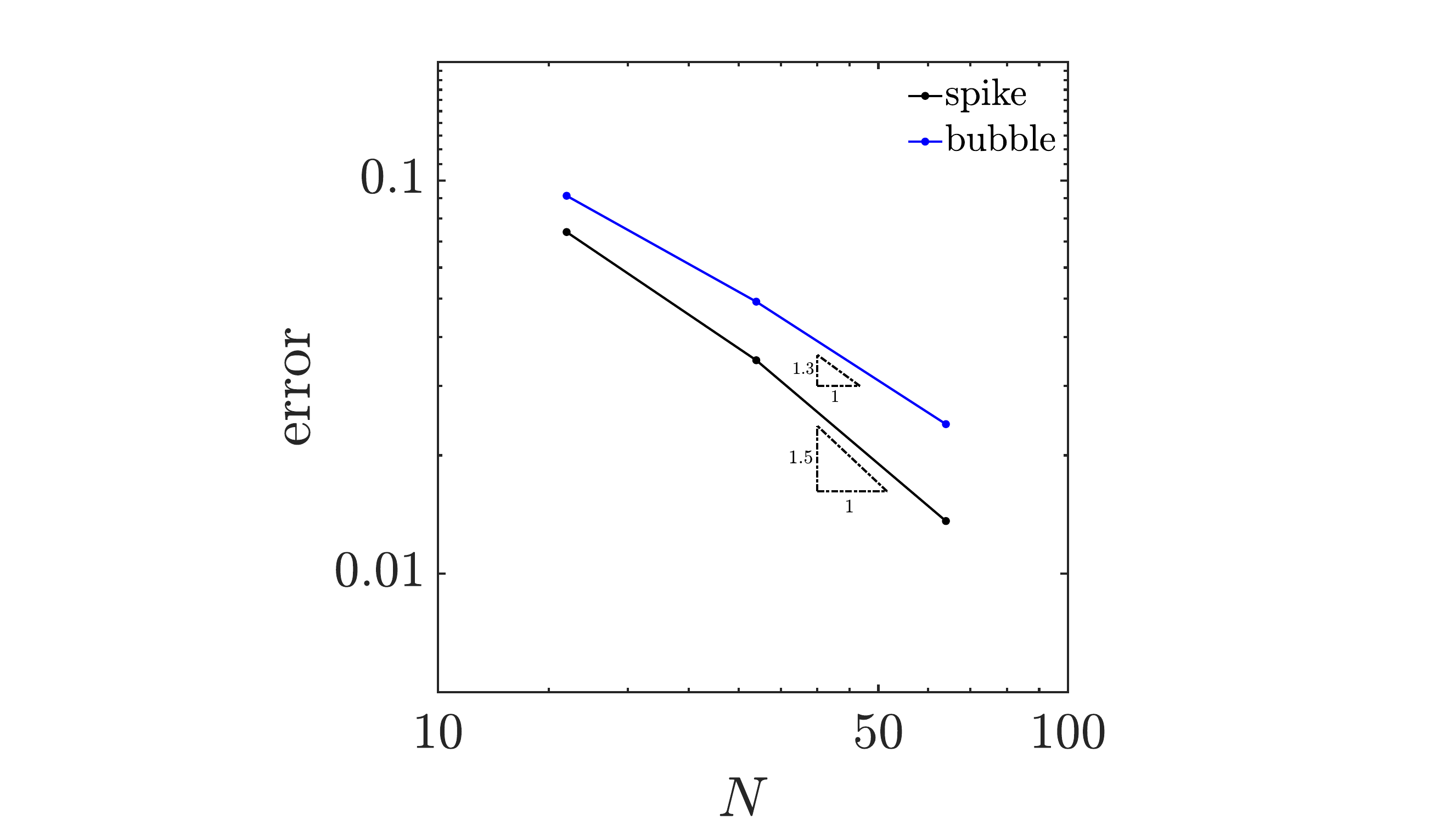}}
\caption{\label{fig:RT1} (\textit{a}) The interface of the Rayleigh--Taylor instability with a density ratio of 3 ($\mathrm{At}=0.5$) at $t\sqrt{\mathrm{At}}=0, 1, 1.25, 1.5, 1.75, 2, 2.25$, and 2.5, from left to right, for four different mesh resolutions. (\textit{b}) The transient location of the spike tip and bubble is depicted during the simulation for four different mesh resolutions of $16\times64$, $32\times128$, $64\times256$, and $128\times512$. The results are compared to four different studies by Huang \textit{et al.}~\cite{huang2020}, Ding \textit{et al.}~\cite{ding2007}, Guermond and Quartapelle~\cite{guermond2000}, and Tryggvason~\cite{tryggvason2011}. (\textit{c}) $L_2$ error in spike tip and bubble locations during the Rayleigh--Taylor simulation with the density ratio of three for the mesh resolutions of $16\times64$, $32\times128$, and $64\times256$.}
\end{figure*}

Figure~\ref{fig:RT1}(a) depicts the results for the four different mesh resolutions at time $t\sqrt{\mathrm{At}}=0$ to 2.5 with an increment of $0.25$. It is visually evident that even for the coarsest mesh resolution case, the main features of the Rayleigh--Taylor instability are captured. However, refining the mesh allows for a better representation of the evolution and growth of bubbles and spikes in the solution. For a more comprehensive quantitative analysis, we validated our results with four different previous studies by Huang \textit{et al.}~\cite{huang2020}, Ding \textit{et al.}~\cite{ding2007}, Guermond and Quartapelle~\cite{guermond2000}, and Tryggvason~\cite{tryggvason2011}. To this end, the transient location of the spike tip and the interface location at the left (right) edge during the simulation are compared to results from previous studies, shown in Fig.~\ref{fig:RT1}(b). As can be observed from Fig.~\ref{fig:RT1}(b), by increasing the mesh resolution, our results converge to ones from earlier studies, especially for the mesh resolution of $128\times512$, the results closely match those reported in the literature. Note that the study performed by Tryggvason~\cite{tryggvason2011} only considered the inviscid case, which explains the slight deviation from the other results. Assuming the solution of the finest grid, $128\times512$, as an analytical solution, the convergence rate of the solution is calculated. To this end, the $L_2$ norm of the spike and bubble locations during the simulation is computed. Figure~\ref{fig:RT1}(c) displays the order of accuracy, which is around 1.5. The obtained convergence rate is close to the expected second-order, confirming that the solver is robust and accurate for more complex problems as well.

\begin{figure*}[]
\centering
{\includegraphics[width = 1\textwidth]{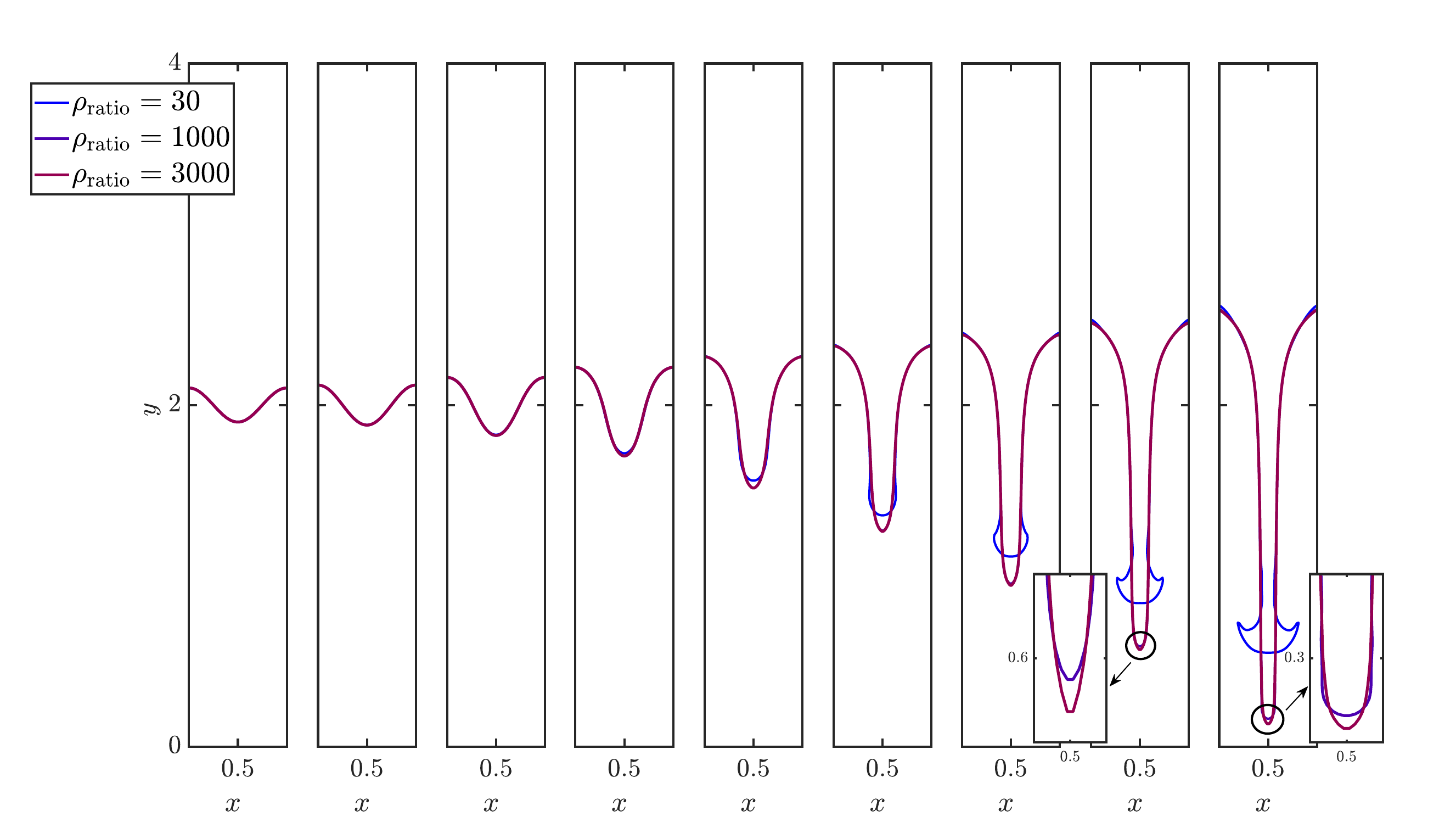}}
\caption{The interface of the Rayleigh--Taylor instability for different density ratios of 30 ($\mathrm{At}=0.935$), 1000 ($\mathrm{At}=0.998$), and 3000 ($\mathrm{At}=0.999$) at $t\sqrt{\mathrm{At}}=0, 0.25, 0.5, 0.75, 1, 1.25, 1.5, 1.75$, and $2$, from left to right, for the mesh resolution of $64\times256$.}
\label{fig:RT2}
\end{figure*}

Finally, the performance of the solver is also evaluated by simulating the Rayleigh--Taylor instability for higher density ratios of 30, 1000, and 3000. The mesh resolution is set as $64\times256$ and $\Delta t= 5 \times 10^{-4}/\sqrt{\mathrm{At}}$. Figure \ref{fig:RT2} displays the results for density ratios of 30 ($\mathrm{At}=0.935$), 1000 ($\mathrm{At}=0.998$), and 3000 ($\mathrm{At}=0.999$), at time $t\sqrt{\mathrm{At}}=0$ to 2 with an increment of 0.25. As expected, the interface evolves faster for higher density ratios, and the interface structures are simpler. Therefore, for the minimum density ratio of 30, the rate of penetration of the heavier fluid to the lighter one is less as a result of the reduced growth rate of the Rayleigh--Taylor instability, and the mushroom structure of the Rayleigh--Taylor instability can be observed. Since the Atwood number is comparable for two cases of density ratios of $1000$ and $3000$, the results of the surface evolution are similar (see Fig.~\ref{fig:RT2}). However, in the case of the density ratio of $3000$, the tip of the spike has penetrated the lower density region a bit further, and its structure is sharper at the tip (see the insets of Fig.~\ref{fig:RT2}). The greater density ratio of 3000 indicates a more significant disparity in fluid densities, leading to stronger buoyancy forces. This stronger buoyancy causes the spikes at the interface to penetrate farther into the lower-density region due to the accelerated downward motion of the denser fluid. Additionally, a greater density ratio corresponds to a higher Atwood number, which accelerates the exponential growth of Rayleigh--Taylor instability, further promoting the deeper penetration of spike tips into the low-density region. However, because the Atwood numbers for these cases are relatively close, the difference in results remains relatively small.

\section{\label{sec:conductingTwophase} Implementation of two-phase solver for magnetic flows}

In this section, the previously introduced two-phase solver is extended to account for magnetic flows. To achieve this, the effect of the Lorentz force is incorporated into the momentum equation, taking into consideration the magnetic permeability jump across the interface. The governing equations of two-phase flows along with the magnetostatic equation are solved, and the detailed numerical discretization is presented.

\subsection{\label{sec:conductingTwophaseDiscretization} Discretization of governing equations for incompressible flows under magnetic fields} 

The Lorentz force quantifies the force experienced by fluids due to electromagnetic fields, given as $\mathbf{J} \times \mathbf{B}$, where $\mathbf{J}$ and $\mathbf{B}$ are electric current and magnetic flux densities, respectively. This force can be written in the form of the Maxwell stress tensor, $\tau^{\mathrm{M}}$, given as~\citep{davidson2002}
\begin{equation}
    \nabla \cdot \tau_{ij}^{\mathrm{M}} = \nabla \cdot \left(\frac{B_i B_j}{\mu_\mathrm{m}} - \frac{\mathbf{B}^2}{2 \mu_\mathrm{m}}\delta_{ij} \right),
\end{equation}
where variable $\mu_\mathrm{m}$ denotes the magnetic permeability. This force can be treated as a body force acting on a fluid, and, hence, the updated momentum equation is expressed as
\begin{equation}
    \frac{\partial \mathbf{u}}{\partial t} + \nabla \cdot \left(\mathbf{u} \mathbf{u} \right) = -\frac{1}{\rho} \nabla p + \frac{1}{\rho} \nabla \cdot \mu \left(\nabla \mathbf{u} + \nabla \mathbf{u}^\mathrm{T} \right) + \mathbf{g} + \frac{1}{\rho} \mathbf{F}_\mathrm{sv} + \frac{1}{\rho} \nabla \cdot \tau_{ij}^{\mathrm{M}}.
    \label{eq:twophase1}
\end{equation}

When the field quantities do not change with time, Maxwell's equations are reduced to the electrostatic and magnetostatic case, which are given as

\begin{subequations}
\begin{equation}
   \nabla \cdot \mathbf{B} = 0, \       \  \          \ \        \ \         \ \         \ (\text{Gauss's law}) 
   \label{48a}
\end{equation}
\begin{equation}
    \nabla \times \mathbf{E} = 0,  \          \ \        \ \         \  \            \ (\text{Faraday's law})
\end{equation}
\begin{equation}
    \nabla \times \mathbf{H} = \mathbf{J},  \          \ \        \ \         \  \            \ (\text{Maxwell--Amp\`ere law})
    \label{48c}
\end{equation}
\text{and}
\begin{equation}
    \nabla \cdot \mathbf{J} = 0,  \          \ \        \ \         \  \            \ \  \  (\text{equation of continuity})
\end{equation}
\end{subequations}
where variables $\mathbf{H}$ and $\mathbf{E}$ denote the magnetic field and electric field intensities, respectively. The magnetic flux density, $\mathbf{B}$, is related to the magnetic field intensity, $\mathbf{H}$, using the magnetic permeability, $\mu_\mathrm{m} = \mathbf{B}/\mathbf{H}$. In the magnetostatic case, the behaviour of the magnetic field can be studied in the absence of electric currents, since the electric charges are either at rest or moving very slowly, so that the magnetic field induced by them can be neglected. Consequently, there is no interaction between electric and magnetic fields, and an electrostatic case or a magnetostatic case can be studied separately.

Under the magnetostatic assumption, Eqs.~(\ref{48a}) and (\ref{48c}) explain the evolution of the magnetic field. One approach to solving the Maxwell--Amp\`ere equation while satisfying the magnetic field divergence-free constraint is the vector potential formulation. In this method, the magnetic field, $\mathbf{B}$, is defined as the curl of an auxiliary vector, $\mathbf{A}$, with the gauge condition of $\nabla \cdot \mathbf{A} = 0$, as $\mathbf{B} = \nabla \times \mathbf{A}$. As a result, Gauss's law of magnetism is automatically satisfied, and Eq.~(\ref{48c}) is recast as 
\begin{equation}
    \nabla \times \mathbf{H} = \nabla \times \left(\frac{1}{\mu_\mathrm{m}} \mathbf{B}\right) = \nabla \times \left(\frac{1}{\mu_\mathrm{m}} \nabla \times \mathbf{A} \right).
    \label{eq:ampere1}
\end{equation}
For the two-dimensional case, $\mathbf{B}=\left(B_x, B_y, 0 \right)$, the vector potential is reduced to $\mathbf{A} = \left(0, 0, A_z \right)$, and Eq.~(\ref{eq:ampere1}) will be simplified as 
\begin{equation}
     \nabla \times \left(\frac{1}{\mu_\mathrm{m}} \mathbf{B}\right)= \nabla \times \left(\frac{1}{\mu_\mathrm{m}} \nabla \times \mathbf{A} \right) = \nabla \times \left(\frac{1}{\mu_\mathrm{m}} \frac{\partial A_z}{\partial y} \mathbf{i} - \frac{1}{\mu_\mathrm{m}} \frac{\partial A_z}{\partial x} \mathbf{j} \right)=\nabla \left(-\frac{1}{\mu_\mathrm{m}} \nabla A_z \right) = J_z. 
     \label{eq:ampere2}
\end{equation}        
Equation~(\ref{eq:ampere2}) is a Poisson equation with variable coefficients and can be discretized similarly to the pressure Poisson equation discussed earlier, Eq.~(\ref{eq:Poisson}), provided that $\mu_\mathrm{m}$ and $A_z$ are defined at cell centers (see Fig.~\ref{fig:grid}). 

Employing the obtained $A_z$ from Eq.~(\ref{eq:ampere2}), the components of the magnetic field are given as 
\begin{subequations}
    \begin{equation}
        B_x = \frac{\partial A_z}{\partial y} = \frac{\delta_\mathrm{2nd} \, \, A_z}{\delta_{\mathrm{2nd}} \, \, y},
    \end{equation}
    and
    \begin{equation}
        B_y = - \frac{\partial A_z}{\partial x} = - \frac{\delta_\mathrm{2nd} \, \, A_z}{\delta_{\mathrm{2nd}} \, \, x},
    \end{equation}
\end{subequations}
at cell faces in $y-$ and $x-$directions, respectively (see Fig.~\ref{fig:grid}). Finally, for the two-dimensional case of $\mathbf{x} = \left(x, y\right)$, the components of the Lorentz force are discretized as
\begin{subequations}
    \begin{equation}
        \left[\nabla \cdot \tau_{ij}^{\mathrm{M}} \right]_{x-\mathrm{comp}} = \frac{\delta_{2\mathrm{nd}}}{\delta_{2\mathrm{nd}} \, x} \left(\frac{\left(\overline{B_x}^{2\mathrm{nd} \, \,  y}\right)^2 - \left(\overline{B_y}^{2\mathrm{nd} \, \, x}\right)^2}{2 \mu_\mathrm{m}} \right) + \frac{\delta_{2\mathrm{nd}}}{\delta_{2\mathrm{nd}} \, y} \left(\frac{\overline{B_x}^{2\mathrm{nd} \, \, x} \, \, \overline{B_y}^{2\mathrm{nd} \, \, y}}{\overline{\overline{\mu_\mathrm{m}}^{2\mathrm{nd} \, \, x}}^{2\mathrm{nd} \, \, y}} \right), 
    \end{equation}
    \begin{equation}
        \left[\nabla \cdot \tau_{ij}^{\mathrm{M}} \right]_{y-\mathrm{comp}} = \frac{\delta_{2\mathrm{nd}}}{\delta_{2\mathrm{nd}} \, x} \left(\frac{\overline{B_x}^{2\mathrm{nd} \, \, x} \, \, \overline{B_y}^{2\mathrm{nd} \, \, y}}{\overline{\overline{\mu_\mathrm{m}}^{2\mathrm{nd} \, \, x}}^{2\mathrm{nd} \, \, y}} \right) + \frac{\delta_{2\mathrm{nd}}}{\delta_{2\mathrm{nd}} \, y} \left(\frac{\left(\overline{B_y}^{2\mathrm{nd} \, \,  x}\right)^2 - \left(\overline{B_x}^{2\mathrm{nd} \, \, y}\right)^2}{2 \mu_\mathrm{m}} \right). 
    \end{equation}
\end{subequations}

\begin{figure}
\centering
\includegraphics[scale=2.0]{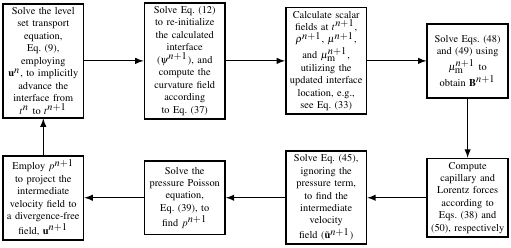}
\caption{\label{fig:agorithmProcedure} A flowchart representing the complete procedure for implementing the two-phase incompressible numerical toolkit for magnetic flows.}
\end{figure}

The solution procedure introduced in the previous section for two-phase nonmagnetic flows can be applied to the magnetic case as well with some modifications. In step (2), the magnetic permeability field at $t^{n+1}$, $\mu_\mathrm{m}^{n+1}$, should be computed according to the updated location of the interface as well. The obtained magnetic permeability field will then be employed to calculate the magnetic field at time step $n+1$, solving Eq.~(\ref{eq:ampere2}), which in turn will be utilized to determine the Lorentz force in the momentum equation. Figure~\ref{fig:agorithmProcedure} summarizes the complete procedure for implementing the two-phase incompressible solver for magnetic flows.

\subsection{\label{sec:conductingTwophaseTests} Magnetic two-phase test cases}

Three test cases are conducted in this section, namely, the deformation of both a static and a sheared ferrofluid droplet as well as Rayleigh--Taylor instability in magnetic fluids, to evaluate the performance and accuracy of the implemented solver. The static droplet test case is designed to assess the capability of the solver to accurately simulate the behaviour of the Lorentz force at the interface for various magnetic field strengths. The numerical results are validated by comparing them with experimental and analytical data. In the second test, the deformation of a droplet in a shear flow is investigated, considering both low and high capillary flow regimes under varying magnetic field conditions. This test also involves comparing results with theoretical solutions, particularly in the context of low magnetic field values. Furthermore, within this test case, the impact of the magnetic permeability ratio between the ferrofluid droplet and the surrounding flow on its deformation and rotation is examined. The third benchmark is employed to evaluate the solver's performance in modelling the evolution of a complex interface in the presence of different magnetic field densities and high magnetic permeability jumps across the interface. Additionally, the impact of the magnetic field on the growth rate of the Rayleigh--Taylor instability is investigated and compared with the results obtained from linear analysis. It is noteworthy to mention that since ferrofluids do not conduct electric current, and in our test cases, no external current is imposed, the right-hand side value of Eq.~(\ref{eq:ampere2}), $J_z$, is set to zero in the following numerical simulations.

\subsubsection{Deformation of a stationary magnetic droplet}
In this test case, a liquid droplet of diameter $D = 1$ is considered at the center of a two-dimensional domain of $[0, 4] \times [0, 4]$ filled with gas, in a stationary velocity field. In the case of zero gravity, similar to the static droplet test case presented in Appendix~\ref{appendixE}, the droplet remains at rest since the pressure and capillary forces are balanced. However, if the gas and liquid phases have different values of magnetic permeability, in the presence of a magnetic field, the induced Lorentz force at the interface affects the deformation of the droplet; the competition between the Lorentz force and the surface tension force dictates the evolution of the droplet's interface. Suppose the density and viscosity are constant for both phases, $\rho_1 = \rho_2 = 1$ and $\mu_1=\mu_2=0.001$. A uniform magnetic field, $\mathbf{B}_0 = 2~\mathrm{mT}$, is imposed from bottom to top and $\mu_{\mathrm{m,l}}=9 \, \mu_{\mathrm{m,g}}$. The capillary force attempts to maintain the interface of the droplet in its initial shape. Nonetheless, since the magnetic field lines will be distorted around the droplet's interface, the created Lorentz force causes the droplet to deform and stretch. In order to better interpret the deformation of the droplet for different scenarios and magnetic strengths, scale analysis is employed. To this end, the following non-dimensional variables are introduced
\begin{equation}
    \mathbf{u^*} = \mathbf{u}/u_0, \      \ \mathbf{B^*}=\mathbf{B}/B_0, \       \, \mathbf{l}^*=\mathbf{l}/l_0, \      \ t^*=t\,u_0/l_0, \     \ p^*=p/\rho_0 \, u_0^2,
\end{equation}
where the zero subscripts refer to the initial value, and $l_0$ is the length scale of the problem. Rewriting the momentum equation in the non-dimensional form will then result in
\begin{eqnarray}
    \frac{\partial \mathbf{u}^*}{\partial t^*} + \frac{\partial}{\partial \mathbf{l}^*} \left(\mathbf{u}^* \mathbf{u}^* \right) = -\frac{\partial p^*}{\partial \mathbf{l}^*} + \overbrace{\frac{\mu}{u_0 \, l_0 \, \rho_0}}^{\frac{1}{\mathrm{Re}}} \frac{\partial}{\partial \mathbf{l}^*} \left(\frac{\partial \mathbf{u}^*}{\partial \mathbf{l}^*} + \frac{\partial {\mathbf{u}^*}^\mathrm{T}}{\partial \mathbf{l}^*} \right) - \overbrace{\frac{\sigma}{\rho_0 \, u_0^2 \, l_0} \left(\nabla \cdot \mathbf{n} \right) \nabla \psi}^{\text{surface tension}} \nonumber \\ + \underbrace{\frac{{B_0 ^2}}{\rho_0 \, \mu_\mathrm{m} \, {u_0}^2} \frac{\partial}{\partial \mathbf{l}^*} \left(\mathbf{B}^* \mathbf{B}^* - \frac{|\mathbf{B^*}|^2}{2} \delta_{i,j} \right)}_{\text{Lorentz force}}.
\end{eqnarray}
The ratio between the Lorentz force and the surface tension force can be qualified by a non-dimensional number defined as
\begin{equation}
    \frac{\text{Lorentz force}}{\text{surface tension force}}=\frac{\frac{B_0^2}{\rho_0 \, \mu_\mathrm{m} \, {u_0}^2}}{\frac{\sigma}{\rho_0 \, u_0^2 \, l_0}}=\frac{B_0^2 \, l_0}{\mu_\mathrm{m} \sigma}.
    \label{eq:bond}
\end{equation}
Hence, for the case of $B_0^2 \, l_0/\mu_\mathrm{m} \, \sigma \ll 1$, the surface tension force overcomes the Lorentz force and the droplet is expected to retain its shape. Conversely, in the case of $B_0^2 \, l_0/\mu_\mathrm{m} \, \sigma \gg 1$, since the Lorentz force is greater than the capillary force, the droplet deforms and stretches. If $B_0^2 \, l_0/\mu_\mathrm{m} \, \sigma$ is of the order of one, the Lorentz and surface tension forces are of the same magnitude. As a consequence, oscillations in the deformation of the droplet could be observed, as the Lorentz force stretches the interface and the surface tension force tries to prevent the deformation.

\begin{figure*}[]
\hspace*{-2cm}
{\includegraphics[width = 1.2\textwidth]{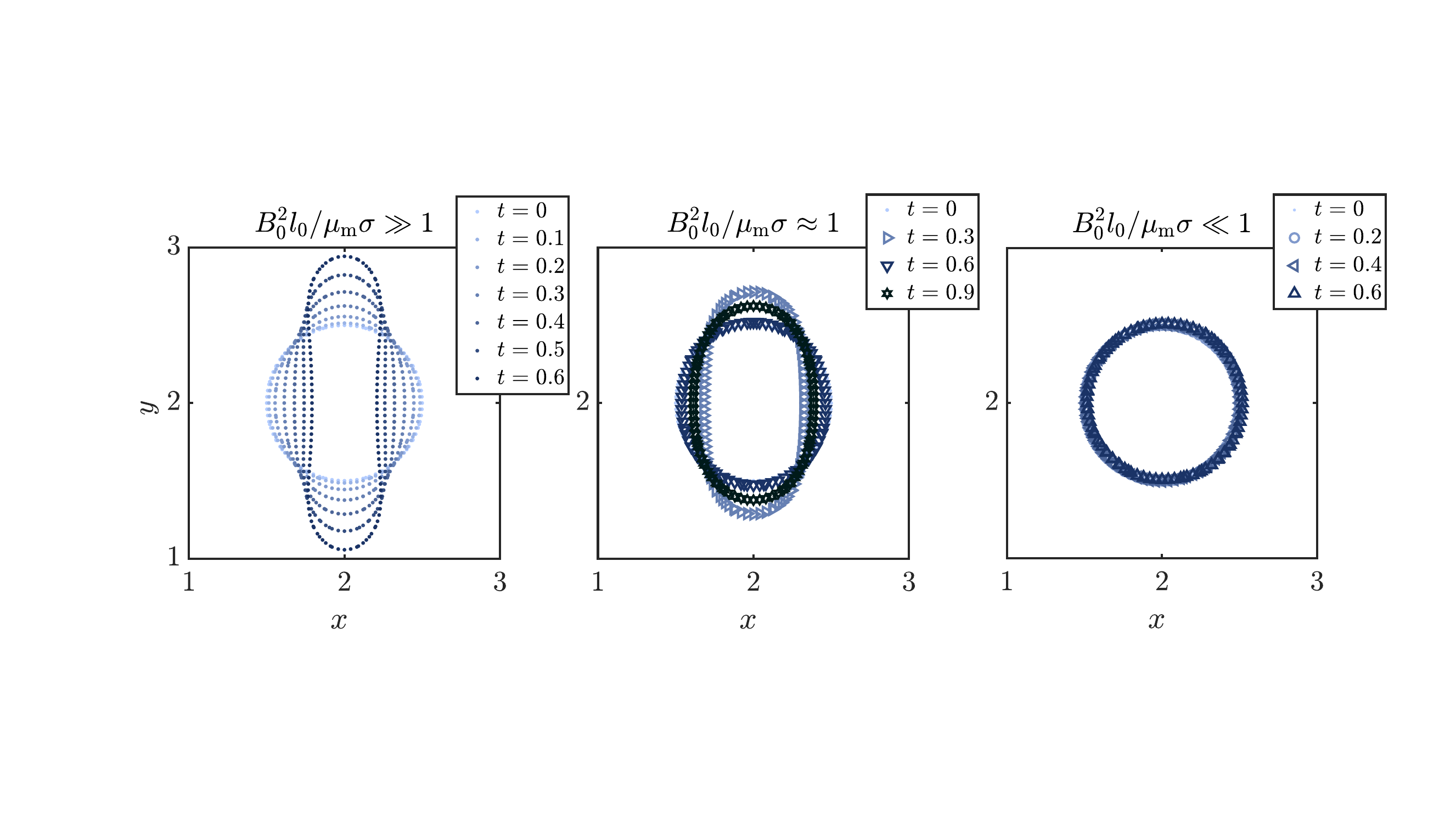}}\\
\vspace*{-2cm}
\caption{The evolution of the droplet under the constant magnetic field imposed from bottom to top of the computational domain, with $\mu_{\mathrm{m,l}}/\mu_{\mathrm{m,g}}=9$. Three cases of $B_0^2 \, l_0/\mu_\mathrm{m} \, \sigma \gg 1$, $B_0^2 \, l_0/\mu_\mathrm{m} \, \sigma \approx 1$, and $B_0^2 \, l_0/\mu_\mathrm{m} \, \sigma \ll 1$ are represented from left to right, respectively.}
\label{fig:dropletDeformation}
\end{figure*}

This finding is then used to validate the behaviour of the implemented solver. In the numerical setup, the thickness of the interface is set to $\epsilon=\Delta x/2$, with the grid resolution of $101 \times 101$, where a no-slip boundary is imposed at all boundaries. Three cases of $B_0^2 \, l_0/\mu_\mathrm{m} \, \sigma \ll 1$, $B_0^2 \, l_0/\mu_\mathrm{m} \, \sigma \gg 1$, and $B_0^2 \, l_0/\mu_\mathrm{m} \, \sigma \approx 1$ are investigated, where the surface tension coefficient is set to 0.01, 1, and 100, respectively, and the results are represented in Fig.~\ref{fig:dropletDeformation}. It can be appreciated in Fig.~\ref{fig:dropletDeformation} that for the case of  $B_0^2 \, l_0/\mu_\mathrm{m} \, \sigma \gg 1$, the Lorentz force is vertically stretching the droplet while for the $B_0^2 \, l_0/\mu_\mathrm{m} \, \sigma \ll 1$ case, the droplet remains unchanged in time. Additionally, when the Lorentz and capillary forces are of the same order of magnitude, there is an oscillation in the deformation of the droplet's interface, evident in Fig.~\ref{fig:dropletDeformation}.

The non-dimensional number introduced in Eq.~(\ref{eq:bond}) is the magnetic Bond number, denoted as $\mathrm{Bo_m}$, mostly introduced as $\mathrm{Bo}_\mathrm{m}=l_0 \, \mu_\mathrm{m,0} \, H_0^2/2 \sigma$ in the literature. This parameter plays a critical role in various applications, including the study of the dynamics and deformation of ferrofluid droplets in the presence of a magnetic field. Ferrofluids are colloidal suspensions of nanoscale magnetic particles, typically around 10~nm in size, dispersed in a base fluid~\citep{majidi2022}. They were initially introduced by NASA in 1963, and since then, ferrohydrodynamics has become a subject of significant interest in the field of fluid mechanics~\citep{rosensweig2013ferrohydrodynamics}. Ferrofluids have found applications across various fields, including microfluidics~\citep{bijarchi2021experimental}, biomedical applications such as the treatment of retinal detachment and targeted drug delivery\citep{mefford2007field, voltairas2001elastic}, droplet generation from nozzles~\citep{bijarchi2020experimental}, and heat transfer augmentation~\citep{zarei2019visualization}. Understanding the behavior of ferrofluid droplets in the presence of magnetic fields is essential for their practical applications. When subjected to a uniform magnetic field, a ferrofluid droplet suspended in a viscous medium elongates along the direction of the field, ultimately reaching a stable equilibrium configuration. Researchers have conducted numerous studies to investigate the deformation and dynamics of single ferrofluid droplets under the influence of magnetic fields, employing analytical solutions, numerical simulations, and experimental observations~\citep{bacri1982study, li2021magnetic,afkhami2008field, afkhami2010deformation}.  

\begin{figure*}[]
\centering
\vspace*{-1.6cm}
\subfloat[]{\includegraphics[width = 0.8\textwidth]{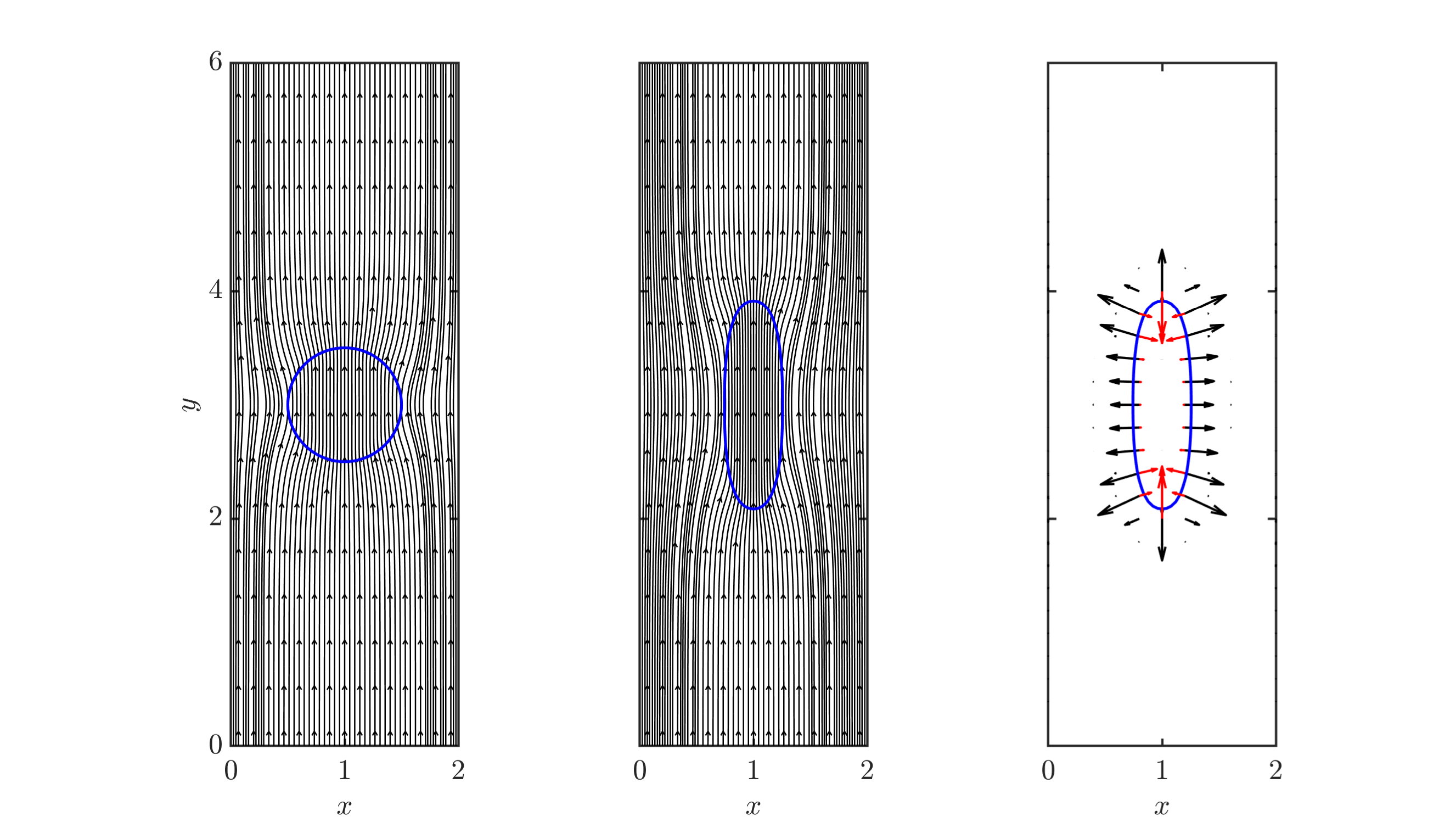}}\\
\hspace*{-2.2cm}
\subfloat[]{\includegraphics[trim=0 0 0 25,width = 1.2\textwidth]{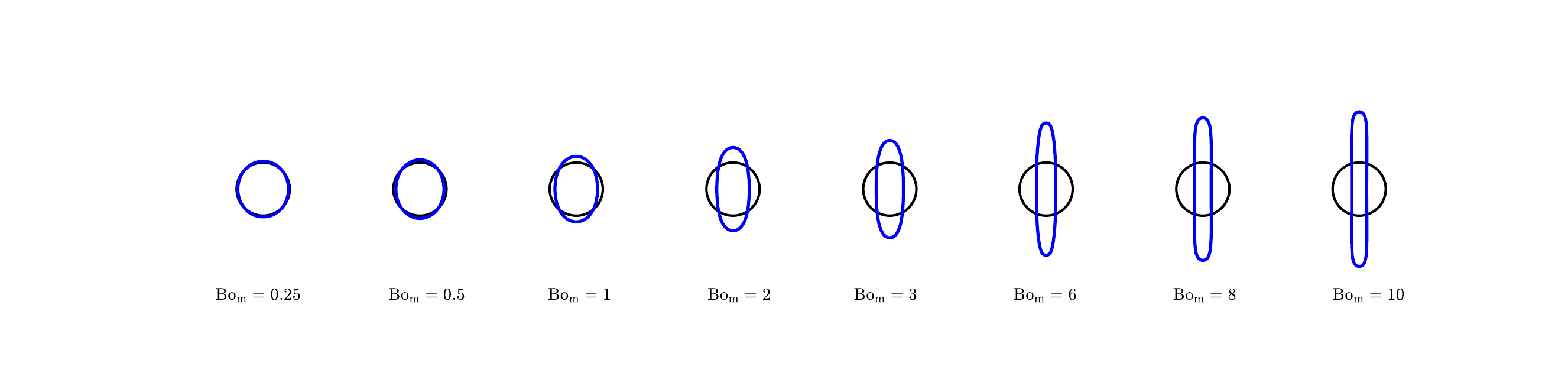}}\\
\subfloat[]{\includegraphics[width = 0.8\textwidth]{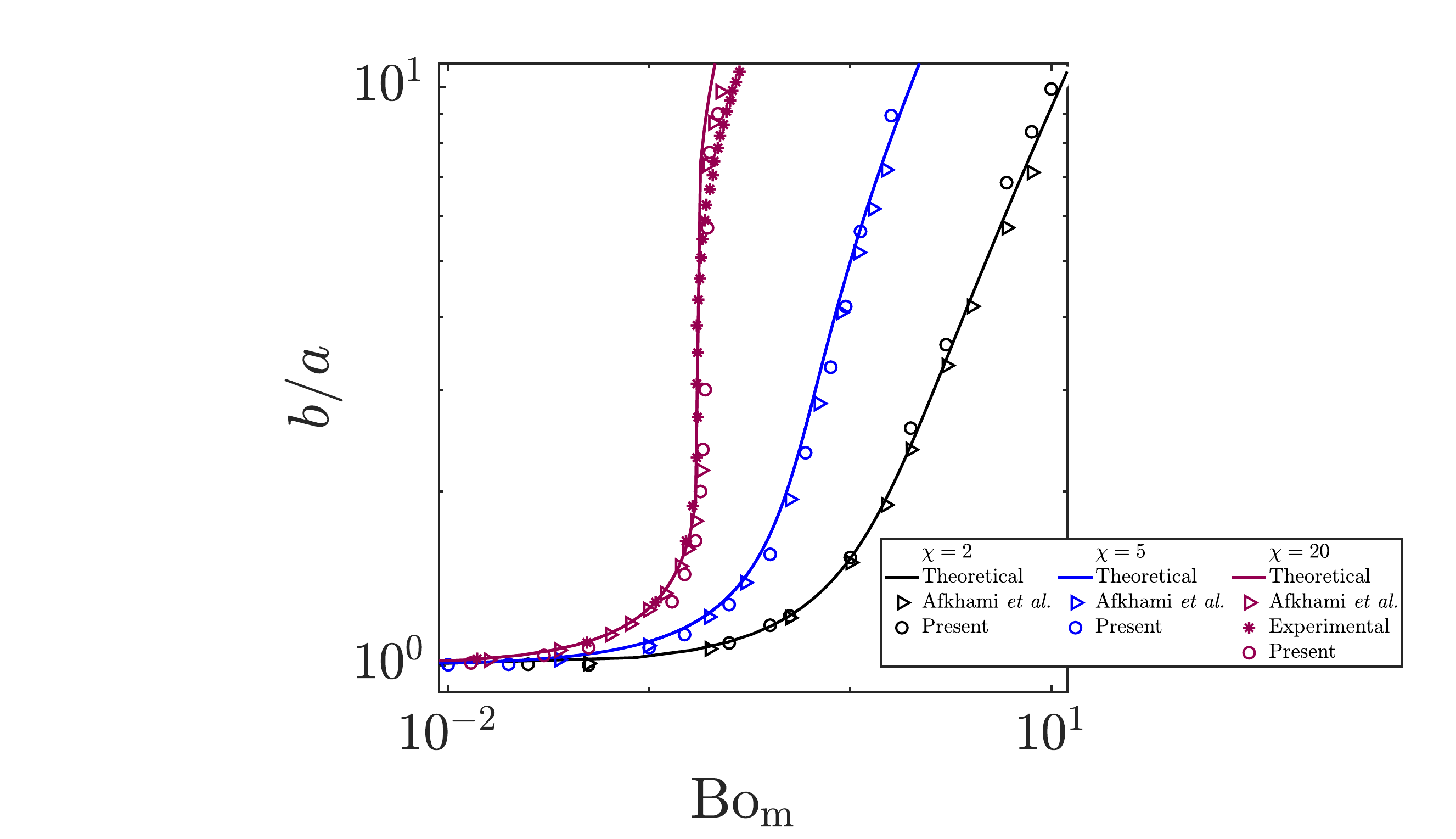}}
\caption{(\textit{a}) Configuration of the ferrofluid droplet along with magnetic field at [left]~$t=0$, [middle]~steady state, and [right]~depiction of the magnetic force (in black) and capillary force (in red) acting at the interface for the case $\chi=2$ and $\mathrm{Bo}_\mathrm{m} = 3$. (\textit{b}) droplet deformation across eight different magnetic Bond numbers with a susceptibility value of $\chi = 2$. (\textit{c}) Comparison of analytical, numerical, and experimental results for three different magnetic susceptibility values of $\chi = 2$, 5, and 20, under varying magnetic Bond numbers.}
\label{fig:ferrofluidDroplet}
\end{figure*}

To further validate our implemented two-phase solver for magnetic flows quantitatively, we have leveraged the theoretical work presented by Afkhami \textit{\textit{et al.}}~\citep{afkhami2010deformation}, who explored the deformation of a ferrofluid droplet in a quiescent fluid subjected to a uniform magnetic field. In their study, they established a relation between the deformation of a ferrofluid and the magnetic Bond number. This theoretical solution is derived under the assumption that the droplet maintains its ellipsoidal shape during elongation due to the presence of the magnetic field~\citep{afkhami2010deformation}. Consequently, the extent of deformation of the droplet is quantified by introducing the aspect ratio denoted as $b/a$, where $2a$ and $2b$ represent the major and minor axes of the droplet, respectively, after it has undergone deformation and reached a steady state. Afkhami \textit{\textit{et al.}}~\citep{afkhami2010deformation} modeled the magnetization of the ferrofluid droplet, denoted as $\mathbf{M}_\mathrm{ferrofluid}$, as a linear function of the applied magnetic field, given as $\mathbf{M}_\mathrm{ferrofluid} = \chi \mathbf{H}$, where $\mathbf{H}$ represents the external uniform magnetic field strength, and $\chi$ is the magnetic susceptibility of the ferrofluid droplet~\citep{majidi2022,afkhami2010deformation}. Magnetic susceptibility, $\chi$, is a material property that quantifies the magnetization response of the ferrofluid droplet to an applied magnetic field. It is expressed as $\chi = \left(\mu_\mathrm{m,droplet}/\mu_{\mathrm{m},0} - 1 \right)$ and is assumed to remain constant in each phase in the analytical solution of Afkhami \textit{\textit{et al.}}~\citep{afkhami2010deformation}. Consequently, the magnetic induction field is calculated as $\mathbf{B}=\mu_{\mathrm{m},0} \left(\mathbf{H} + \mathbf{M}_\mathrm{ferrofluid} \right)=\mu_{\mathrm{m},0} \left(1 + \chi \right)\mathbf{H}$~\citep{afkhami2010deformation}. The theoretical finding regarding the droplet deformation in an external magnetic field reported by Afkhami \textit{\textit{et al.}}~\citep{afkhami2010deformation} is presented as
\begin{equation}
    \mathrm{Bo}_\mathrm{m} = \left(\frac{1}{\chi} + k \right)^2 \left(\frac{b}{a} \right)^\frac{1}{3} \left[ 2\frac{b}{a} - \left(\frac{b}{a} \right)^{-2} - 1 \right],
    \label{eq:analsol}
\end{equation}
where $k$ is the demagnetizing factor that is calculated as
\begin{equation}
    k = \left(\frac{1-E^2}{2 E^3} \right) \left(\ln{\frac{1 + E}{1 - E}} - 2E\right),\    \ \  \ \mathrm{with} \  \ \  \ E = \sqrt{1- a^2/b^2}.
\end{equation}
Here, a test case similar to the previous test is conducted to compare our results with the theoretical solution of Eq.~(\ref{eq:analsol}) and other numerical and experimental results existing in the literature. In this test case, a circular ferrofluid droplet with a radius of $R_0=0.5$ is placed at the center of a computational domain of $[0,2] \times [0,6]$ filled with gas. The mesh resolution of $100 \times 300$ and time step $\Delta t = 0.001$ were used for all simulations. The density and viscosity are set to be the same for both phases, $\rho_1=\rho_2=1$ and $\mu_1=\mu_2=0.01$, and a constant surface tension value of $\sigma=1$ was employed. The magnetic permeability of the gas is set to $\mu_\mathrm{m,g}=\mu_{\mathrm{m},0}=1$, therefore, to change the magnetic permeability ratio between two phases, the magnetic permeability of the ferrofluid droplet is altered. Figure~\ref{fig:ferrofluidDroplet}(a)[left] illustrates the initial magnetic field configuration for the case where $\chi=2$ and $\mathrm{Bo}_\mathrm{m} = 3$, along with the initial shape of the ferrofluid droplet at $t=0$, as an example. As can be observed in this figure, the magnetic field lines near the droplet interface are distorted due to the varying magnetic permeability of the two phases. These distorted magnetic field lines induce magnetic forces at the droplet interface, resulting in the deformation of the ferrofluid droplet. The deformed droplet for this specific case at steady state along with the corresponding magnetic field are represented in Fig.~\ref{fig:ferrofluidDroplet}(a)[middle]. As expected, the droplet has elongated along the magnetic field lines due to the presence of magnetic forces. Figure~\ref{fig:ferrofluidDroplet}(a)[right] demonstrates the forces acting on the ferrofluid droplet interface, namely, the magnetic force (depicted in black) and the capillary force (shown in red). It is visually evident from this figure that the magnetic force exhibits a higher amplitude, enabling it to overcome the capillary force and induce elongation. The surface tension force, primarily concentrated at the poles of the droplet where high curvature is present, opposes the magnetic forces, attempting to preserve the initial shape of the droplet.

Figure~\ref{fig:ferrofluidDroplet}(b) displays the droplet deformation for eight different magnetic Bond numbers of $\mathrm{Bo}_\mathrm{m} = 0.25$, 0.5, 1, 2, 3, 6, 8, and 10, with the susceptibility value of $\chi = 2$ for all cases. It can be seen that for higher magnetic Bond numbers, the droplet deformation is more pronounced, owing to the increased magnetic forces, since the surface tension coefficient remains consistent across all cases. As the magnetic Bond number increases, indicating stronger magnetic forces, the magnetic force becomes more effective in overcoming capillary forces, further deforming the droplet.  

In Fig.~\ref{fig:ferrofluidDroplet}(c), the deformation of the ferrofluid droplet is explored under three different magnetic susceptibility values of $\chi = 2$, 5, and 20, across various magnetic Bond numbers. The results are compared with analytical solutions, demonstrating a close agreement between numerical and theoretical predictions. As expected, an increase in the magnetic Bond number leads to more significant changes in the aspect ratio. The maximum error between the numerical and analytical results in the studied cases is approximately 4.2\%. This discrepancy can be attributed to the theoretical assumption that the droplet is axisymmetric and retains an elliptical shape during deformation~\citep{afkhami2010deformation}. However, this assumption may not hold true, particularly for cases with higher susceptibility and magnetic Bond numbers~\citep{afkhami2010deformation}, where more significant discrepancies with the analytical solution are evident. Additionally, the obtained results for $\chi = 2$, 5, and 20 cases are compared with numerical results from Afkhami \textit{\textit{et al.}}~\citep{afkhami2010deformation}, showing good agreement, with some discrepancy due to the differences in the nature of simulations (axisymmetric in~\citep{afkhami2010deformation} vs. two-dimensional in this study). 

Furthermore, in the case where $\chi=20$, we compare our results with the experimental data obtained by Bacri and Salin~\citep{bacri1982instability}. While a reasonable agreement was observed, discrepancies, particularly for higher magnetic Bond numbers, can be attributed to the three-dimensional nature of the experiments and the assumption of constant surface tension in our simulations. In reality, it is observed that the interfacial tension of the ferrofluid droplet depends on the magnetic field, especially for high magnetic field strengths~\citep{afkhami2010deformation}.

\subsubsection{Deformation of a sheared magnetic droplet}

As previously mentioned, ferrofluid droplets have garnered significant attention from researchers due to their diverse applications in areas such as biomedicine, microfluidics, and rheology~\citep{cunha2018}. In some applications, these droplets are not only subjected to an external magnetic field, as discussed in the previous test case, but also experience the presence of hydrodynamic flow. This flow can influence the droplet's deformation, inclination, and potential breakup into smaller droplets. Consequently, studying the deformation of sheared droplets becomes crucial across various industrial applications that utilize emulsions. The deformation of the sheared droplet depends on various parameters, such as surface tension, shear rate, magnetic field strength, and the viscosity ratio between the droplet and the suspending fluid. In addition to the magnetic Bond number, $\mathrm{Bo}_\mathrm{m}$, another non-dimensional parameter known as the capillary number, $\mathrm{Ca}$, which quantifies the ratio between shear and surface tension forces, also influences the droplet deformation. Hassan \textit{\textit{et al.}}~\citep{hassan2018} conducted a numerical study on the deformation of sheared droplets under a uniform external magnetic field using the finite element method. Their study investigated the effects of shear rate, magnetic field strength, and magnetic field direction on ferrofluid deformation. They found that in the low capillary regime, magnetic forces dominate over shear forces, exerting primary control over droplet dynamics~\citep{hassan2018}. Additionally, increasing the magnetic field strength, i.e., increasing the magnetic Bond number, enhances droplet deformation~\citep{hassan2018}. In a subsequent study by Hassan and Wang~\citep{hassan2019magnetic}, the impact of viscosity ratio and magnetic field direction on droplet deformation and breakup was explored under the constraint of a low Reynolds number ($\mathrm{Re}\leq0.03$). Their results indicated that when the magnetic field is applied at an angle of $45 ^{\circ}$, droplet elongation is more pronounced, leading to accelerated droplet breakup. Conversely, varying the magnetic field angle to $0^{\circ}$ or $135^{\circ}$ suppresses droplet breakup. To our knowledge, no previous study has investigated sheared droplet deformation by varying the susceptibility value of the ferrofluid droplet. Consequently, in this test case, we investigate sheared droplet deformation for low capillary number to validate the implemented numerical solver. Additionally, we explore the effect of magnetic susceptibility ratio between the ferrofluid droplet and its surrounding medium on the dynamics and deformation of the ferrofluid droplet in both low and high capillary regimes.

\begin{figure*}[]
\centering
{\includegraphics[width = 0.7\textwidth]{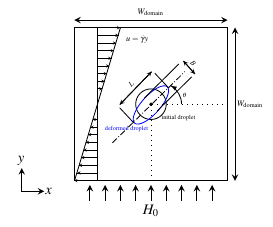}}\\
\caption{Schematic of a ferrofluid droplet suspended in another nonmagnetic viscous fluid in a simple shear flow under a uniform external magnetic field, $\mathbf{H}_0$.}
\label{fig:ScehmaticShearedDroplet}
\end{figure*}

Figure~\ref{fig:ScehmaticShearedDroplet} illustrates the schematic of this test case, including a ferrofluid droplet with radius $R_0=0.5$ suspended in another nonmagnetic viscous fluid. The droplet is positioned at the center of a square computational domain with dimension $W_\mathrm{domain}=6$, and a velocity profile of $u=\Dot{\gamma}y$ is imposed, where $\Dot{\gamma}$ represents the corresponding shear rate. A uniform magnetic field of $\mathbf{H}_0$ at the angle of $90 ^{\circ}$ is applied. It is worth noting that numerical studies model simple shear flows by moving two confining walls, which can introduce confinement effects on the deformation of the droplet. Studies by Kennedy~\citep{kennedy1994motion} and Guido~\citep{guido1998three} on the droplet deformation in a simple shear flow suggest that the confinement effect on droplet deformation is negligible when $2R_0/W_\mathrm{domain}<0.4$. In our simulations, $2R_0/W_\mathrm{domain}$ is equal to 0.16, effectively eliminating the confinement effect on droplet deformation. In the presented simulations, the density and viscosity of both phases are assumed to be equal, $\rho=1$ and $\mu=0.1$. The top and bottom boundaries of the computational domain are treated as moving walls with velocities of $+\Dot{\gamma} W_\mathrm{domain}/2$ and $-\Dot{\gamma} W_\mathrm{domain}/2$, respectively, producing constant shear rate of $\Dot{\gamma}$, while the left and right boundaries are periodic. By setting surface tension coefficient and shear rate to 1 and 0.4, respectively, the capillary number is adjusted to $\mathrm{Ca}= \mu_{\mathrm{m},0} \, R_0 \, \Dot{\gamma}/\sigma=0.02$, and the deformation of the droplet is investigated for different values of $\mathrm{Bo_m}$ and $\chi$. To quantify droplet deformation, Taylor's deformation parameter~\citep{taylor1932viscosity,taylor1934formation} is calculated, given as 
\begin{equation}
    D=\frac{L-B}{L+B},
\end{equation}
where $L$ and $B$ represent the major and minor axes of the deformed droplet (see Fig.~\ref{fig:ScehmaticShearedDroplet}). The pioneering work of Taylor indicating the deformation parameter of a buoyant droplet suspended in another viscous fluid under a shear flow in the Stokes flow limit~\citep{taylor1932viscosity,taylor1934formation} was extended by Jesus \textit{\textit{et al.}}~\citep{jesus2018deformation} to a sheared ferrofluid droplet under an external magnetic field in the limit of both small $\mathrm{Ca}$ and $\mathrm{Bo}_\mathrm{m}$ numbers. According to the asymptotic theory derived by Jesus \textit{\textit{et al.}}~\citep{jesus2018deformation}, the deformation of a sheared ferrofluid droplet in the presence of a uniform magnetic field is given as
\begin{equation}
    D = \frac{\sqrt{\left[\alpha\left(\nu_\mathrm{r}\right)\mathrm{Ca} \right]^2 + \left[\beta \left(\chi_\mathrm{droplet}\right) \mathrm{Bo}_\mathrm{m}\right]^2}}{2 + \frac{1}{3} \beta\left(\chi_\mathrm{droplet} \right) \mathrm{Bo}_\mathrm{m}},
\end{equation}
where $\alpha \left(\nu_\mathrm{r}\right)$ and $\beta \left(\chi_\mathrm{droplet}\right)$ are the functions of the viscosity ratio between the droplet and the surrounding fluid, $\nu_\mathrm{r}$, and the susceptibility of the ferrofluid droplet, $\chi_\mathrm{droplet}$, respectively. These functions are calculated as
\begin{subequations}
    \begin{equation}
        \alpha \left(\nu_\mathrm{r}\right) = \frac{19 \nu_\mathrm{r} + 16}{8(\nu_\mathrm{r} + 1)}
    \end{equation}
    and
    \begin{equation}
        \beta \left(\chi_\mathrm{droplet}\right)=\frac{3 \, \chi_\mathrm{droplet} \left(2 \, \chi_\mathrm{droplet}+ 1 \right)}{4 \left(\chi_\mathrm{droplet} + 3 \right)^2}.
    \end{equation}
\end{subequations}

\begin{figure*}[]
\centering
\hspace*{-2cm}
{\includegraphics[width = 1.2\textwidth]{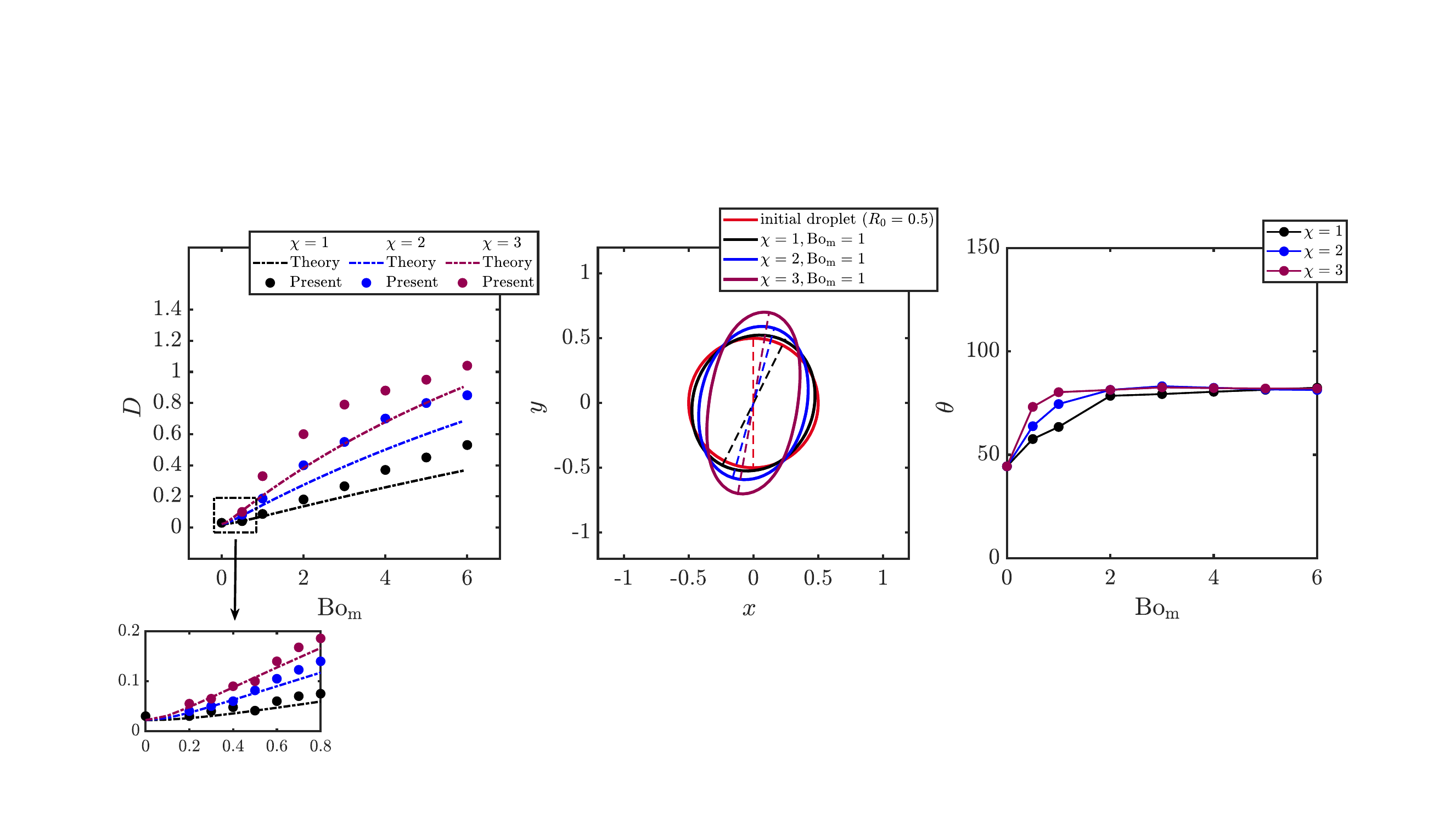}}\\
\caption{[left]~Comparison of analytical and numerical results for three different magnetic susceptibility values of $\chi=1$, 2, and 3 under varying magnetic Bond numbers and fixed capillary number, $\mathrm{Ca=0.02}$. [middle]~Ferrofluid droplet at steady state for three cases of $\chi=1$, 2, and 3, with capillary number and magnetic Bond number of 0.02 and 1, respectively. The dashed line connects the poles of the droplet for each case. [right]~Ferrofluid droplet inclination, $\theta$, for three different magnetic susceptibility values of $\chi=1$, 2, and 3 under varying magnetic Bond numbers and fixed capillary number, $\mathrm{Ca=0.02}$.}
\label{fig:ShearedDroplet1}
\end{figure*}

\begin{figure*}[]
\centering
\hspace*{-1cm}
{\includegraphics[width = 1.2\textwidth]{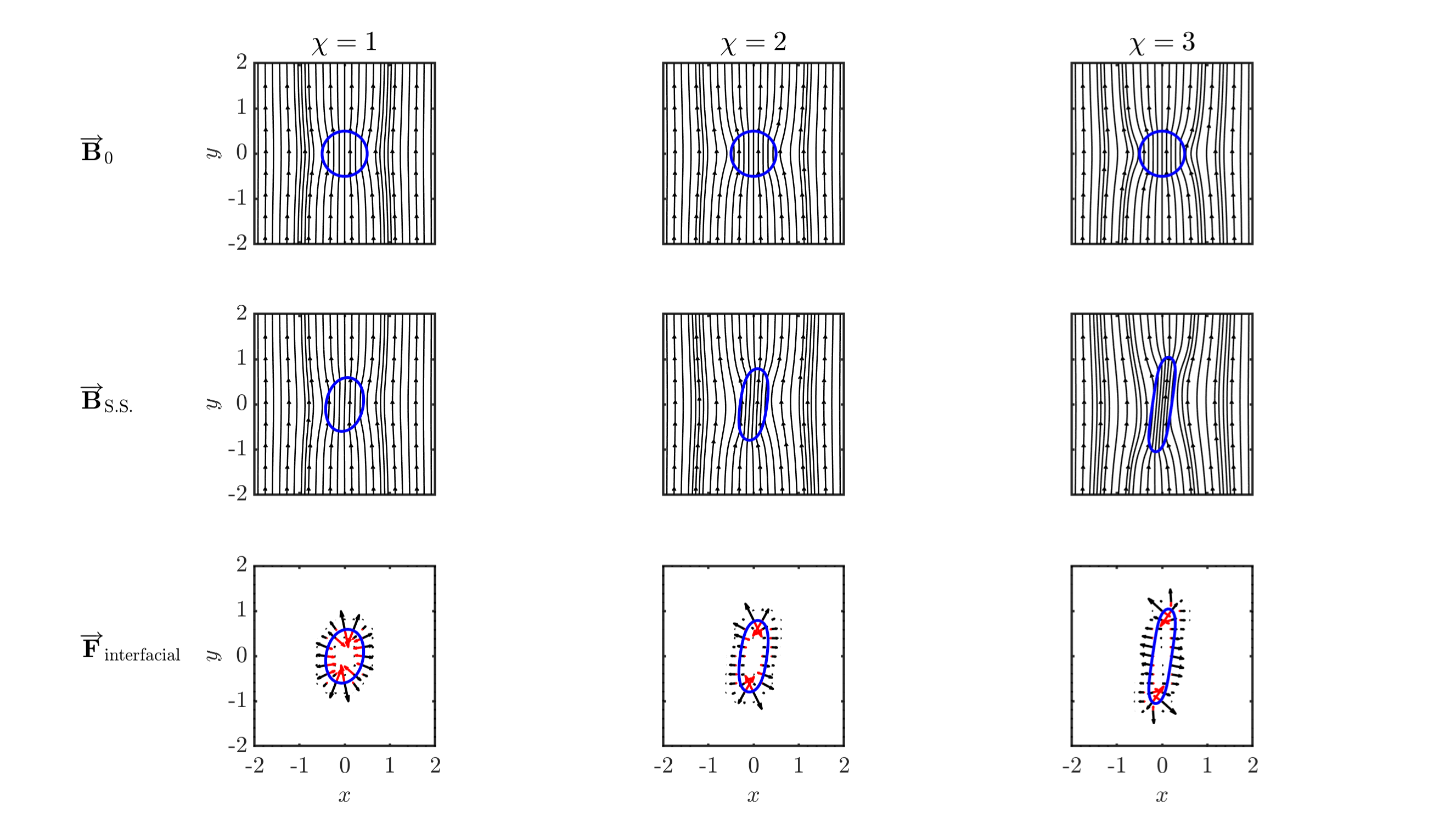}}\\
\caption{[top]Configuration of the ferrofluid droplet along with magnetic field at $t=0$, [middle]~steady state, and [bottom]~depiction of the magnetic force (in black) and capillary force (in red) acting at the droplet interface for three cases of $\chi=1$, 2, and 3, with $\mathrm{Bo}_\mathrm{m}=2$ and $\mathrm{Ca}=0.02$.}
\label{fig:ShearedDroplet2}
\end{figure*}

Figure~\ref{fig:ShearedDroplet1}[left] presents the droplet deformation parameter with $\mathrm{Ca}=0.02$ for different magnetic bond numbers, ranging from 0 to 6, for three different droplet susceptibility values of $\chi=1$, 2, and 3. Since the simulations are performed in the low capillary regime, it is expected that magnetic forces dominate over shear forces and as the magnetic Bond number increases, the droplet deformation is anticipated to increase. Figure~\ref{fig:ShearedDroplet1}[left] visually confirms that the numerical results capture this expected behaviour. As shown in this figure, the results for $\mathrm{Bo_m}<1$ are in good agreement with the theoretical solution. Please refer to the inset in this figure, which depicts analytical and numerical solutions for the magnetic Bond number ranging from 0 to 0.8. However, as the magnetic Bond number increases, the numerical results deviate from the theoretical solution derived by Jesus \textit{\textit{et al.}}~\citep{jesus2018deformation}. This behavior is expected since the proposed theoretical solution is only valid for small magnetic Bond numbers ($\mathrm{Bo_m} \ll 1$). Furthermore, it is evident that by increasing the susceptibility value, the numerical results differ more pronouncedly from the small perturbation analysis of the theoretical solution. According to the numerical results, increasing the magnetic susceptibility value of the droplet significantly affects its deformation. For example, for $\chi=3$ and $\mathrm{Bo}_\mathrm{m}=6$, the deformation parameter is approximately 2 times higher than the same case with $\chi=1$. Figure~\ref{fig:ShearedDroplet1}[middle] depicts the steady-state droplet at $\mathrm{Bo}_\mathrm{m}=1$ for the three different magnetic susceptibility values of $\chi=1$, 2, and 3. As can be seen from this figure, increasing the magnetic susceptibility value by three times results in an elongated, thin droplet shape compared to the $\chi=1$ case. This behavior can be explained by investigating the existing Maxwell stresses at the droplet interface for different susceptibility values. In Fig.~\ref{fig:ShearedDroplet2}, the magnetic field for three different values of $\chi=1$, 2, and 3 is shown at $t=0$ and the steady state. According to this figure, a larger $\chi$, or in other words, a larger magnetic permeability discontinuity between the ferrofluid droplet and the surrounding fluid, will result in a greater distortion of magnetic field lines around the droplet interface. This higher bending in the magnetic field lines induces a larger Lorentz force at the interface, which in turn causes greater droplet elongation, as the surface tension (restoring force) is constant for all cases. This conclusion emphasizes that changing the susceptibility of the ferrofluid droplet also plays an important role in its dynamics and deformation. In the study by Hu \textit{\textit{et al.}}~\cite{hu2018phase}, it was also shown that the magnetic field lines inside the ferrofluid droplet at $t=0$ are aligned with the externally imposed magnetic field. However, the magnetic field will be distorted near the interface of the droplet. Hu \textit{\textit{et al.}}~\cite{hu2018phase} found the analytical solution for the magnetic field intensity inside and outside of the circular cylinder within an externally vertical uniform magnetic field. According to their analytical solution, the magnetic field intensity inside the cylinder will be given as $\mathbf{H}= A \mathbf{e}_y$, where $A=H_0 \, (2 \mu_{\mathrm{m},2})/(\mu_{\mathrm{m},1} + \mu_{\mathrm{m},2})$. Variables $\mu_{\mathrm{m},1}$ and $\mu_{\mathrm{m},2}$ denote the magnetic permeability of the surrounding gas and the ferrofluid droplet, respectively. We measured the numerical magnetic field intensity for the three different cases of $\chi=1$, $2$, and $3$, shown in Fig.~\ref{fig:ShearedDroplet2}, and compared it to the analytical solution. According to our results, the magnetic field intensity for the cases $\chi=1$, $2$, and $3$ are $1.32 H_0$, $1.51 H_0$, and $1.62 H_0$, respectively. These results are in good agreement with the numerical findings of $1.33 H_0$, $1.5 H_0$, and $1.6 H_0$ for the cases $\chi=1$, $2$, and $3$, respectively.

In Fig.~\ref{fig:ShearedDroplet2}, the acting magnetic and surface tension forces at the interface are shown in black and red, respectively, at the steady-state situation. Since we are in the low capillary regime, shear forces play a minor role in droplet deformation and are neglected in the figure. For $\chi=1$, since the droplet is still in a circular shape, the surface tension is present around the entire interface. However, for $\chi=2$ and 3, as the droplet has elongated, the surface tension force is mainly focused at the poles, competing with the magnetic force. Increasing the magnetic Bond number requires a longer simulation time for the droplet reaches a steady state, as previously reported by Jesus \textit{\textit{et al.}}~\citep{jesus2018deformation}.

Another important parameter to consider when investigating ferrofluid droplet deformation is the rotation of the droplet under various conditions, quantified by the angle measured counterclockwise from the positive $x-$direction to the major axis of the droplet, $\theta$. Previous studies have explored the rotational angle of the droplet under different magnetic Bond numbers in both low and high capillary regimes. For instance, Hassan \textit{\textit{et al.}}~\citep{hassan2018} demonstrated that for $\mathrm{Ca}\approx0.2$ and an imposed perpendicular magnetic field, increasing the magnetic Bond number leads to an increase in $\theta$ angle. This effect is attributed to the combined influence of the magnetic field and shear flow. However, the impact of magnetic susceptibility on droplet rotation has not yet been discussed in detail. In Fig.~\ref{fig:ShearedDroplet1}[right], the rotation of the droplet for Ca=0.02 under different magnetic Bond numbers is presented. For a susceptibility value of 1, similar to the behavior reported in previous studies~\citep{hassan2018}, an increase in the magnetic Bond number leads to an increase in the angle $\theta$ until it reaches a constant value. The shear flow attempts to rotate the droplet at an angle of 45$^\circ$, while the external magnetic field aims to elongate the droplet along its vertical field lines. The superposition of these two effects results in the rotation of the droplet. In the low capillary regime, as the magnetic Bond number increases, magnetic forces overcome the shear effect, causing the $\theta$ angle to increase along with the droplet deformation parameter. According to the numerical results, for a constant magnetic Bond number less than 2, the higher the droplet's susceptibility, the greater the rotational angle. This can be attributed to the increased Maxwell stress for higher susceptibility values, which further prevents the droplet from deflecting at the 45$^\circ$~angle. However, for magnetic Bond numbers of 3 and greater, the rotational angle becomes almost the same for all different susceptibility values, with no noticeable difference observed between them. These results suggest that not only does increasing the susceptibility value significantly affect droplet deformation, but also for lower magnetic Bond numbers, higher susceptibility leads to a greater rotational angle, causing the droplet to become more vertical. This is an important observation that can be further utilized when studying sheared droplet deformation in various applications.

\begin{figure}
\hspace*{-2.5cm}
\subfloat[]{\includegraphics[width = 0.72\textwidth]{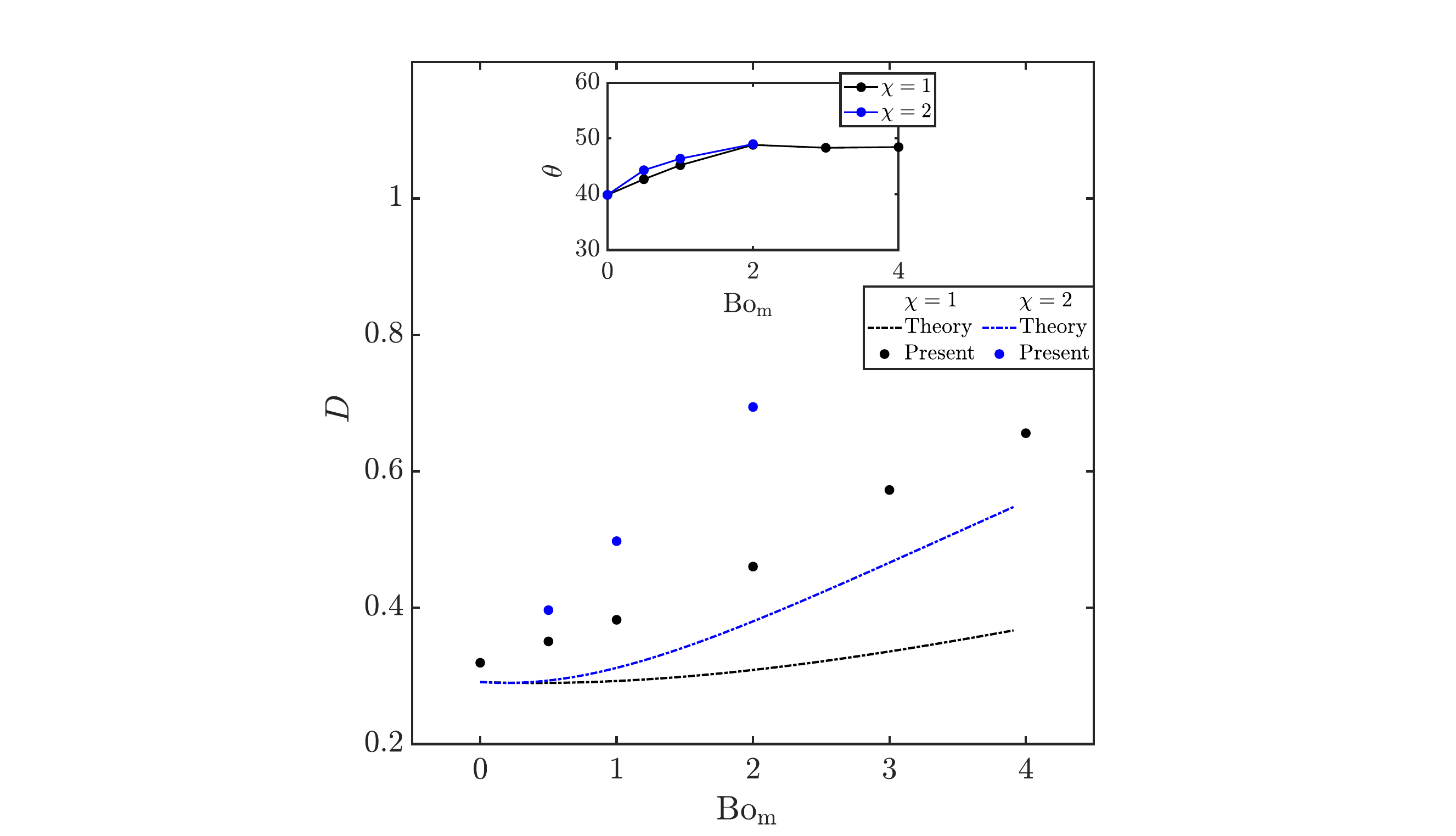}}\hspace*{-3.0cm}
\subfloat[]{\includegraphics[width=0.7\textwidth]{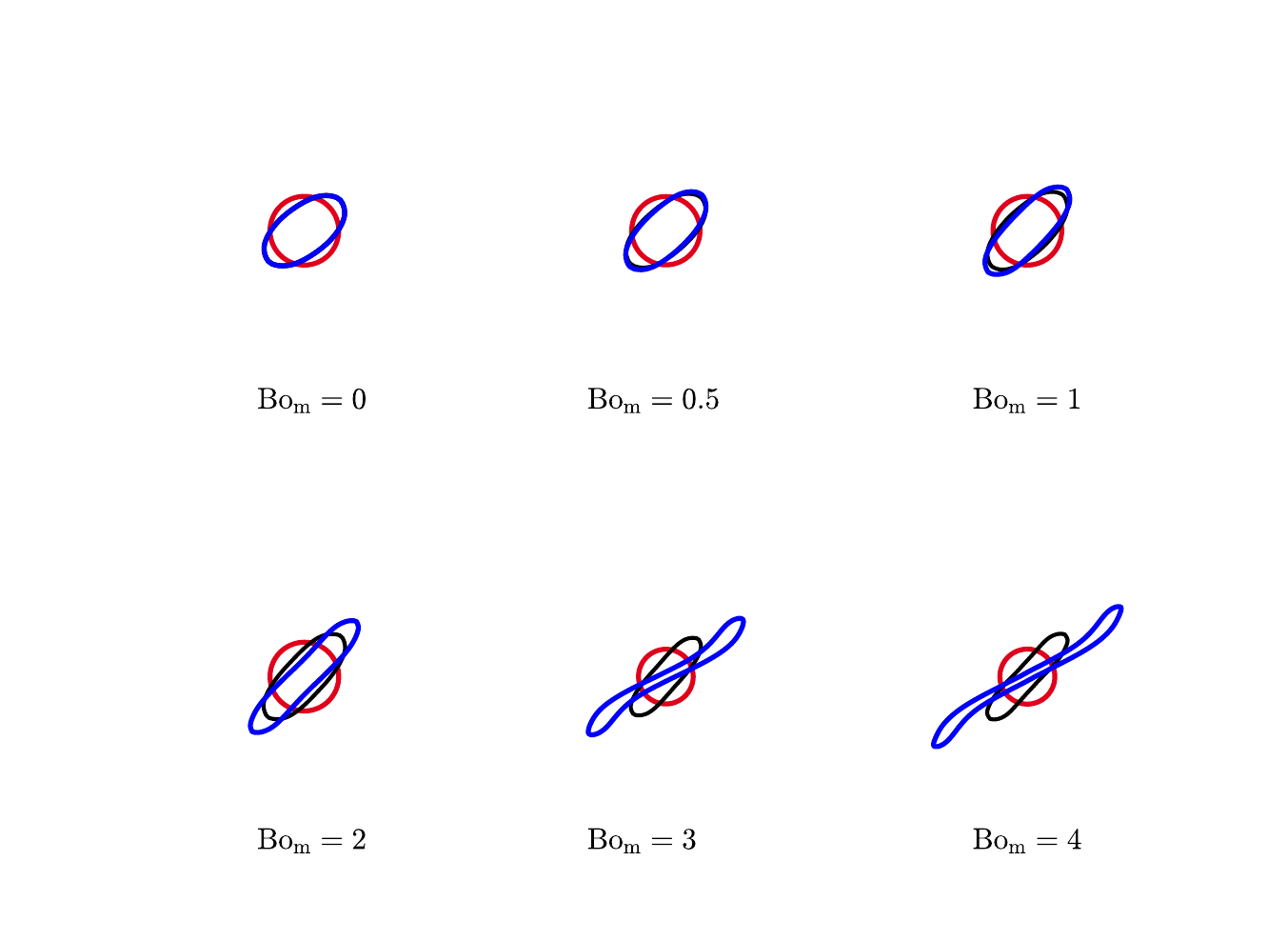}}\\
\hspace*{-3.7cm}
\subfloat[]{\includegraphics[width=1.4\textwidth]{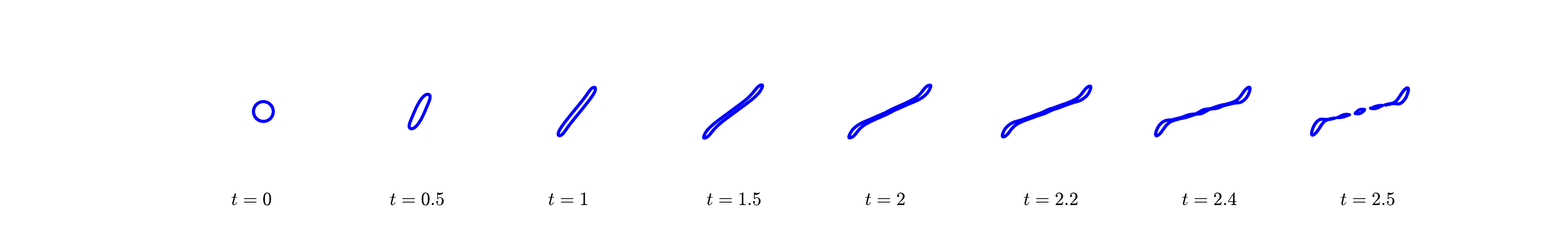}}
\caption{(a) Comparison of theoretical and numerical results for two different magnetic susceptibility values, $\chi=1$ and 2, under varying magnetic Bond numbers, while maintaining a fixed capillary number, $\mathrm{Ca=0.2}$. The inclination of the ferrofluid droplet, $\theta$, for different cases is also included in this figure. (b) Visualization of a ferrofluid droplet at steady state for two cases with $\chi=1$ and 2, each at a capillary number of 0.2, across six different magnetic Bond numbers (0, 0.5, 1, 2, 3, and 4). The red contour represents the initial droplet configuration at $t=0$. Notably, for the case with $\chi=2$ and magnetic Bond numbers of 3 and 4, the ferrofluid droplet has not reached a steady state, leading to a loss of its ellipsoidal shape as it continues to stretch and thin in the middle. (c) The sheared droplet deformation and breakup for the case $\chi=2$ with a magnetic Bond number of 4 at $t=0$, 0.5, 1, 1.5, 2, 2.2, 2.4, and 2.5, presented from left to right, respectively.}
\label{fig:ShearedDroplet3}
\end{figure}

The effect of the value of susceptibility on the deformation of a ferrofluid droplet is also investigated in the high capillary regime, Ca=0.2, achieved by increasing the shear rate. In Fig.~\ref{fig:ShearedDroplet3}(a), the deformation parameters of deformed droplets with two different susceptibilities, $\chi=1$ and 2, are represented for magnetic Bond numbers ranging from 0 to 4, while keeping Ca=0.2 constant. The theoretical results are also included in this figure. As expected, since we are in the higher capillary regime, the theoretical approach does not accurately predict the $D$ values. The theoretical analysis is valid only for $\mathrm{Ca} \ll 1$, and its predictions become less reliable as the magnetic Bond number increases.
According to this figure, for a constant magnetic Bond number, increasing the $\chi$ values results in an increase in the deformation parameter, similar to the observations in the previous section. However, it is notable that the obtained deformation parameters in this case are greater than those in the previous case with Ca=0.02. This is because the greater shear rate allows the external shear flow to more effectively influence the droplet's deformation. Moreover, it is evident that by increasing the magnetic Bond number, the angle $\theta$ increases and approaches an almost constant value. However, in this case, due to the stronger shear flow, the droplet tends to deflect more toward the 45$^\circ$ shear angle (please compare Fig.\ref{fig:ShearedDroplet1}[right] and Fig.~\ref{fig:ShearedDroplet3}(a)). The calculated deflection angles in the case of $\chi=2$ for magnetic Bond numbers below 2, are higher than those for $\chi=1$. This is because for $\chi=2$, the magnetic field lines are deflected further near the droplet's interface due to a higher magnetic permeability discontinuity across the interface, resulting in a stronger force to elongate the droplet along the field lines and leading to higher $\theta$ angles. It is worth noting that the results for $\chi=2$ under magnetic Bond numbers of 3 and 4 are absent in this figure. This is because for these magnetic Bond number values, the ferrofluid droplet has not reached a steady-state condition, as shown in Fig.~\ref{fig:ShearedDroplet3}(b), and it continues deforming until it undergoes breakup. Figure~\ref{fig:ShearedDroplet3}(c) illustrates the sheared droplet's deformation and breakup for the case where $\chi=2$ and $\mathrm{Bo_m}=4$ at eight different time steps, up to the moment when breakup occurs. The magnetic Bond number of approximately 3 is the critical value below which the droplet does not rupture. In previous studies, other properties such as the effect of viscosity ratio on droplet breakup were investigated~\citep{hassan2019magnetic}. In this test case, we have demonstrated that increasing the magnetic susceptibility value of the droplet can also cause droplet breakup, introducing the concept of a critical susceptibility value above which, for a constant Ca and $\mathrm{Bo_m}$ values, breakup occurs. Therefore, the magnetic permeability ratio between the droplet and its surrounding flow not only affects its deformation but can also lead to a breakup mechanism.

\color{black}
\subsubsection{Rayleigh--Taylor instability in magnetic fluids}
The Rayleigh--Taylor instability, which occurs when the interface between fluids of different densities is accelerated towards the direction of the heavy fluid, is widely studied for conducting fluids in the presence of a magnetic field as well. Magneto-Rayleigh--Taylor instability is of importance in different applications, e.g., Z-pinch fusion reactors~\citep{velikovich1996}, magnetic flux compression~\citep{harris1962}, and laser-driven inertial confinement fusion~\citep{velikovich2015}. In this test case, we investigate the Rayleigh--Taylor instability for magnetic fluids in the presence of a tangential quasi-static magnetic field to validate the performance of the implemented solver. By employing the linear analysis on the Rayleigh--Taylor instability, Awasthi~\cite{awasthi2014} derived the following quadratic dispersion relation for two viscous, incompressible, and magnetic fluids in the presence of a tangential magnetic field   

\begin{equation}
\overbrace{\left[\rho_1 + \rho_2 \right]}^{\alpha}\omega^2 + \overbrace{\left[4 k^2 (\mu_1 + \mu_2) \right]}^{\beta}\omega + \overbrace{\left[\left(\rho_1 - \rho_2 \right)g k + \sigma k^3 +  \frac{k^2 H_0^2 \left(\mu_{\mathrm{m,2}} - \mu_{\mathrm{m,1}}\right)^2}{ \left( \mu_{\mathrm{m,1}} + \mu_{\mathrm{m,2}} \right)} \right]}^{\gamma} = 0.
\label{eq:dispersion}
\end{equation}
In Eq.~(\ref{eq:dispersion}), $\omega$ and $k$ variables are the growth rate and wave number of a small perturbation at the interface, respectively, and subscript 1 (2) refers to the lighter (heavier) fluid. According to the instability criteria of Routh--Hurwitz, the stability condition is given as $\alpha>0$, $\beta>0$, and $\gamma>0$~\citep{awasthi2014}. Coefficients $\alpha$ and $\beta$ are always positive, and, therefore, the stability condition is reduced to $\gamma > 0$. The last term of the coefficient $\gamma$ represents the effect of the imposed magnetic field on the growth rate of the Rayleigh--Taylor instability. This term is always positive since the value of all the components in this term is positive. Hence, the tangential magnetic field always has a stabilizing effect on the Rayleigh--Taylor instability growth, similar to the effect of surface tension. For the sake of a more consistent analysis, by using the following dimensionless variables as
\begin{equation}
    \hat{k}=k L, \         \ \hat{\rho}=\frac{\rho_1}{\rho_2}, \        \ \hat{\mu}=\frac{\mu_1}{\mu_2}, \       \ \hat{\mu}_\mathrm{m}=\frac{\mu_\mathrm{m, 1}}{\mu_{\mathrm{m, 2}}}, \        \ \hat{\sigma}=\frac{\sigma}{\rho_2 \, g \, L^2}, \        \ \hat{H}= \frac{H_0 \sqrt{\hat{\rho}}}{\sqrt{\left(1-\hat{\rho}\right) \, g \, L}} \sqrt{\frac{\mu_{\mathrm{m,2}}}{\rho_2}},
\end{equation}
where $L$ is the characteristic length scale of the domain, the non-dimensional stability condition is expressed as
\begin{equation}
    1-\frac{\hat{\sigma} \hat{k}^2}{\left(1-\hat{\rho} \right)}-\frac{\hat{k} \hat{H}^2 \left(1-\hat{\mu}_\mathrm{m} \right)^2}{\hat{\rho} \left(1+\hat{\mu}_\mathrm{m} \right)} \leq 0.
    \label{eq:stability}
\end{equation}
For the state of the marginal stability, when Eq.~(\ref{eq:stability}) is zero, the critical value of the transverse magnetic field, $\hat{H}_\mathrm{c}$, is given as
\begin{equation}
  {\hat{H}_\mathrm{c}}^2 = \frac{\left[\left(1-\hat{\rho} \right) - \hat{\sigma} \hat{k}^2 \right] \hat{\rho} \left(1+\hat{\mu}_\mathrm{m} \right)}{\hat{k} \left(1-\hat{\rho} \right) \left(1-\hat{\mu}_\mathrm{m} \right)^2}.
\end{equation}
Hence, depending on whether the imposed transverse magnetic field is larger or smaller than the critical magnetic field, the system becomes stable or unstable, respectively.

\begin{figure*}[]
\centering
\hspace*{-1.5cm}
\subfloat[]{\includegraphics[width = 0.94\textwidth]{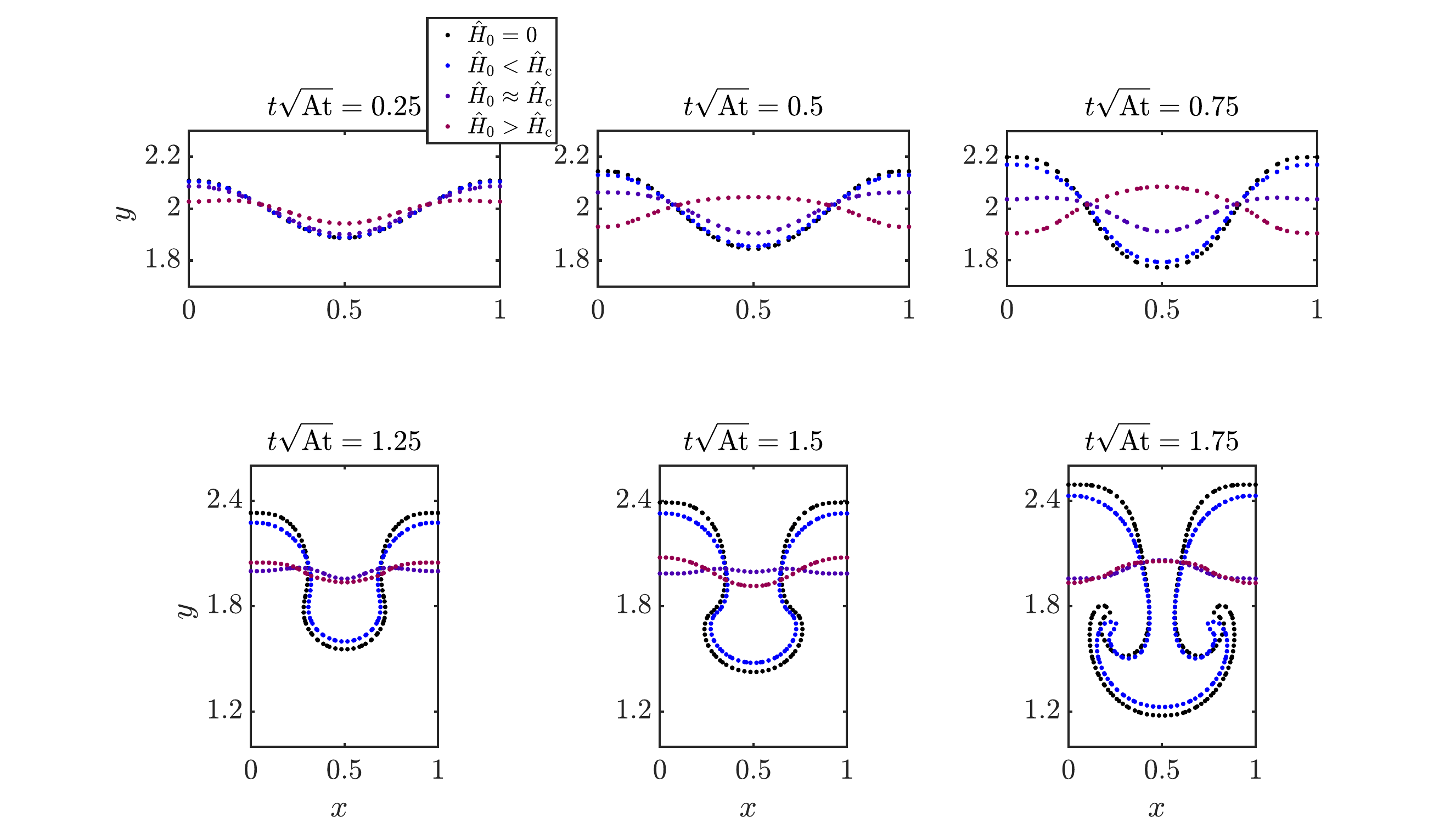}}\\
\vspace*{-0.4cm}
\subfloat[]{\includegraphics[width = 0.94\textwidth]{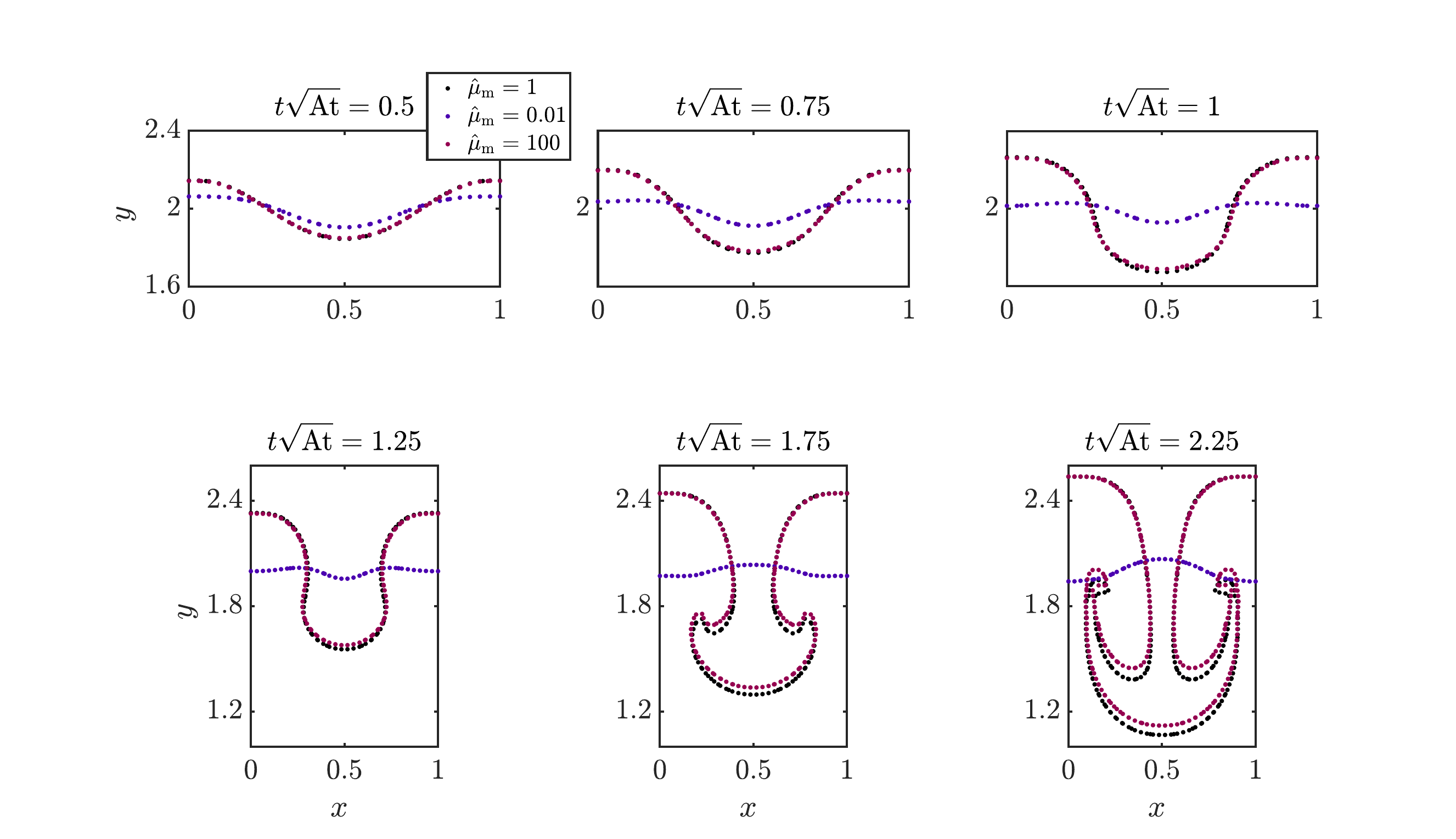}}
\caption{ (\textit{a}) The interface evolution of the Rayleigh--Taylor instability with $\hat{\rho}=1/3$ and $\hat{\mu}_\mathrm{m}=0.01$, at $t\sqrt{\mathrm{At}}=0.25, 0.5, 0.75, 1.25, 1.5,$ and 1.75, for three conditions of  $\hat{H}_\mathrm{0} < \hat{H}_\mathrm{c}$, $\hat{H}_\mathrm{0} \approx \hat{H}_\mathrm{c}$, and $\hat{H}_\mathrm{0} > \hat{H}_\mathrm{c}$, along with the nonmagnetic case. (\textit{b}) The interface evolution for three conditions of $\hat{\mu}_\mathrm{m}=0.01$, $\hat{\mu}_\mathrm{m}=1$, and $\hat{\mu}_\mathrm{m}=100$, with a constant $H_0 = 5$ and $\mu_\mathrm{m,1}=0.001$, at $t\sqrt{\mathrm{At}}=0.5, 0.75, 1, 1.25, 1.75,$ and 2.25. The simulations are performed on $32\times 128$ mesh resolution and time step $\Delta t= 5 \times 10^{-4}/\sqrt{\mathrm{At}}$.}
\label{fig:MRT}
\end{figure*}

To investigate the performance of the implemented solver for two-phase magnetic flows, the evolution of the Rayleigh--Taylor instability under different magnetic field intensities is simulated. To this end, the same initial condition as those presented in Sec.~\ref{sec:twophaseTests} for the Rayleigh--Taylor instability test case, with the grid resolution of $32 \times 128$, $\hat{\rho}=1/3 \, (\rho_1=1, \rho_2=3)$, and $\hat{\mu}_\mathrm{m}=0.01 \, (\mu_{\mathrm{m,1}}= 10^{-3}, \mu_\mathrm{m,2}=10^{-1})$ is considered. Three cases of $\hat{H}_\mathrm{0} < \hat{H}_\mathrm{c} \, ({H}_\mathrm{0} = 2 )$, $\hat{H}_\mathrm{0} \approx \hat{H}_\mathrm{c} \, ({H}_\mathrm{0} = 5)$, and $\hat{H}_\mathrm{0} > \hat{H}_\mathrm{c} \, ({H}_\mathrm{0} = 10)$ are examined \footnote{The critical value of the transverse magnetic field, ${H}_\mathrm{c}$, is expected to be around 4.}. The purpose is to qualitatively compare the numerical results with the ones obtained from linear perturbation analysis. Based on Fig.~\ref{fig:MRT}(a), it is visually evident that the presence of a transverse magnetic field, regardless of its magnitude, tends to suppress the growth of the Rayleigh--Taylor instability. This observation is consistent with the results obtained from other studies~\citep{el1994,awasthi2014}, stating that the Lorentz force at the interface acts as a restoring force. However, depending on the magnitude of the imposed transverse magnetic field, the evolution of the interface under the Rayleigh--Taylor instability varies. In the case of $\hat{H}_\mathrm{0} < \hat{H}_\mathrm{c}$, since the magnitude of the magnetic field is small compared to the critical magnetic field, the growth of the Rayleigh--Taylor instability is expected to be similar to the non-magnetic case, but with a slightly reduced growth rate over time. This behaviour is captured by the implemented numerical solver, as represented in Fig.~\ref{fig:MRT}(a). On the other hand, when the applied magnetic field is of the order of or greater than the calculated critical magnetic field, the Rayleigh--Taylor instability is anticipated to reach a stable regime. Consequently, the interface between the two fluids oscillates, and the penetration of the heavier fluid to the lighter one is prevented. The solver successfully reproduced these expected results, as demonstrated in Fig.\ref{fig:MRT}(a). Additionally, it is apparent from Fig.\ref{fig:MRT}(a) that for higher values of the applied magnetic field, the interface oscillates at a faster rate, consistent with the dispersion relation described by Eq.~(\ref{eq:dispersion}).

The magnetic permeability ratio between the two fluids, $\hat{\mu}_\mathrm{m}$, also influences the evolution of the Rayleigh--Taylor instability. By keeping the imposed transverse magnetic field constant and by varying $\hat{\mu}_\mathrm{m}$, the critical magnetic field, $\hat{H}_\mathrm{c}$, changes. Consequently, if the Rayleigh--Taylor instability becomes stable for a given $\hat{\mu}_\mathrm{m}$ and $\hat{H}_{0}$, altering the value of $\hat{\mu}_\mathrm{m}$ can render it unstable. It is for this reason that the magnetic permeability ratio has a dual role of destabilizing and stabilizing the system, as was mentioned by Awatshi~\cite{awasthi2014}. To assess the solver's capability to accurately handle different magnetic permeability ratios across the interface, three cases of $\hat{\mu}_\mathrm{m}=0.01$, $\hat{\mu}_\mathrm{m}=1$, and $\hat{\mu}_\mathrm{m}=100$ are investigated for a constant imposed $\hat{H}_0$, shown in Fig.~\ref{fig:MRT}(b). It can be visually confirmed that the case of $\hat{\mu}_\mathrm{m}=1$ is the most unstable one, a finding that is consistent with Eq.~(\ref{eq:dispersion}). The Rayleigh--Taylor instability transitions to a stable regime when $\hat{\mu}_\mathrm{m}=0.01$. However, by changing the magnetic permeability ratio to $100$, the system becomes unstable, demonstrating the destabilizing effect of the $\hat{\mu}_\mathrm{m}$.

\section{Conclusion}

This paper presents a numerical development of a two-phase incompressible solver for magnetic flows. The proposed numerical toolkit couples the Navier--Stokes equations of hydrodynamics with Maxwell's equations of electromagnetism, enabling us to model the behaviour of magnetic flows in the presence of a static magnetic field. In order to achieve this goal, first, a detailed implementation of a second-order two-phase solver for incompressible nonmagnetic flows is introduced. This two-phase solver utilizes a fifth-order conservative level set method within the finite-difference framework to capture the interface evolution during the simulation. The accuracy, robustness, and mass conservation properties of the level set solver are verified by conducting three test cases, namely, a rotating circle, a circle in a deformation field, and Zalesak's disk. Additionally, the incompressible Navier--Stokes equations are modelled based on the projection method and solved in the conservative form. The order of accuracy for the implemented two-phase nonmagnetic solver is verified through the capillary wave test case. The Rayleigh--Taylor instability test is also performed to evaluate the performance of the solver for more complicated interface evolution and higher density ratios. For the magnetic case, the reduced Maxwell equations under the magnetostatic assumption are solved based on the vector potential formulation. Subsequently, the primary two-phase solver is extended to account for magnetic flows by incorporating the Lorentz force in the momentum equation. The developed solver demonstrated the capability to handle high magnetic permeability jumps across the interface. Three benchmarks were conducted to evaluate the performance and robustness of the implemented two-phase solver for magnetic flows in the presence of a static magnetic field. The first test case explored droplet deformation under the influence of an external magnetic field in a quiescent flow, examining the interplay between the Lorentz force and capillary force at the interface across varying magnetic field amplitudes. These findings were validated by comparing them with analytical solutions and existing numerical and experimental results from the literature. In the sheared ferrofluid droplet test, the impact of magnetic susceptibility on droplet dynamics was studied. The results revealed that an increase in the susceptibility of the ferrofluid droplet could influence its deformation and rotation in low capillary regimes. However, in higher capillary flows, it was observed that, under constant capillary and magnetic Bond number values, increasing magnetic susceptibility could lead to droplet breakup. Furthermore, the solver was employed to simulate Rayleigh--Taylor instability in magnetic fluids. Numerical results were compared with linear stability analysis across different magnetic field intensities and magnetic permeability ratios at the interface. This analysis facilitated a discussion regarding the effects of these variables in either stabilizing or destabilizing the Rayleigh--Taylor instability.


\section*{Acknowledgments}
The authors would like to thank Victor Boniou for his valuable comments on the surface tension implementation and curvature calculations. Paria Makaremi-Esfarjani further acknowledges the helpful suggestions of Khashayar F. Kohan on the initial draft of the manuscript. The authors thank Mathias Larrouturou for useful feedback on the manuscript.  This study was supported
by the Natural Science and Engineering Research Council of Canada
(NSERC) through a NSERC Discovery Grant and Mitacs through the Mitacs Accelerate program.

\appendix
\section{\label{appendixA}Fifth-order weighted essentially non-oscillatory (WENO) interpolation}

The computation of the level set convective flux requires the interpolation of cell center $\psi$ values to cell faces using an upwind-based scheme. In this study, a fifth-order WENO interpolation is implemented in our numerical solver, explained in greater detail in this section. The concept of the WENO interpolation is similar to the WENO reconstruction. This method involves using a weighted combination of three sub-stencils to calculate the flux values, resulting in a high-order scheme. The weights are chosen to minimize the contribution of stencils with discontinuities, thereby avoiding numerical instability and providing non-oscillatory interpolation.

\begin{figure}
\centering
\includegraphics[scale=1.5]{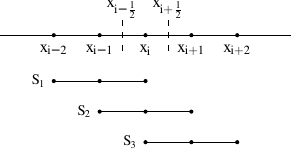}
\caption{\label{fig:WENOSstencil} The stencil for the fifth-order WENO interpolation scheme to find $\tilde{\psi}_{\mathrm{L}, \, x_{i+1/2}}$ and $\tilde{\psi}_{\mathrm{R}, \, x_{i-1/2}}$ values at $x_{i+1/2}$ and $x_{i-1/2}$ cell faces, respectively.}
\end{figure}
In Fig.~\ref{fig:WENOSstencil}, the one-dimensional stencil for the fifth-order interpolation in $x_i-$direction is shown. Subscripts $\mathrm{L}$ (left) and $\mathrm{R}$ (right) are employed to denote the interpolated values at each cell face based on upwinding considerations. As a result, cell values at  $\tilde{\psi}_{\mathrm{L}, x_{i+1/2}}$ and $\tilde{\psi}_{\mathrm{R}, x_{i-1/2}}$ have the same five-points stencil, shown in Fig.~\ref{fig:WENOSstencil}, and can be divided into three sub-stencils
\begin{subequations}
\begin{equation}
\mathrm{S}_1 = \left\{x_{i-2}, \, x_{i-1}, \, x_i \right \},
\end{equation}
\begin{equation}
 \mathrm{S}_2 = \left \{x_{i-1},\,  x_i, \, x_{i+1} \right \},
\end{equation}
and
\begin{equation}
 \mathrm{S}_3 = \left \{x_i, \, x_{i+1}, \, x_{i+2} \right \}.
\end{equation}
\end{subequations}
Using the third-order interpolation, the cell face values can be obtained for each sub-stencil as
\begin{subequations}
\begin{equation}
 \psi ^{(1)}_{\mathrm{L}, \, x_{i+1/2}} = \frac{3}{8} \psi_{x_{i-2}} - \frac{5}{4} \psi_{x_{i-1}} + \frac{15}{8} \psi_{x_i},       
\end{equation}
\begin{equation}
 \psi ^{(2)}_{\mathrm{L}, \, x_{i+1/2}} = -\frac{1}{8} \psi_{x_{i-1}} + \frac{3}{4} \psi_{x_{i}} + \frac{3}{8} \psi_{x_{i+1}},       
\end{equation}
\begin{equation}
 \psi ^{(3)}_{\mathrm{L}, \,  x_{i+1/2}} =  \frac{3}{8} \psi_{x_{i}} + \frac{3}{4} \psi_{x_{i+1}} - \frac{1}{8} \psi_{x_{i+2}},              
\end{equation}
\end{subequations}
and
\begin{subequations}
\begin{equation}
 \psi ^{(1)}_{\mathrm{R}, \, x_{i-1/2}} = -\frac{1}{8} \psi_{x_{i-2}} + \frac{3}{4} \psi_{x_{i-1}} + \frac{3}{8} \psi_{x_i},       
\end{equation}
\begin{equation}
 \psi ^{(2)}_{\mathrm{R}, \, x_{i-1/2}} = \frac{3}{8} \psi_{x_{i-1}} + \frac{3}{4} \psi_{x_{i}} - \frac{1}{8} \psi_{x_{i+1}},       
\end{equation}
\begin{equation}
 \psi ^{(3)}_{\mathrm{R}, \, x_{i-1/2}} =  \frac{15}{8} \psi_{x_{i}} - \frac{5}{4} \psi_{x_{i+1}} + \frac{3}{8} \psi_{x_{i+2}}.     \end{equation}
\end{subequations}
The fifth-order interpolation is then obtained by combining the calculated third-order interpolation for three sub-stencils using non-linear weights, defined as
\begin{equation}
    \tilde{\psi}_{\mathrm{L}, \, x_{i+1/2}} = \sum_{k=1}^{3} \omega_{\mathrm{L}, \,k} \, \psi^{(k)}_{\mathrm{L},\, x_{i+1/2}}, \quad  \textrm{and} \quad \tilde{\psi}_{\mathrm{R}, \, x_{i-1/2}} = \sum_{k=1}^{3} \omega_{\mathrm{R}, \, k} \, \psi^{(k)}_{\mathrm{R}, \, x_{i-1/2}}, 
\end{equation}
where $\omega$ variables are the nonlinear weights adopted from Jiang and Shu~\cite{jiang1996}, given as
\begin{equation}
    \omega_{\mathrm{L}, \, k}=\frac{\beta_{\mathrm{L}, \, k}}{\sum_{m=1}^{3} \beta_{\mathrm{L}, m}}, \quad \textrm{and} \quad \omega_{\mathrm{R}, \, k}=\frac{\beta_{\mathrm{R}, \, k}}{\sum_{m=1}^{3} \beta_{\mathrm{R}, m}},
\end{equation}
where
\begin{equation}
    \beta_{\mathrm{L}, \, k}=\frac{\bar{\omega}_{\mathrm{L}, \, k}}{\left( \epsilon + \mathrm{IS}_k\right)^2},  \quad \textrm{and} \quad \beta_{\mathrm{R}, \, k}=\frac{\bar{\omega}_{\mathrm{R}, \, k}}{\left( \epsilon + \mathrm{IS}_k\right)^2}.
\end{equation}
Here parameter $\epsilon$ is used to prevent the denominators to become zero. The value of $\epsilon$ is typically chosen to be between $10^{-5}$ and $10^{-7}${~\cite{jiang1996}. In this study, we used a value of $\epsilon=10^{-6}$. Variables $\mathrm{IS}_k$ are the smooth indicators for each sub-stencil, defined as
\begin{subequations}
\begin{eqnarray}
    \mathrm{IS}_1 = \frac{1}{3} \left[\psi_{x_{i-2}} \left(4\psi_{x_{i-2}} - 19\psi_{x_{i-1}} + 11\psi_{x_{i}} \right) + \psi_{x_{i-1}} \left(25 \psi_{x_{i-1}} - 31 \psi_{x_{i}} \right) + 10 \psi^{2}_{x_{i}} \right],
\end{eqnarray}
\begin{eqnarray}
    \mathrm{IS}_2 = \frac{1}{3} \left[\psi_{x_{i-1}} \left(4\psi_{x_{i-1}} - 13\psi_{x_{i}} + 5 \psi_{x_{i+1}} \right) + 13 \psi_{x_{i}} \left(25 \psi_{x_{i}} - \psi_{x_{i+1}} \right) + 4 \psi^{2}_{x_{i+1}} \right],
\end{eqnarray}
and
\begin{eqnarray}
    \mathrm{IS}_3 = \frac{1}{3} \left[\psi_{x_{i}} \left(10\psi_{x_{i}} - 13\psi_{x_{i+1}} + 11 \psi_{x_{i+2}} \right) + \psi_{x_{i+1}} \left(25 \psi_{x_{i+1}} - 19 \psi_{x_{i+2}} \right) + 4 \psi^{2}_{x_{i+2}} \right],
\end{eqnarray}
\end{subequations}
where $\bar{\omega}_{\mathrm{L}, \, k}$ and $\bar{\omega}_{\mathrm{R}, \, k}$ are the linear optimal weights for the fifth-order interpolation, calculated using the Lagrange interpolation, and are determined as
\begin{equation}
    \bar{\omega}_{\mathrm{L}, \, k}= \left \{\frac{1}{16}, \frac{10}{16}, \frac{5}{16} \right \}, \quad \textrm{and} \quad \bar{\omega}_{\mathrm{R}, \, k}= \left \{\frac{5}{16}, \frac{10}{16}, \frac{1}{16} \right \}.
\end{equation}
The same procedure can be applied in each direction for multi-dimensional problems as well.

\section{\label{appendixB}Third-order Runge--Kutta scheme}

The third-order, total variation diminishing (TVD) Runge--Kutta scheme introduced by Gottlieb and Shu~\cite{gottlieb1998} is an explicit temporal integration scheme, which can attenuate spurious oscillations appearing in the solution.

Consider a system of differential equations given by
\begin{equation}
    \frac{\partial \mathbf{U}}{\partial t}=\mathbf{F}(\mathbf{U}, t)\\,
\end{equation}
where $\mathbf{U}$ is the vector of conservative variables and $\mathbf{F}$ is a right-hand-side operator. For a known solution, $\mathbf{U}^n$, the third-order Runge--Kutta scheme approximates the solution at the next time step, $\mathbf{U}^{n+1}$, using an intermediate step 
\begin{equation}
        \mathbf{U}^1=\mathbf{U}^n + \Delta t \mathbf{F}(\mathbf{U}^n,t),
    \end{equation}
followed by
\begin{equation}
        \mathbf{U}^2=\frac{3}{4}\mathbf{U}^n + \frac{1}{4} \mathbf{U}^1+\frac{1}{4} \Delta t \mathbf{F}(\mathbf{U}^1,t),
    \end{equation}
which in turn is followed by the final step    
\begin{equation}
        \mathbf{U}^{n+1}=\frac{1}{3}\mathbf{U}^n + \frac{2}{3} \mathbf{U}^2+\frac{2}{3} \Delta t \mathbf{F}(\mathbf{U}^2,t),
        \\
    \end{equation}
where $\Delta t$ denotes the time step.
 
\section{\label{appendixC} Evaluation of level set solver numerical accuracy: Rotating circle test}

The rotating circle test is examined to study the numerical order of accuracy of the level set solver. In this test case, a circular interface of radius $r=0.15$, with its center initially located at $(x_0, y_0)=(0.25, 0.25)$ is considered. The circle rotates in the two-dimensional computational domain, $[x,y]\in [-0.8, 0.8] \times [-0.8, 0.8]$, under the constant velocity field $\left(u,v \right)=(y, -x)$. According to the initial condition of the problem, the circle should return to its initial place after one revolution at $t=2\pi$.
The initial velocity field and the evolution of the rotating circle from $t=0$ to $t=2\pi$ is represented in Fig.~\ref{fig:ls_accuracy}(a).
The initial thickness of the level set profile is set to be $\epsilon = (\Delta x)^{0.8}/2$, with the constant time step equal to $\Delta t = 0.001$ and $\Delta \tau = 0.0005$ for transport and re-initialization equations, respectively. The re-initialization process is executed every $20$ steps, with a maximum of two iterations per step. The simulation is performed for five different grid resolutions, and $L_2$ and $L_\infty$ errors are determined for each case. The values of $L_2$ and $L_\infty$ errors are calculated as

\begin{equation}
    L_2 \, \mathrm{error} = \sqrt{\frac{\Sigma_{i=1} ^N \left(\psi_{\mathrm{anal,}i} - \psi_{\mathrm{num},i} \right)^2}{N}} \ \       \         \ \      \ \mathrm{and} \  \  \      \ \       \ L_\infty \, \mathrm{error} = \max \big| \psi_{\mathrm{anal,}i} - \psi_{\mathrm{num},i} \big|,
\end{equation}
\noindent where $\psi_{\mathrm{anal,}i}$ and $\psi_{\mathrm{num},i}$ denote the analytical and numerical results at grid point $i$, respectively, and $N$ is the total number of the grid points.
In Fig.~\ref{fig:ls_accuracy}(b), the error values are plotted versus the number of nodes in the logarithmic scale, and, hence, the slope of the plot depicts the estimated convergence rate of the implemented numerical scheme. As can be observed from Fig.~\ref{fig:ls_accuracy}(b), the order of accuracy is between $4$ and $4.5$, which is pretty close to the excepted fifth-order accuracy of the implemented method, even for the defined sharp interface case ($\epsilon<\Delta x/2$). However, the re-initialization step is expected to affect the global accuracy of the level set solver.\\     

\begin{figure}
\hspace*{-2.0cm}
\subfloat[]{\includegraphics[width = 0.7\textwidth]{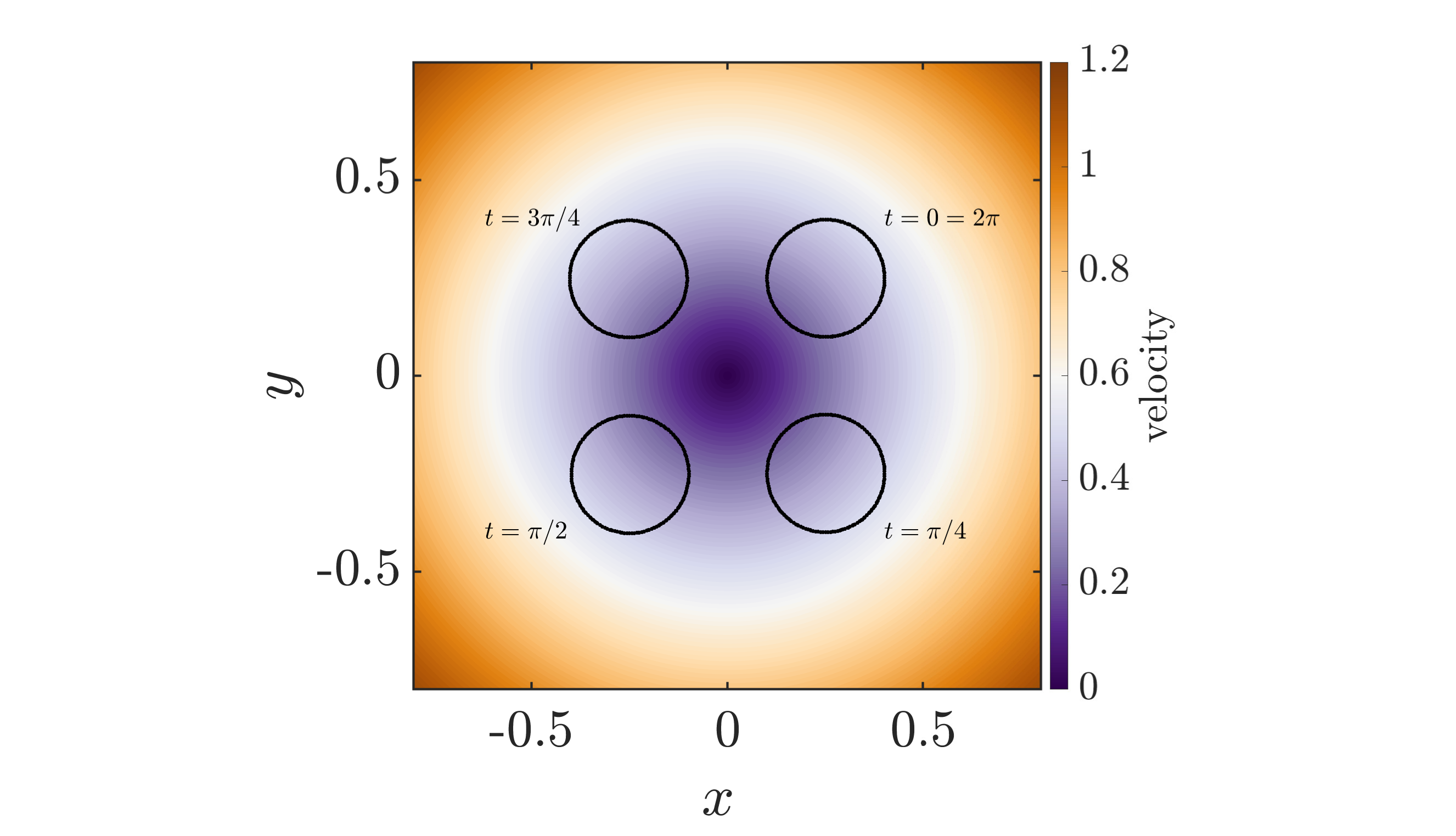}}\hspace*{-2.0cm}
\subfloat[]{\includegraphics[width = 0.7\textwidth]{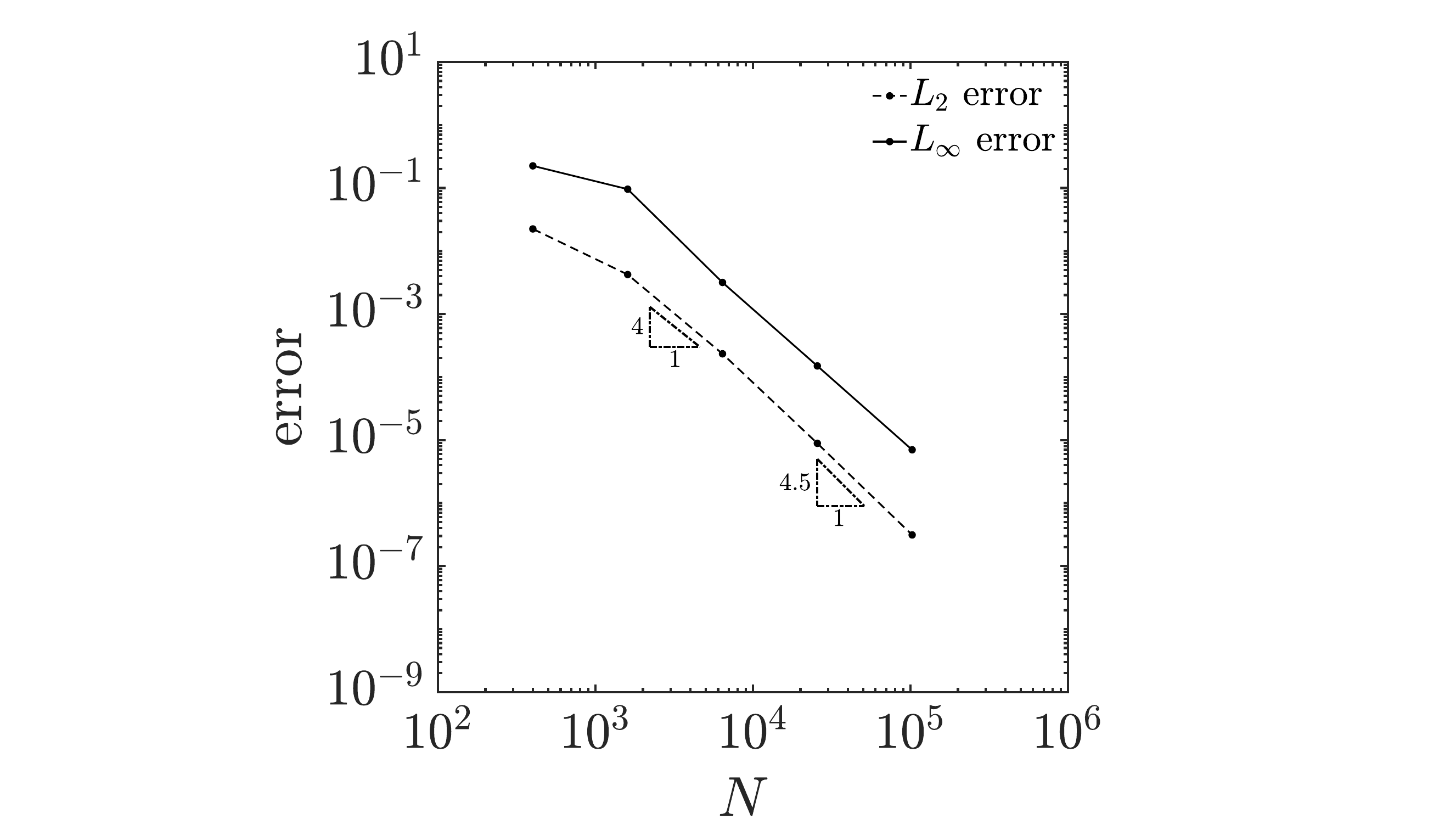}}
\caption{\label{fig:ls_accuracy} (\textit{a}) The velocity field of the rotating circle test case along with the $\psi=0.5$ location at $t=0, \pi/4, \pi/2, 3\pi/4,$ and $2\pi$. (\textit{b}) Order of accuracy analysis for the implemented level set solver using the rotating circle test case. The error values are computed for five different grid resolutions, $20 \times 20$, $40 \times 40$, $80 \times 80$, $160 \times 160$, and $320 \times 320$, with a constant time step for all simulations.}
\end{figure}

\section{\label{appendixD} Incompressible solver test cases}

In this section, two test cases are presented, the Taylor--Green vortex and the lid-driven cavity, to examine the order of accuracy and robustness of the implemented incompressible solver.

\subsection{Decay of a Taylor--Green vortex}
The Taylor--Green vortex is a well-known test case in the literature to examine the ability of an incompressible Navier--Stokes numerical solver to simulate transient problems. In this study, the Taylor--Green test is investigated to evaluate the order of accuracy of the implemented incompressible solver. The analytical solution of this benchmark is given as
\begin{subequations}
\begin{equation}
    u = \sin \left(k x \right) \cos \left(k y \right) \exp \left(-2 \mu k^2 t/\rho \right),
\end{equation}
\begin{equation}
    v = -\cos \left(k x \right) \sin \left(k y \right) \exp \left(-2 \mu k^2 t/\rho \right),
\end{equation}
and
\begin{equation}
    p = \frac{1}{4} \rho \left[\cos \left(2 k x \right) + \cos \left( 2 k y \right) \right]  \exp \left(-4 \mu k^2 t/\rho \right),
\end{equation}
\end{subequations}
where $k$ is the wavenumber, set to $k=1$. The magnitude of the initial velocity field is represented in Fig.~\ref{fig:incomp_accuracy}(a), which will be exponentially damped due to the presence of the viscosity. The simulation is performed on the computational domain $[x, y] \in [0, 2\pi] \times [0, 2\pi]$, with a periodic boundary condition implemented at all boundaries. The values of density and dynamic viscosity are set to $1$ and $0.01$, respectively. The simulation is run for five different grid resolutions, $16 \times 16, 32 \times 32, 64\times64, 128\times128$, and $256 \times 256$, with the constant time step, $\Delta t=0.001$. The maximum error of the velocity field is calculated at $t=0.2$ for each case. In Fig.~\ref{fig:incomp_accuracy}(b), the error values are plotted against the number of nodes in the logarithmic scale, demonstrating the estimated convergence rate of approximately three for the implemented incompressible solver. It should be noted that the global accuracy of the velocity field is affected by the approximation of the pressure gradient and the discretization of the diffusion term, which are second-order accurate herein. In Table~\ref{tab:table1}, second-order convergence is observed for the pressure field as expected. Also, the maximum calculated error for the velocity divergence is quite small, $\mathcal{O}\left(10^{-13} \right)$, showing the proper divergence-free property of the velocity field.
\begin{figure}
\hspace*{-2.0cm}
\subfloat[]{\includegraphics[width = 0.72\textwidth]{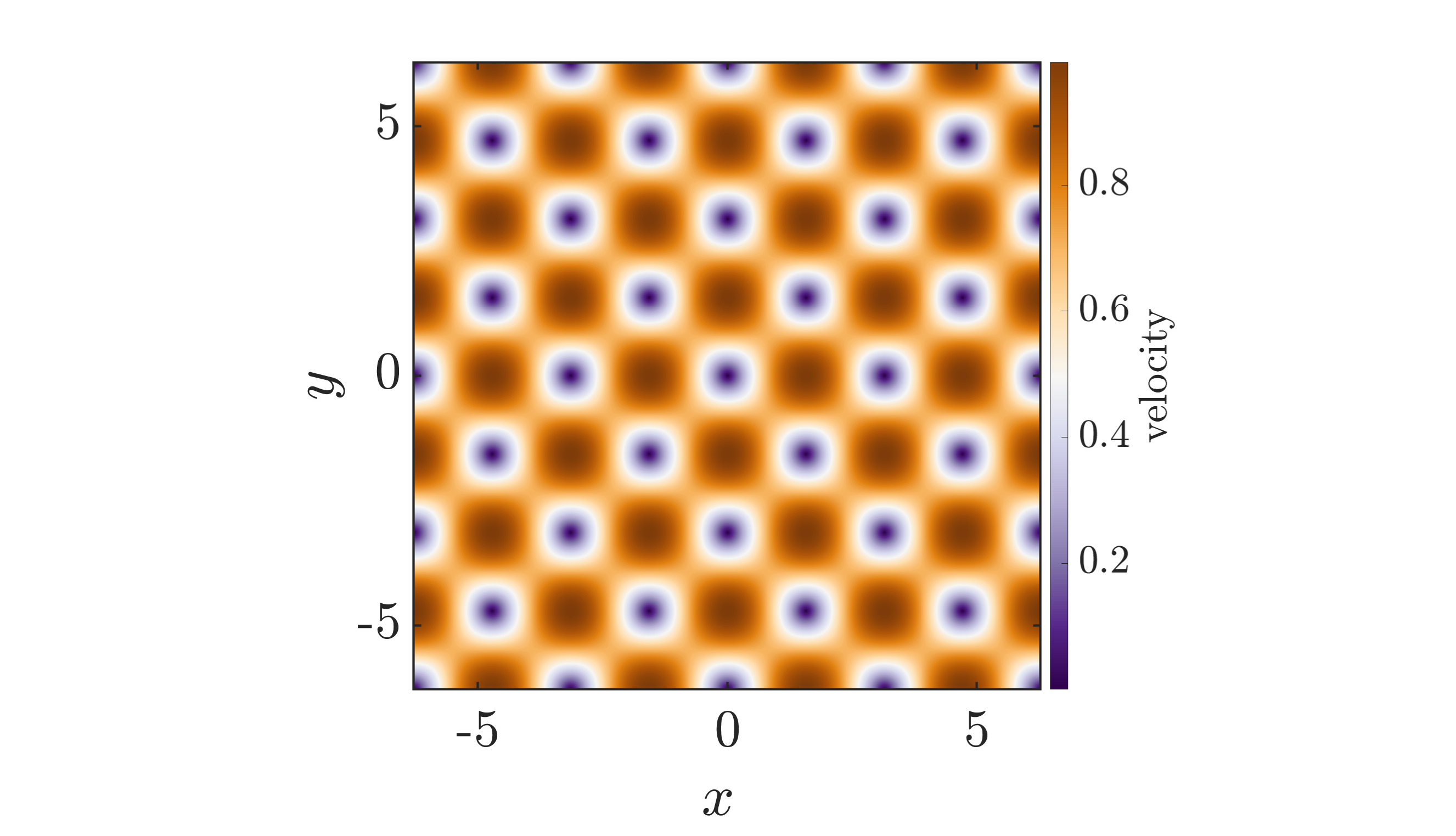}}\hspace*{-2.0cm}
\subfloat[]{\includegraphics[width=0.7\textwidth]{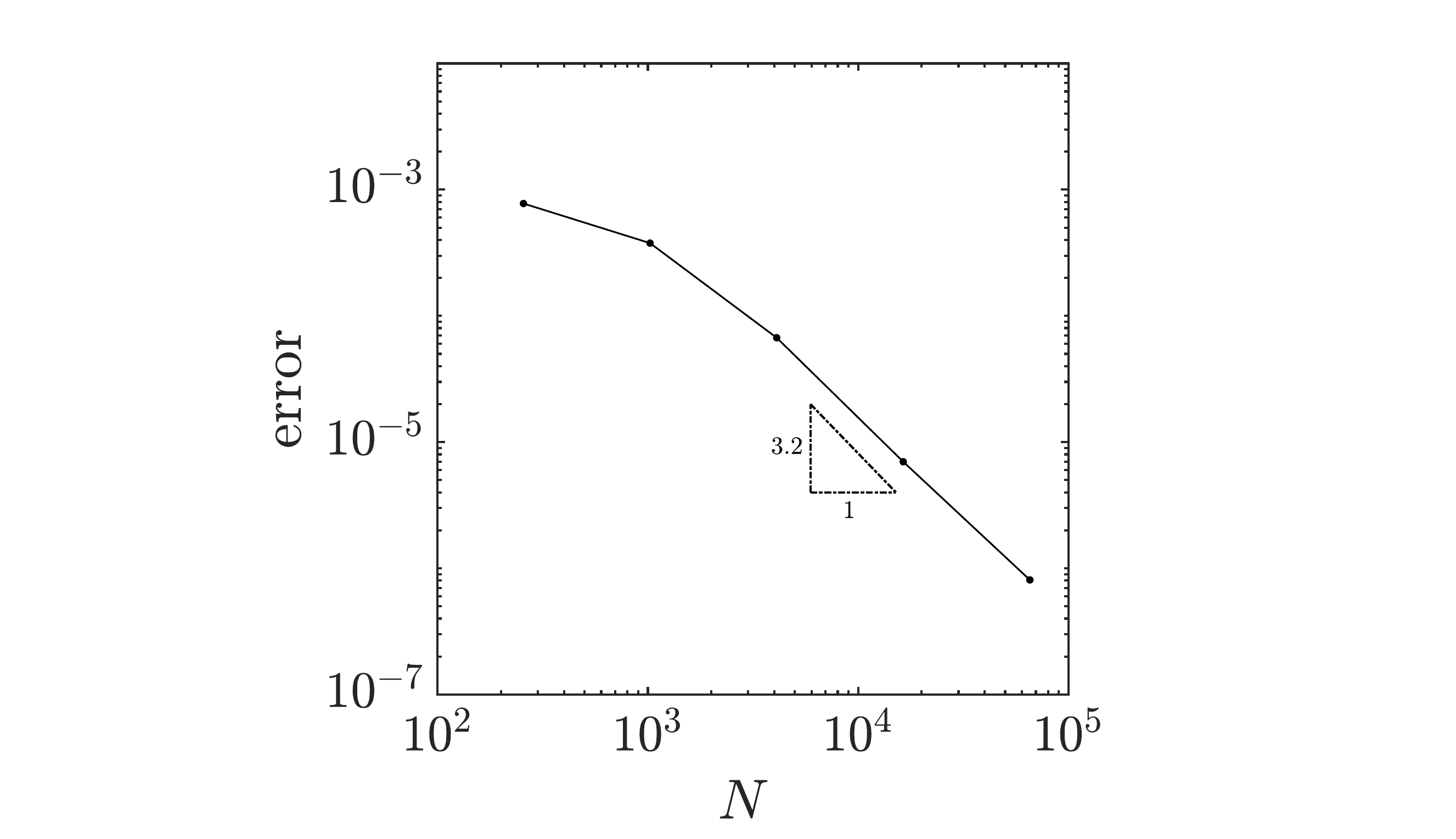}}
\caption{\label{fig:incomp_accuracy} (\textit{a}) Initial velocity field of the Taylor--Green test case. (\textit{b}) Order of accuracy analysis for the implemented incompressible solver using the decaying Taylor--Green vortex test case. The error values are computed for five different grid resolutions, $16 \times 16$, $32 \times 32$, $64 \times 64$, $128 \times 128$, and $256 \times 256$, with the constant time step, $\Delta t=0.001$, for all the simulations.}
\end{figure}
\begin{table*}
\centering
\caption{\label{tab:table1} Calculated $L_{\infty}$ error of the pressure and velocity divergence for different grid resolutions of the Taylor--Green test case.}
\begin{tabular}{cccc}
\hline
\hline
 Mesh resolution & Error of $p$ & Rate & Error of $\left(\nabla \cdot \mathbf{u} \right)$\\ \hline
$16 \times 16$ & $5.28 \times 10^{-2}$ & \textbf{------} & $9.92 \times 10^{-14}$ \\
$32 \times 32$ & $1.69 \times 10^{-2}$ & 1.64 & $2.50 \times 10^{-13}$ \\
$64 \times 64$ & $5.44 \times 10^{-3}$ & 1.63 & $3.96 \times 10^{-13}$ \\
$128 \times 128$ & $1.50 \times 10^{-3}$ & 1.85 & $7.54 \times 10^{-13}$ \\
$256 \times 256$ & $3.91 \times 10^{-4}$ & 1.94 & $9.85 \times 10^{-13}$ \\
\hline
\hline
\end{tabular}
\end{table*}

\subsection{Lid-driven cavity}
The lid-driven cavity flow test case is used to evaluate the performance of the implemented solver for complicated flow fields and its capability to reach a steady state solution. This test case includes the square cavity, $[x, y] \in [0, 1] \times [0, 1]$, initially filled with an incompressible fluid, with the top wall moving with the constant velocity, $u=1$, while the other three walls are stationary. The simulation is performed for two cases of Reynolds numbers of $100$ and $1000$, with a no-slip boundary condition imposed at all boundaries. The two simulations are run for the grid resolution of $100 \times 100$ and $\Delta t=0.001$. Figure~\ref{fig:CavityTest}(a) depicts the streamlines of the velocity fields for $\mathrm{Re}=100$ and $\mathrm{Re}=1000$ cases at $t=10$ and $t=50$, respectively. This figure visually confirms that the calculated velocity field follows the expected behavior, which is the formation of a primary vortex towards the center of the cavity and the generation of smaller corner vortices at the bottom corners. For a better evaluation, the computed results for the $u$ and $v$ velocities at the vertical and horizontal centerlines, respectively, are compared with the corresponding results from the study by Huang \textit{et al.}~\cite{huang2019}, shown in Fig.~\ref{fig:CavityTest}(b) and (c). As can be observed in this figure, there is a good agreement between the results of the two studies for both cases of $\mathrm{Re}=100$ and $\mathrm{Re}=1000$.

\begin{figure}[]
\centering
\subfloat[]{\includegraphics[width = 1\textwidth]{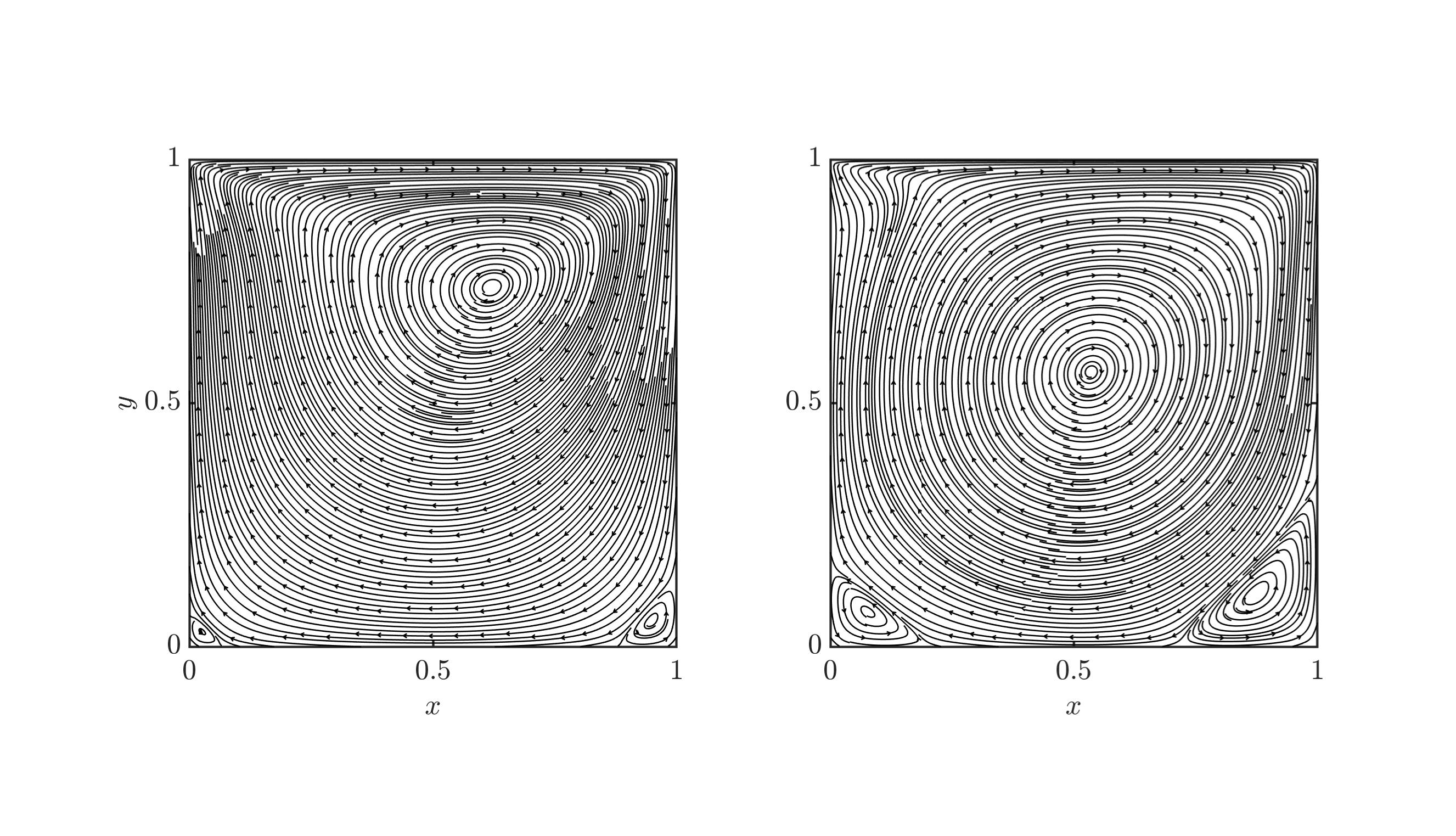}}\\
\hspace*{-0.95cm}
\subfloat[]{\includegraphics[width = 0.65\textwidth]{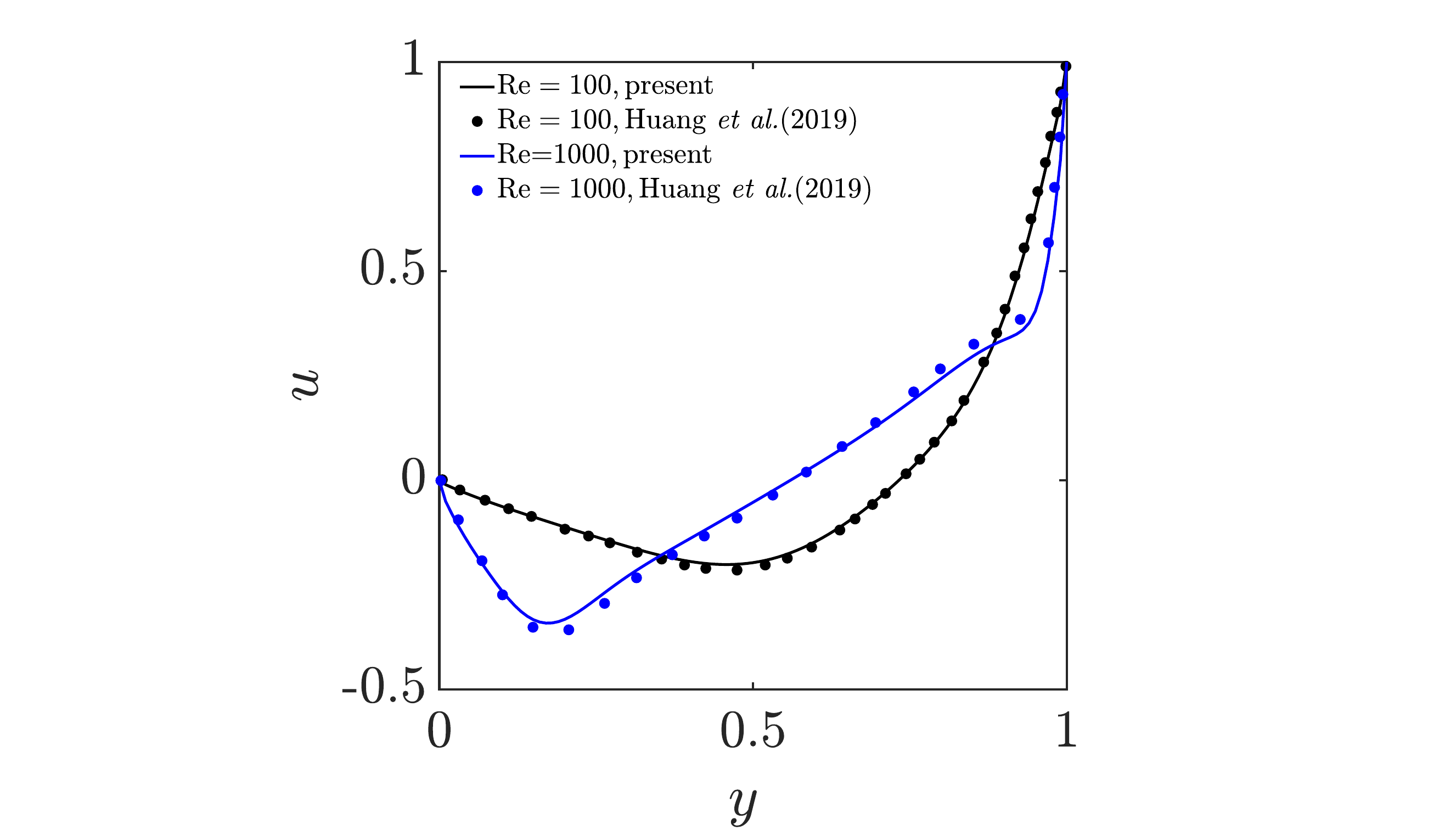}}\hspace*{-2.5cm}
\subfloat[]{\includegraphics[width = 0.65\textwidth]{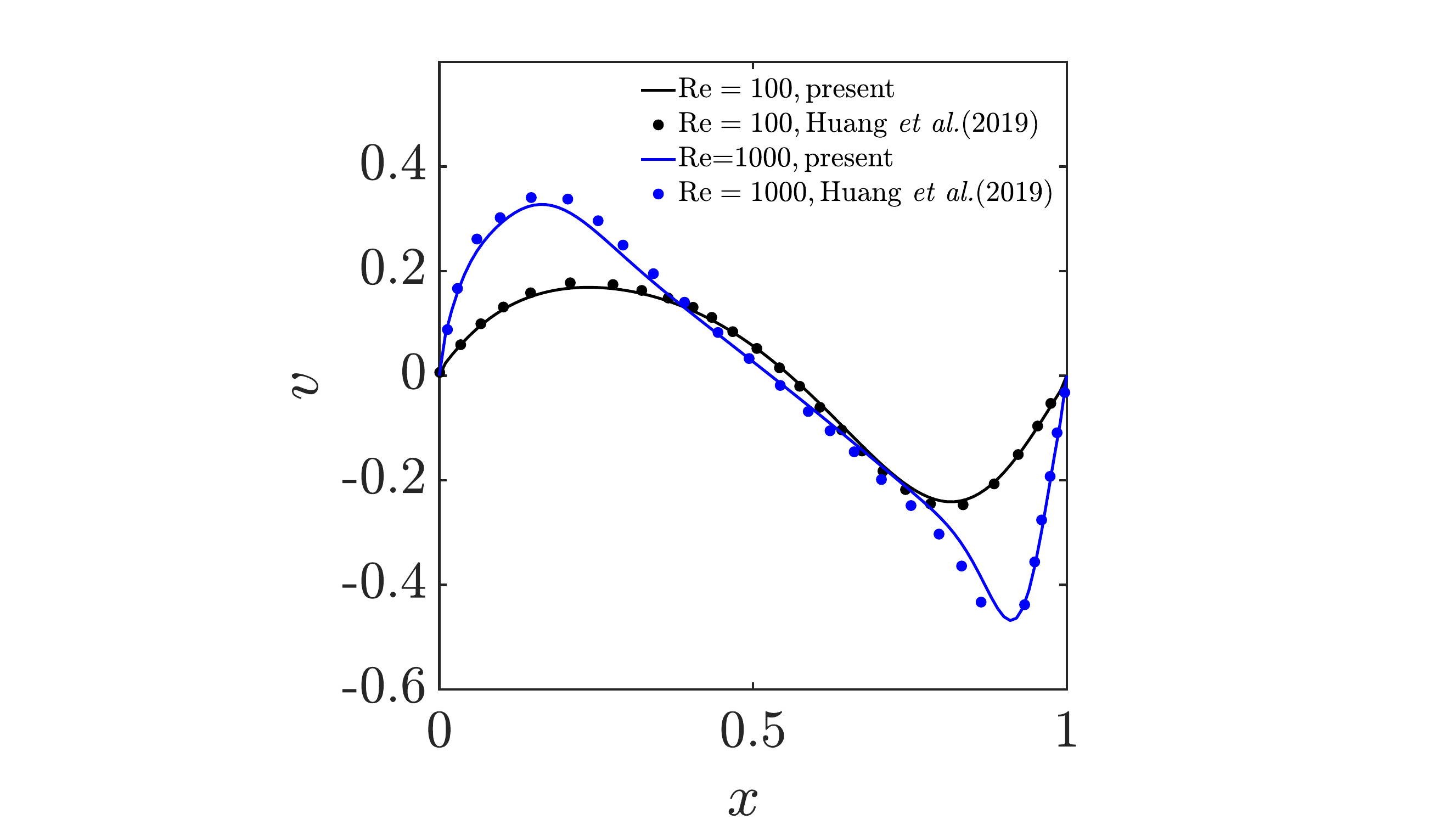}}
\caption{(\textit{a}) Velocity streamlines of the lid-driven cavity test case for [left]~$\mathrm{Re}=100$ and [right]~$\mathrm{Re}=1000$ cases, at $t=10$ and $t=50$, respectively. Calculated results of the lid-driven cavity test case for (\textit{b}) $u$ along the vertical line passing through the center, and (\textit{c}) $v$ along the horizontal line passing through the center, for $\mathrm{Re}=100$ (black) and $\mathrm{Re}=1000$ (blue) cases, at $t=10$ and $t=50$, respectively. Results obtained from this study and the study by Huang \textit{et al.}~\cite{huang2019} are plotted using a solid line and a circle symbol, respectively.}
\label{fig:CavityTest}
\end{figure}

\section{\label{appendixE} Formation of spurious currents: static droplet test}

In this test case, a two-dimensional droplet is considered in a stationary velocity field, $\mathbf{u}=0$. By setting the velocity field to zero and ignoring gravity, the equilibrium solution of the Navier--Stokes equation will result in $\Delta p = \mathbf{F}_\sigma = \sigma \kappa$, which satisfies the pressure jump condition, $[p] = \sigma \kappa$. This equilibrium condition is the well-known Laplace's relation between pressure and surface tension forces of a droplet in an equilibrium condition~\citep{popinet2018}. Therefore, the droplet is expected to remain at rest since the pressure force balances the capillary force. If the numerical solver cannot accurately calculate the curvature and, hence, fails to recover the equilibrium solution, quasi-stationary velocity patterns known as parasitic or spurious currents will appear in the solution~\citep{boniou2022}. Thus, this test case has been widely used by researchers to investigate the capability of their solver to properly balance pressure and surface tension forces across the interface and its accuracy in computing the interface curvature~\citep{boniou2022}. Additionally, the viscosity discontinuity between two flows results in discontinuity of the velocity derivative across the interface, which is more pronounced for larger viscosity ratios. This discontinuity can cause numerical errors in the divergence calculation of the velocity field, which leads to a source of error for pressure~\citep{huang2019}. The inaccuracy in pressure calculation will generate errors when balancing forces across the interface and leads to the formation of unphysical spurious currents. Therefore, the spurious velocity formation is also investigated for different viscosity ratios across the interface in this benchmark. 

In the initial condition, a circle of diameter ${D}=0.4$ is considered at the center of the two-dimensional computational domain $[0, 1] \times [0, 1]$, where a no-slip boundary condition is imposed at all boundaries. The thickness of the interface is set to $\epsilon=\Delta x/2$. In order to study the magnitude of the spurious flow that will be generated due to the inaccuracies while computing the interface curvature, the proposed test case by  Desjardins and Pitsch~\cite{desjardins2009} is followed. In their test case, the surface tension coefficient is set to $\sigma=1$, with the same viscosity of $\mu=0.1$ for both fluids and a constant density ratio of one, i.e., $\rho_1=\rho_2$. The Laplace number, represented by $\mathrm{La}=\sigma \rho D/\mu^2$, is then altered by changing the density of both fluids to examine the performance of the solver for various ratios of surface tension and viscous forces. The intensity of the unwanted spurious currents appearing in the solution can be quantified by measuring the capillary number, $\mathrm{Ca}=|u_\mathrm{max}|\mu/\sigma$, for different Laplace numbers. The mesh resolution is set to $32 \times 32$, and the calculated capillary numbers at a non-dimensional time $t \sigma/\left(\mu D\right) = 250$ for five different Laplace numbers are reported in Table~\ref{tab:table2}. For each Laplace number, the time step should be determined according to the CFL constraint introduced in Eq.~(\ref{eq:cfl}). According to the calculated capillary numbers, the magnitude of spurious currents, even after a long simulation time, is minimal and independent of the Laplace number, confirming the well-balanced results between pressure and surface tension forces across the interface in the solver. It is worth mentioning that the order of calculated capillary numbers for different Laplace numbers is close to the ones obtained by Desjardins and Pitsch~\cite{desjardins2009}, showing that the magnitude of the spurious errors is consistent with other studies in the literature.

\begin{table*}
\centering
\caption{\label{tab:table2} Calculated Ca values for different La numbers for the static droplet test case on the mesh resolution of $32 \times 32$, with density and viscosity ratios of one.}
\begin{tabular}{ccc}
\hline
\hline
$\rho_1 (\rho_2)$ & La & Ca  \\ \hline
$0.3$ & $12$ & $5.029 \times 10^{-5}$ \\
$3$ & $120$ & $3.378 \times 10^{-5}$  \\
$30$ & $1200$ & $3.612 \times 10^{-5}$  \\
$300$ & $12000$ & $9.428 \times 10^{-5}$ \\
$3000$ & $120000$ & $2.657 \times 10^{-4}$ \\
\hline
\hline
\end{tabular}
\end{table*}

In order to investigate the influence of the shear stress discontinuity across the interface due to the viscosity jump on the formation of spurious velocity, a benchmark akin to that of Huang \textit{et al.}~\cite{huang2019} is also adopted. In this test case, a circle with the diameter of ${D}=2$ in a computational domain $[0, 8] \times [0, 8]$ with the grid resolution of $101 \times 101$ is considered. For the initial condition of $\sigma=730$, $\rho_1=\rho_2=1$, and $\mu_1=0.001$, the calculated $L_2$ and $L_\infty$ error norms of the velocity field after $1$ and $1000$ time steps are given in Table~\ref{tab:table3} for four different viscosity ratios of $1$, $10$, $100$, and $1000$, along with the inviscid case. It is appreciated from Table~\ref{tab:table3} that even for large viscosity ratios, the intensity of spurious currents is small and is close to the inviscid case. In other words, the strength of the parasite current is independent of the viscosity jump, and the solver accurately balances pressure and surface tension forces across the interface in the presence of a shear stress jump. Also, the calculated $L_2$ and $L_\infty$ errors for both cases, after 1 and 1000 time steps, are in the same order as those reported by Huang \textit{et al.} study~\cite{huang2019}.

\begin{table*}
\centering
\caption{\label{tab:table3} Calculated $L_2$ and $L_\infty$ norms of the velocity field for four different viscosity ratios along with the inviscid case for the static droplet test case.}
\begin{tabular}{ccccc}
\hline
\hline
 &\multicolumn{2}{c}{at $1^\mathrm{st}$ time step}&\multicolumn{2}{c}{at $1000^\mathrm{th}$ time step}\\
 $\mu_2/\mu_1$&$L_2$&$L_\infty$&$L_2$&$L_\infty$\\ \hline
 0&$2.198\times 10^{-5}$&$2.286\times 10^{-4}$&$1.534\times 10^{-3}$&$7.919\times 10^{-3}$ \\
 1&$6.608\times 10^{-6}$&$6.903\times 10^{-5}$&$4.619\times 10^{-4}$&$2.522\times 10^{-3}$\\
 10&$6.608\times 10^{-6}$&$6.903\times 10^{-5}$&$4.618\times 10^{-4}$&$2.521\times 10^{-3}$\\
 100&$6.608\times 10^{-6}$&$6.903\times 10^{-5}$&$4.613\times 10^{-4}$&$2.509\times 10^{-3}$\\
 1000&$6.608\times 10^{-6}$&$6.903\times 10^{-5}$&$4.562\times 10^{-4}$&$2.394\times 10^{-3}$\\
 \hline
 \hline
\end{tabular}
\end{table*}


\bibliographystyle{elsarticle-num.bst}  
\bibliography{main}

\end{document}